\documentclass{article}

\usepackage{arxiv}

\usepackage[utf8]{inputenc} % allow utf-8 input
\usepackage[T1]{fontenc}    % use 8-bit T1 fonts
\usepackage{hyperref}       % hyperlinks
\usepackage{url}            % simple URL typesetting
\usepackage{booktabs}       % professional-quality tables
\usepackage{array}
\usepackage{amsfonts}       % blackboard math symbols
\usepackage{nicefrac}       % compact symbols for 1/2, etc.
\usepackage{graphicx}
\usepackage{verbatim}
\usepackage{subcaption}% for subfigs
\usepackage{adjustbox}
\usepackage[numbers,sort&compress]{natbib}
\usepackage{doi}
\usepackage{amsmath}
\usepackage{amssymb}
\usepackage{float}
\usepackage{algorithm}
\usepackage{algpseudocode}
\usepackage[section]{placeins} % \FloatBarrier at each \section
\usepackage{etoolbox}
\pretocmd{\subsection}{\FloatBarrier}{}{} % also barrier at each \subsection: floats stay within their subsection
\pretocmd{\subsubsection}{\FloatBarrier}{}{} % also barrier at each \subsubsection: floats stay within their subsubsection
\usepackage{xcolor}
\usepackage{mathtools}
\usepackage{fancyhdr}
\usepackage{amsthm}
\usepackage{cleveref}
\usepackage{bm}
\usepackage[mathscr]{euscript}
\usepackage{textcomp}
\usepackage{stmaryrd}

\DeclareMathAlphabet{\pazocal}{OMS}{zplm}{m}{n}

\usepackage{lscape}		
\usepackage{fancyvrb}

\newtheoremstyle{remarkstyle}  % Name
  {5pt}                        % Space above
  {5pt}                        % Space below
  {}                           % Body font (upright)
  {}                           % Indent amount
  {\bfseries}                  % Theorem head font (bold)
  {.}                          % Punctuation after theorem head
  { }                          % Space after theorem head
  {}                           % Theorem head spec (empty = `normal`)
\newtheorem{theorem}{Theorem}[section]

\theoremstyle{remarkstyle}
\newtheorem{remark}[theorem]{Remark}

\crefname{figure}{Fig.}{Figs.}
\Crefname{figure}{Figure}{Figures}
\crefname{algorithm}{Algorithm}{Algorithms}
\crefname{equation}{Eq.}{Eqs.}
\crefname{section}{Section}{Sections}
\crefname{subsection}{Section}{Sections}
\crefname{subsubsection}{Section}{Sections}
\crefname{appendix}{Appendix}{Appendices}
\crefname{subappendix}{Appendix}{Appendices}

\usepackage[normalem]{ulem}

\title{
A Differentiable Framework for Inverse Identification of Heterogeneous Cohesive Laws
}

\author{
Sida Hao$^{1}$, 
Jinkyo Han$^{1}$, 
Bahador Bahmani$^{1,2}$\thanks{Corresponding author: \texttt{bahador.bahmani@northwestern.edu}} \\
\\
$^{1}$Department of Mechanical Engineering, Northwestern University, Evanston, IL 60208, USA \\
$^{2}$Theoretical and Applied Mechanics, Northwestern University, Evanston, IL 60208, USA
}

\renewcommand{\shorttitle}{Inverse Identification of Heterogeneous Cohesive Laws}

\hypersetup{
    colorlinks=true,      % Enable colored links
    linkcolor=green,      % Color for internal links
    urlcolor=blue,        % Color for URLs
    citecolor=blue        % Color for citations
}

\date{}

\begin{document}
\maketitle

\vspace{-10pt}

\begin{abstract}
This work develops a differentiable finite element framework based on an intrinsic cohesive-zone formulation for identifying traction--separation laws from global force--displacement histories and spatially sparse strain measurements. The framework employs physics-constrained neural networks to represent flexible, thermodynamically structured cohesive laws, with sensitivities computed through automatic differentiation and a discrete adjoint that accounts for the history-dependent evolution of interfacial damage. To enable identification of spatially heterogeneous cohesive behavior, we extend learned constitutive priors to irreversible interfacial damage by constructing a shared, low-dimensional family of cohesive laws from representative traction--separation histories. This representation exploits common constitutive structure and parameter sharing to reduce the number of unknowns and regularize the inverse problem, while allowing local cohesive behavior to vary spatially through latent coordinates. Numerical investigations involving double-cantilever beams and fiber--matrix composites demonstrate that accurate reproduction of global force--displacement responses does not necessarily imply accurate recovery of the underlying cohesive laws. Incorporating spatially resolved strain measurements substantially improves constitutive identification, even when the admissible responses are constrained by a learned prior. Comparisons with conventional bilinear cohesive models and Bayesian optimization further illustrate the benefits of the learned constitutive representation and adjoint-based identification for high-dimensional inverse problems.
\end{abstract}


\section{Introduction}

The local traction--separation response of an interface is rarely accessible to direct experimental measurement, even though it governs fracture and debonding in bonded structures, composites, and heterogeneous materials.
Instead, experiments typically provide structural-scale observations, such as force--displacement histories and deformation fields, from which the underlying interfacial constitutive behavior must be inferred.
Cohesive zone models (CZMs) represent this behavior through traction--separation laws that describe the transfer of forces between separating surfaces \citep{Dugdale1960,Barenblatt1962,hillerborg1976analysis,park2011}.
Identifying these laws from indirect structural measurements is therefore essential for developing predictive models of interfacial failure.

The information available for cohesive law identification depends strongly on the specimen geometry, loading conditions, and size of the cohesive process zone.
When the process zone is small relative to the relevant structural dimensions, the global response may be governed predominantly by the fracture energy, with limited sensitivity to the detailed shape of the traction--separation law \citep{bao1992remarks,alfano2006influence}.
In contrast, under large-scale bridging, the distribution of cohesive tractions can significantly influence the macroscopic response \citep{sills2015cohesive,Jensen2019}.
In the former case, different cohesive laws may produce nearly indistinguishable force--displacement responses, making the recovery of their local strength and softening behavior potentially severely ill-posed \citep{Jensen2019}.
Spatially resolved deformation measurements, particularly those obtained using digital image correlation (DIC), provide complementary information about local deformation and load transfer that may improve constitutive identification \citep{fedele2009identification,richefeu2012toward,ruybalid2018mixed,becker2024extracting}.
Nevertheless, the extent to which global and full-field observations can constrain the complete cohesive response remains dependent on the mechanical problem and the admissible constitutive representation.

Several approaches have been developed to identify cohesive laws from structural measurements.
For specific experimental configurations, direct extraction methods based on the $J$-integral or displacement fields can recover cohesive behavior \citep{sorensen1998large,sorensen2003determination,wei2022deep,Wei2026}.
More generally, finite element model updating (FEMU) identifies constitutive parameters by minimizing discrepancies between measured structural responses and finite element predictions.
Derivative-free optimization methods, including genetic algorithms, Nelder--Mead optimization, and Bayesian inference, have been widely employed for this purpose \citep{xu2014inverse,Shen2010,albuquerque2018bayesian,tang2025bayesian}.
However, their computational cost can increase substantially with the number of unknown parameters.
Gradient-based approaches offer an alternative by exploiting sensitivities of the structural response to the constitutive parameters.
In particular, adjoint sensitivity analysis enables the gradient of a scalar objective to be computed with a number of adjoint solves independent of the number of unknowns \citep{oberai2003,bendsoe1988generating,arora1977design}, making it particularly attractive for high-dimensional inverse problems.
Increasingly combined with automatic differentiation, adjoint methods have enabled efficient inverse identification and optimization in computational mechanics \citep{thakolkaran2022nn,xue2023jax,han2024topology,chandrasekhar2021auto,pundir2026versatile}.
For cohesive interfaces, such an approach must additionally account for the nonlinear, history-dependent evolution of interfacial damage and its influence on subsequent structural responses.

Beyond optimization efficiency, inverse identification also depends on the assumed constitutive representation.
Conventional cohesive models generally employ prescribed analytical traction--separation laws whose parameters control quantities such as strength, fracture energy, and softening behavior.
While effective for many applications, these representations may not capture more complex constitutive responses.
Recent developments in neural cohesive modeling have provided greater flexibility in representing interfacial behavior \citep{wang2019meta,wei2023data,tao2024reconstruction, gavris2026discovering}.
In particular, our previous work \citep{hao2026data} introduced a neural cohesive formulation within the generalized standard materials (GSM) framework, in which the elastic energy and damage resistance are represented by physics-constrained networks.
While that work focused on learning cohesive behavior from prescribed local traction--separation histories, the present study addresses the distinct problem of identifying such behavior from indirect structural observations.

The inverse problem becomes particularly underdetermined when cohesive behavior varies spatially, as in bonded and architected interfaces where local variations in adhesion influence load transfer, crack propagation, and effective toughness \citep{li2020spatial,hernandez2017toughness,fulco2026tailoring}.
In principle, each interface segment could be assigned an independently parameterized cohesive law, but the number of unknowns would then grow with both the number of segments and the complexity of the constitutive representation.
More importantly, such an approach neglects the potential relationships among cohesive responses belonging to a given material system.
Variations in cohesive strength, characteristic separation, softening behavior, and mode dependence may be correlated through common material characteristics and failure mechanisms.
Consequently, the physically plausible cohesive laws may occupy only a restricted subset of the broader space of thermodynamically admissible constitutive responses.
Restricting the inverse search space to such a family offers a natural form of data-driven regularization, incorporating constitutive information that may not be recoverable from structural observations alone.

This perspective is closely related to parameter sharing and representation learning in machine learning, where common structures are learned across related tasks or instances \citep{caruana1997multitask,park2019deepsdf}.
Similarly, learned generative priors have been used to regularize inverse problems by restricting admissible solutions to the range of a trained model \citep{bora2017compressed}.
Building on these ideas, our previous work introduced constitutive priors for inverse design, in which a low-dimensional family of nonlinear elastic responses was learned from representative material data and used to optimize spatially varying constitutive behavior \citep{han2027constitutive}.
Here, we extend this concept to irreversible cohesive damage by learning a shared representation of the elastic energy and damage resistance from representative traction--separation histories.
The common network parameters capture the shared constitutive structure and relationships among the reference responses, while low-dimensional latent coordinates distinguish individual cohesive laws.
This representation exploits parameter sharing to reduce the number of unknowns while restricting the admissible constitutive responses without imposing spatial uniformity.

In this work, we develop a differentiable inverse identification framework that integrates the GSM-based neural cohesive formulation with discrete adjoint sensitivity analysis and learned constitutive priors.
We investigate two complementary settings.
For homogeneous interfaces, the neural constitutive parameters are identified directly from structural observations without prescribing an analytical traction--separation law.
For heterogeneous interfaces, the constitutive family is learned offline, after which its shared network parameters remain fixed and only the element-wise latent coordinates are identified.
The discrete adjoint accounts for the history-dependent evolution of interfacial damage, enabling gradient-based identification from global force--displacement records, spatially resolved strain measurements, or their combination.

We investigate the framework through numerical examples involving double-cantilever-beam (DCB) and fiber--matrix composite specimens.
The homogeneous examples examine the influence of cohesive process-zone size and mixed-mode loading on the recovery of traction--separation laws.
The heterogeneous examples assess identification within a learned constitutive family and compare the contributions of global force and spatial strain observations to local cohesive law recovery.
Further comparisons with conventional bilinear cohesive models and Bayesian optimization investigate the roles of constitutive representation and differentiable optimization.
All investigations use synthetic structural observations, with the heterogeneous reference laws generated within the learned constitutive family.

The remainder of this paper is organized as follows.
\Cref{sec:czm_priors} presents the inverse formulation, constitutive prior, differentiable finite element solver, and discrete adjoint sensitivity analysis.
\Cref{sec:examples} presents the numerical investigations, and \cref{sec:conclusions} summarizes the main findings.

\section{
%Prior-Informed Inverse Identification of Cohesive Laws
Differentiable Inverse Identification of Cohesive Laws
}
\label{sec:czm_priors}

This section formulates the identification of spatially heterogeneous cohesive
behavior as a PDE-constrained inverse problem. We first state the unknown
constitutive behavior, the available observations, and the role of a learned
constitutive prior. We then define the interface kinematics and forward
boundary-value problem, introduce a
thermodynamically admissible family of cohesive laws within the generalized
standard material (GSM) framework, construct a low-dimensional constitutive
prior from reference traction--separation histories, and describe its structure-preserving neural representation. Finally, we present the discrete finite-element formulation and derive the corresponding discrete adjoint used
to identify the latent cohesive coordinates.

\subsection{
Inverse Problem Formulation
}
\label{sec:inverse_problem_general}

\begin{figure}[htbp]
    \centering
    \includegraphics[width=0.60\textwidth]{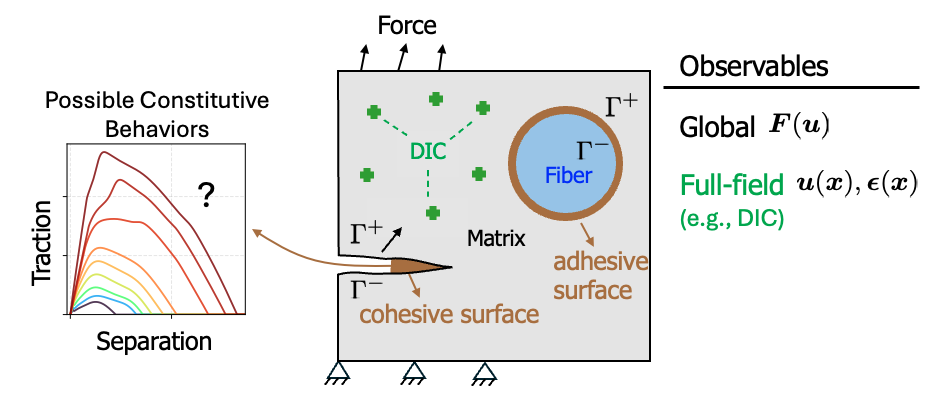}
    \caption{
    Schematic of the inverse cohesive law identification problem. The cohesive
    response along an internal fracture surface or a bonded material interface
    is unknown and may vary spatially along the interface. The local
    traction--separation behavior is inferred from observations of the
    structural response, which may include global force--displacement data and
    spatially resolved displacement or strain fields obtained from digital
    image correlation (DIC).
    }
    \label{fig:overview}
\end{figure}

The objective of this work is to identify spatially varying, history-dependent cohesive laws from indirect measurements of the structural response.
The specimen geometry, bulk elastic properties, loading conditions, measurement configuration, and cohesive surface $\Gamma_{\mathrm c}$ are assumed known.
As illustrated in \cref{fig:overview}, $\Gamma_{\mathrm c}$ may represent a potential fracture surface within a solid or a bonded interface between dissimilar materials.
The unknown cohesive laws characterize the local traction--separation behavior along this prescribed surface.

To represent spatial variability, the cohesive surface is divided into
$N_{\mathrm c}$ segments. Let $\mathcal A_{\mathrm{coh}}$ denote the
admissible family of history-dependent cohesive laws. Each segment is
assigned a local law,
\begin{equation}
    \mathcal T_e\in\mathcal A_{\mathrm{coh}},
    \qquad e=1,\ldots,N_{\mathrm c}.
\label{eq:local_law_membership}
\end{equation}
The unknown is therefore the spatial collection of constitutive functions
\begin{equation}
    \boldsymbol{\mathcal T}
    \coloneqq
    \left(
        \mathcal T_1,\ldots,\mathcal T_{N_{\mathrm c}}
    \right)
    \in
    \mathcal A_{\mathrm{coh}}^{N_{\mathrm c}}.
\label{eq:cohesive_law_collection}
\end{equation}
A homogeneous interface is recovered when
$\mathcal T_1=\cdots=\mathcal T_{N_{\mathrm c}}$.

The available data comprise $N_{\mathrm{exp}}$ experiments or loading paths.
For each experiment $j$, the loading program $\boldsymbol\ell_j$ and the
observations $\boldsymbol y_j$ are known. These observations may include
reaction-force histories under prescribed displacement, spatially resolved
displacement or strain fields from digital image correlation (DIC), or both.
Reaction forces constrain the overall structural response, whereas full-field
measurements retain information about the spatial distribution of deformation.

For a prescribed loading program $\boldsymbol\ell$, let $\boldsymbol S$
denote the complete forward state history, including the displacement field
and all constitutive internal variables. The equilibrium equations, boundary
conditions, and constitutive evolution laws are written abstractly as
\begin{equation}
    \mathcal F
    \left(
        \boldsymbol S,
        \boldsymbol{\mathcal T};
        \boldsymbol\ell
    \right)
    =
    \boldsymbol 0.
\label{eq:general_residual}
\end{equation}
An observation operator $\mathcal O_j$ maps the state history and cohesive
laws to the measured quantities. The functional inverse problem is
\begin{equation}
    \begin{aligned}
        \min_{\substack{
            \boldsymbol{\mathcal T}\in\mathcal A_{\mathrm{coh}}^{N_{\mathrm c}},
            \,\{\boldsymbol S_j\}_{j=1}^{N_{\mathrm{exp}}}
        }}
        \quad&
        \sum_{j=1}^{N_{\mathrm{exp}}}
        \mathcal L_j
        \left(
            \mathcal O_j(\boldsymbol S_j,\boldsymbol{\mathcal T}),
            \boldsymbol y_j
        \right)
        \\
        \text{subject to}\quad&
        \mathcal F
        \left(
            \boldsymbol S_j,\boldsymbol{\mathcal T};
            \boldsymbol\ell_j
        \right)
        =\boldsymbol 0,
        \qquad j=1,\ldots,N_{\mathrm{exp}},
    \end{aligned}
\label{eq:functional_inverse_problem}
\end{equation}
where $\mathcal L_j$ measures the discrepancy between predicted and observed
responses. The explicit dependence of $\mathcal O_j$ on
$\boldsymbol{\mathcal T}$ accommodates reaction forces, which depend on the
constitutive response even at a fixed state.

The inverse problem is nonlinear and history-dependent and can be severely ill-posed \cite{Jensen2019}, as different spatial distributions of cohesive laws may produce nearly indistinguishable structural observations.
Although combining global force--displacement measurements with spatially resolved deformation fields provides additional information, even rich observations may be insufficient to uniquely identify the underlying local constitutive behavior.

To further regularize the inverse problem, we adopt the constitutive-prior concept introduced in our previous work \citep{han2027constitutive}.
Rather than allowing arbitrary cohesive laws at each interface location, we learn a low-dimensional family of thermodynamically admissible constitutive responses offline from representative traction--separation histories.
During inverse identification, the learned constitutive model remains fixed, and each interface segment is assigned a latent coordinate that selects its local cohesive law.
This restricts the admissible solution space to constitutive variations represented by the learned family, substantially reducing the number of unknowns while retaining the ability to describe spatially heterogeneous interfacial behavior.

The following subsections present the forward boundary-value problem and generalized-standard-material cohesive formulation, introduce the learned constitutive prior and differentiable finite element solver, and derive the discrete inverse problem and adjoint sensitivities.
The section concludes with the offline--online identification framework.

\subsection{Forward Boundary-Value Problem}
\label{sec:forward_mechanics}

For a prescribed loading history and spatial distribution of cohesive laws, the forward problem couples bulk deformation with the history-dependent traction response along the cohesive interface $\Gamma_{\mathrm c}$.
We consider a two-dimensional, quasi-static formulation under small-strain kinematics.

Spatial heterogeneity is represented by partitioning the cohesive interface into $N_{\mathrm c}$ segments with pairwise disjoint relative interiors,
\begin{equation}
    \Gamma_{\mathrm c}
    =
    \bigcup_{e=1}^{N_{\mathrm c}} \Gamma_{\mathrm c}^{e},
    \qquad
    \operatorname{ri}(\Gamma_{\mathrm c}^{e})
    \cap
    \operatorname{ri}(\Gamma_{\mathrm c}^{e'})
    =
    \emptyset
    \quad (e\neq e').
\label{eq:cohesive_partition}
\end{equation}
Let $\boldsymbol e_n$ and $\boldsymbol e_t$ denote the local unit normal and tangent, with $\boldsymbol e_n$ directed from the $-$ side to the $+$ side of the interface.
The displacement jump is defined as $    \llbracket \boldsymbol u \rrbracket
    \coloneqq
    \boldsymbol u^{+}-\boldsymbol u^{-},$
where $\boldsymbol u^{+}$ and $\boldsymbol u^{-}$ denote the displacement traces on the opposing interface faces.
The local normal and tangential separations are obtained through
\begin{equation}
    \boldsymbol\delta
    \coloneqq
        [\delta_n, \delta_t]^{\top}
    =
    \boldsymbol Q
    \llbracket \boldsymbol u \rrbracket,
    \qquad
\boldsymbol Q \coloneqq
[
    \boldsymbol e_n, \boldsymbol e_t
]^{\top}.
\label{eq:local_separation}
\end{equation}
The corresponding cohesive traction vector in the local and global coordinate systems is
\begin{equation}
    \boldsymbol t
    \coloneqq
    [t_n, t_t]^\top,
    \qquad
    \boldsymbol t_{\mathrm c}
    =
    \boldsymbol Q^{\top}\boldsymbol t
    =
    t_n\boldsymbol e_n+t_t\boldsymbol e_t.
\label{eq:cohesive_components}
\end{equation}
The cohesive law $\mathcal T_e$ assigned to segment $\Gamma_{\mathrm c}^{e}$ relates the local separation history to the current traction,
\begin{equation}
    \boldsymbol t(\boldsymbol X,\tau)
    =
    \mathcal T_e
    \left[
        \left\{
            \boldsymbol\delta(\boldsymbol X,s)
        \right\}_{0\leq s\leq\tau}
    \right],
    \qquad
    \boldsymbol X\in\Gamma_{\mathrm c}^{e},
\label{eq:abstract_local_cohesive_law}
\end{equation}
where $\tau$ parameterizes the loading history.

The bulk occupies the reference domain $\Omega\subset\mathbb R^2$, with the external boundary partitioned into Dirichlet and Neumann portions,
\begin{equation}
    \overline{\partial\Omega}
    =
    \overline{\Gamma_u}\cup\overline{\Gamma_t},
    \qquad
    \Gamma_u\cap\Gamma_t=\emptyset.
\label{eq:boundary_partition}
\end{equation}
The prescribed displacement and traction are denoted by $\bar{\boldsymbol u}(\tau)$ and $\bar{\boldsymbol t}(\tau)$, respectively, and $\boldsymbol b(\tau)$ denotes the body force.
The complete loading program is represented by $\boldsymbol\ell$.

Under small-strain linear elasticity, the bulk strain and stress satisfy
\begin{equation}
    \boldsymbol\varepsilon(\boldsymbol u)
    =
    \nabla^{s}\boldsymbol u
    =
    \frac{1}{2}
    \left[
        \nabla\boldsymbol u
        +(\nabla\boldsymbol u)^{\top}
    \right],
    \qquad
    \boldsymbol\sigma
    =
    \mathbb C:\boldsymbol\varepsilon,
\label{eq:bulk_elasticity}
\end{equation}
where $\mathbb C$ is the known bulk elasticity tensor, specialized to isotropic plane-strain elasticity in the numerical examples.

At each loading step, the governing equilibrium equations and boundary conditions are
\begin{align}
    \nabla\cdot\boldsymbol\sigma+\boldsymbol b
    &=
    \boldsymbol 0
    &&\text{in }\Omega\setminus\Gamma_{\mathrm c},
\nonumber\\
    \boldsymbol u
    &=
    \bar{\boldsymbol u}
    &&\text{on }\Gamma_u,
\nonumber\\
    \boldsymbol\sigma\boldsymbol n
    &=
    \bar{\boldsymbol t}
    &&\text{on }\Gamma_t,
\nonumber\\
    \boldsymbol\sigma^{-}\boldsymbol e_n
    &=
    \boldsymbol t_{\mathrm c},
    \qquad
    \boldsymbol\sigma^{+}(-\boldsymbol e_n)
    =
    -\boldsymbol t_{\mathrm c}
    &&\text{on }\Gamma_{\mathrm c},
\label{eq:strong_form}
\end{align}
where $\boldsymbol n$ is the outward unit normal on the external boundary and $\boldsymbol\sigma^{\pm}$ denote the limiting stresses on the two sides of the cohesive interface.
Equivalently, traction equilibrium across the interface requires
\begin{equation}
    \boldsymbol\sigma^{-}\boldsymbol e_n
    =
    \boldsymbol\sigma^{+}\boldsymbol e_n
    =
    \boldsymbol t_{\mathrm c},
\label{eq:interface_traction_equilibrium}
\end{equation}
where the cohesive traction is determined by the local constitutive law in \cref{eq:abstract_local_cohesive_law}.

For a fixed $\tau$, the admissible displacement and test spaces are
\begin{align}
    \mathcal U(\tau)
    &\coloneqq
    \left\{
        \boldsymbol u
        \in
        [H^1(\Omega\setminus\Gamma_{\mathrm c})]^2
        :
        \boldsymbol u=\bar{\boldsymbol u}(\tau)
        \text{ on }\Gamma_u
    \right\},
\nonumber\\
    \mathcal V_0
    &\coloneqq
    \left\{
        \boldsymbol v
        \in
        [H^1(\Omega\setminus\Gamma_{\mathrm c})]^2
        :
        \boldsymbol v=\boldsymbol 0
        \text{ on }\Gamma_u
    \right\}.
\label{eq:trial_test_spaces}
\end{align}
The weak form seeks $\boldsymbol u(\tau)\in\mathcal U(\tau)$ such that
\begin{align}
    \mathscr R
    \left(
        \boldsymbol u,
        \boldsymbol{\mathcal T};
        \boldsymbol v,
        \tau
    \right)
    \coloneqq{}&
    \int_{\Omega\setminus\Gamma_{\mathrm c}}
        \boldsymbol\sigma(\boldsymbol u)
        :
        \boldsymbol\varepsilon(\boldsymbol v)
        \,\mathrm dV
\nonumber
    +
    \int_{\Gamma_{\mathrm c}}
        \boldsymbol t_{\mathrm c}
        \cdot
        \llbracket\boldsymbol v\rrbracket
        \,\mathrm dS
\nonumber
    -
    \int_{\Omega\setminus\Gamma_{\mathrm c}}
        \boldsymbol b
        \cdot
        \boldsymbol v
        \,\mathrm dV
    -
    \int_{\Gamma_t}
        \bar{\boldsymbol t}
        \cdot
        \boldsymbol v
        \,\mathrm dS
    =
    0
\label{eq:weak_form_residual}
\end{align}
for every $\boldsymbol v\in\mathcal V_0$.
Since the bulk properties are known, the dependence of the forward problem on the unknown cohesive laws enters through the interfacial virtual-work term.

\subsection{Thermodynamically Admissible Cohesive Laws}
\label{sec:CZM-GSM}

We adopt the generalized-standard-material (GSM) cohesive formulation developed in our previous work \citep{hao2026data}, in which the traction--separation response is governed by an elastic energy potential and irreversible damage evolution by a convex dissipation potential.
Here, we summarize the constitutive relations needed to construct the learned cohesive family and formulate the inverse problem.

We introduce the active separation vector
\begin{equation}
    \boldsymbol\delta_{\mathrm a}
    \coloneqq
    (\langle\delta_n\rangle_+,|\delta_t|)^\top,
    \qquad
    \langle x\rangle_+\coloneqq\max(x,0),
    \label{eq:active_separation}
\end{equation}
which separates damageable opening and sliding from compressive contact.
With a scalar damage variable $d\in[0,1]$, the Helmholtz free energy per unit reference area is
\begin{equation}
    \psi(\boldsymbol\delta,d)
    =
    (1-d)\psi_{\mathrm a}^{\mathrm{el}}(\boldsymbol\delta_{\mathrm a})
    +
    \frac12 k_{\mathrm{pen}}\langle-\delta_n\rangle_+^2,
    \label{eq:psi_split}
\end{equation}
where $\psi_{\mathrm a}^{\mathrm{el}}$ is the active elastic surface energy and $k_{\mathrm{pen}}>0$ is the compressive contact penalty stiffness.
The contact contribution remains unaffected by damage, while $d=0$ and $d=1$ correspond to intact and fully decohered interfaces, respectively.
The cohesive traction and thermodynamic force conjugate to damage follow as
\begin{align}
    \boldsymbol t
    &=
    \frac{\partial\psi}{\partial\boldsymbol\delta}
    =
    (1-d)\nabla_{\boldsymbol\delta}
    \psi_{\mathrm a}^{\mathrm{el}}(\boldsymbol\delta_{\mathrm a})
    -
    k_{\mathrm{pen}}\langle-\delta_n\rangle_+(1,0)^\top,
    \label{eq:traction_state_law}\\
    Y
    &=
    -\frac{\partial\psi}{\partial d}
    =
    \psi_{\mathrm a}^{\mathrm{el}}(\boldsymbol\delta_{\mathrm a}).
    \label{eq:damage_driving_force}
\end{align}
The active energy is normalized such that $\psi_{\mathrm a}^{\mathrm{el}}(\boldsymbol 0)=0$.
We characterize the mode mixity by
\begin{equation}
    \beta(\boldsymbol\delta)
    =
    \frac{|\delta_t|}
    {\langle\delta_n\rangle_++|\delta_t|}
    \in[0,1],
    \label{eq:mode_mixity}
\end{equation}
with $\beta=0$ for vanishing active separation, pure opening corresponding to $\beta=0$, and pure sliding to $\beta=1$.

Following the GSM formulation, irreversible damage evolution is governed by the rate-independent dissipation potential
\begin{equation}
    \varphi(\dot d,d,\beta)
    =
    r(d,\beta)\dot d
    +
    I_{\mathcal K(d)}(\dot d),
    \label{eq:dissipation_potential}
\end{equation}
where $r(d,\beta)\geq0$ denotes the damage resistance and $I_{\mathcal K(d)}$ is the indicator function of the admissible damage-rate set.
Here, $\mathcal K(d)=[0,\infty)$ for $0\leq d<1$ and $\mathcal K(1)=\{0\}$.
The generalized normality relation yields the Karush--Kuhn--Tucker conditions
\begin{equation}
    Y-r(d,\beta)\leq0,
    \qquad
    \dot d\geq0,
    \qquad
    \dot d\,[Y-r(d,\beta)]=0,
    \label{eq:kkt_conditions}
\end{equation}
for $d<1$, ensuring non-negative dissipation and irreversible damage evolution.

For a continuous, strictly increasing resistance function at fixed $\beta$, the active consistency relation can be expressed through its inverse,
\begin{equation}
    R(Y,\beta)\coloneqq r^{-1}(Y,\beta),
    \label{eq:inverse_resistance}
\end{equation}
extended by saturation at $0$ and $1$ outside the resistance range.
The damage history for an initially intact interface is then obtained as
\begin{equation}
    d(\tau)
    =
    \max_{0\leq s\leq\tau}R(Y(s),\beta(s)),
    \label{eq:damage_history}
\end{equation}
which enforces irreversibility without requiring a local nonlinear damage solve.

Consequently, a cohesive law within this GSM framework is characterized by the pair
\begin{equation}
    \mathcal T
    \longleftrightarrow
    \left(
        \psi_{\mathrm a}^{\mathrm{el}},R
    \right),
    \label{eq:CZM-general}
\end{equation}
which determines its elastic response, damage evolution, and traction--separation behavior.
For complete decohesion at fixed mode mixity, the corresponding fracture energy is
\begin{equation}
    G_{\mathrm c}(\beta)
    =
    \int_0^1 r(d,\beta)\,\mathrm d d.
    \label{eq:Gc}
\end{equation}
This representation accommodates variations in cohesive strength, characteristic separation, fracture energy, and softening behavior without prescribing a particular analytical traction--separation law.

\subsection{Learned Constitutive Prior}
\label{sec:priors}

Building on the constitutive-prior framework of \citet{han2027constitutive}, we learn a low-dimensional family of cohesive laws from representative traction--separation histories.
Rather than identifying an independent constitutive function at each interface location, the learned family restricts the admissible responses to those represented by the reference data.
Here, the term \emph{prior} denotes a restricted constitutive model class rather than a Bayesian probability distribution.

\subsubsection{Latent Representation of Cohesive Laws}

Each cohesive law is parameterized by a low-dimensional latent coordinate $    \boldsymbol z\in\mathcal Z\subset\mathbb R^{n_z},$
which conditions both the active elastic energy and the inverse damage resistance,
\begin{equation}
    \psi_{\mathrm a}^{\mathrm{el}}
    =
    \mathcal M_{\psi}
    (\boldsymbol\delta_{\mathrm a};
    \boldsymbol z,\boldsymbol\theta_{\psi}),
    \qquad
    R
    =
    \mathcal M_R
    (Y,\beta;\boldsymbol z,\boldsymbol\theta_R).
    \label{eq:prior_family_train}
\end{equation}
The network parameters $\boldsymbol\theta_{\psi}$ and $\boldsymbol\theta_R$ are shared across the family, while $\boldsymbol z$ selects an individual cohesive law.
The resulting admissible family is
\begin{equation}
    \mathcal A_{\mathcal P}
    \coloneqq
    \left\{
        \mathcal T_{\boldsymbol z}
        \mid \boldsymbol z\in\mathcal Z
    \right\}
    \subseteq\mathcal A_{\mathrm{coh}}.
    \label{eq:prior_family}
\end{equation}
The latent coordinate does not represent individual material parameters but instead characterizes the complete constitutive response, allowing cohesive strength, separation scales, softening shape, fracture energy, and mode dependence to vary jointly.

\subsubsection{Structure-Preserving Neural Constitutive Models}
\label{sec:nn_representation}

Following \citet{hao2026data}, the active elastic energy is represented by a partially input-convex neural network (PICNN), while the inverse damage resistance is represented by a monotone neural network (MNN) \citep{kim2024scalable}.
For each fixed $\boldsymbol z$, the PICNN is convex with respect to $\boldsymbol\delta_{\mathrm a}$, whereas the MNN is non-decreasing in the damage-driving force $Y$.
The dependence on $\boldsymbol z$ remains unrestricted, allowing the constitutive response to vary without relaxing these constraints.

The energy representation satisfies
\begin{equation}
    \psi_{\mathrm a}^{\mathrm{el}}(\boldsymbol 0;\boldsymbol z)=0,
    \qquad
    \nabla_{\boldsymbol\delta_{\mathrm a}}
    \psi_{\mathrm a}^{\mathrm{el}}(\boldsymbol 0;\boldsymbol z)
    =\boldsymbol 0,
    \label{eq:picnn_normalization}
\end{equation}
ensuring zero stored energy and active traction in the undeformed state.
We consider both coupled and separable energy representations, with the latter expressed as
\begin{equation}
    \psi_{\mathrm a}^{\mathrm{el}}
    (\boldsymbol\delta_{\mathrm a};\boldsymbol z)
    =
    \psi_n(\langle\delta_n\rangle_+;\boldsymbol z)
    +
    \psi_t(|\delta_t|;\boldsymbol z).
    \label{eq:psi_decoupled}
\end{equation}
Even in the separable representation, mixed-mode coupling is introduced through $R(Y,\beta;\boldsymbol z)$.
The smoothing of the active separation measures used in the coupled representation is described in \cref{app:fe_smoothing}.

The inverse-resistance network satisfies the ideal constitutive conditions
\begin{equation}
    R(0,\beta;\boldsymbol z)=0,
    \qquad
    0\leq R(Y,\beta;\boldsymbol z)\leq1.
    \label{eq:mnn_constraints}
\end{equation}
When additional knowledge of mode dependence is available, we employ a more restrictive full-MNN model in which $R$ is also non-increasing in $\beta$ at fixed $Y$.
This imposes a non-decreasing damage resistance and fracture energy with increasing mode mixity.
The numerical regularization near damage onset is detailed in \cref{app:hyperparams}.

\subsubsection{Offline Training of the Constitutive Prior}

The constitutive family is learned offline by jointly optimizing the shared network parameters and a latent coordinate assigned to each reference cohesive behavior.
Let $\mathcal P=\{\mathcal P_m\}_{m=1}^{N_m}$ denote a library of $N_m$ reference behaviors, where each $\mathcal P_m$ contains $N_p^{(m)}$ ordered traction--separation histories.
These histories may include Mode~I, Mode~II, mixed-mode, unloading, and reloading responses.
For loading path $q$ of reference behavior $m$, the measured separations and tractions at increment $i$ are denoted by $(\boldsymbol\delta_{qi}^{(m)},\boldsymbol t_{qi}^{(m)})$, with $i=1,\ldots,N_s^{(m,q)}$.

For each prescribed separation history, the incremental constitutive update produces the predicted tractions $\widehat{\boldsymbol t}_{qi}^{(m)}$.
The normalized training loss is defined as
\begin{align}
    \mathcal L_{\mathrm{prior}}
    &=
    \frac{1}{N_m}
    \sum_{m=1}^{N_m}
    \frac{1}{N_p^{(m)}}
    \sum_{q=1}^{N_p^{(m)}}
    \frac{1}{N_s^{(m,q)}}
    \sum_{i=1}^{N_s^{(m,q)}}
    \left\|
        \frac{
            \widehat{\boldsymbol t}_{qi}^{(m)}
            -
            \boldsymbol t_{qi}^{(m)}
        }{t_{mq}^{\mathrm{ref}}}
    \right\|_2^2,
    \label{eq:meta_loss_T_discrete}\\
    t_{mq}^{\mathrm{ref}}
    &=
    \max_i\|\boldsymbol t_{qi}^{(m)}\|_2>0.
    \label{eq:prior_traction_scale}
\end{align}
Denoting the shared network parameters by $\boldsymbol\Theta=(\boldsymbol\theta_\psi,\boldsymbol\theta_R)$, the offline optimization problem is
\begin{equation}
    \left(
        \boldsymbol\Theta^\star,
        \{\boldsymbol z_m^\star\}_{m=1}^{N_m}
    \right)
    =
    \operatorname*{arg\,min}_{
        \boldsymbol\Theta,\,
        \{\boldsymbol z_m\in\mathcal Z\}_{m=1}^{N_m}
    }
    \left[
        \mathcal L_{\mathrm{prior}}
        +
        \frac{1}{2\sigma_z^2N_m}
        \sum_{m=1}^{N_m}
        \|\boldsymbol z_m\|_2^2
    \right].
    \label{eq:meta_loss_gauge}
\end{equation}
The quadratic regularization controls the scale of the latent embedding during joint training \citep{park2019deepsdf}.

Once training is complete, the shared parameters $\boldsymbol\Theta^\star$ remain fixed during structural inverse identification.
Each cohesive segment is then assigned an independent latent coordinate that selects its local constitutive law,
\begin{equation}
    \mathcal T_e=\mathcal T_{\boldsymbol z_e},
    \qquad
    \boldsymbol z_e\in\mathcal Z,
    \qquad
    e=1,\ldots,N_{\mathrm c}.
    \label{eq:segment_latent_law}
\end{equation}
The unknown coordinates are collected into
\begin{equation}
    \boldsymbol Z
    =
    (\boldsymbol z_1^\top,\ldots,
    \boldsymbol z_{N_{\mathrm c}}^\top)^\top
    \in\mathcal Z^{N_{\mathrm c}}
    \subseteq\mathbb R^{N_{\mathrm c}n_z}.
    \label{eq:stacked_latent_coordinates}
\end{equation}
The inverse problem is therefore reduced from identifying a spatial collection of constitutive functions to determining $N_{\mathrm c}n_z$ latent variables within the learned family.
Unlike conventional parameterized cohesive models, these coordinates select complete traction--separation laws rather than independently adjusting a prescribed set of material parameters.

\subsection{Differentiable Finite Element Solver}
\label{sec:fe_solver}

We adopt a discretize-then-optimize approach in which the equilibrium equations and constitutive evolution are discretized before computing sensitivities through a discrete adjoint.

The weak form is discretized using a standard Bubnov--Galerkin finite element formulation, with four-node bilinear quadrilateral elements for the bulk and four-node zero-thickness cohesive elements along $\Gamma_{\mathrm c}$.
Each cohesive element corresponds to one constitutive segment $\Gamma_{\mathrm c}^{e}$.
The discrete bulk strain and local cohesive separation are given by
\begin{equation}
    \boldsymbol\varepsilon^h=\boldsymbol B_{\mathrm b}\boldsymbol U,
    \qquad
    \boldsymbol\delta^h=\boldsymbol B_{\mathrm c}\boldsymbol U,
    \label{eq:fe_kinematics}
\end{equation}
where $\boldsymbol U$ is the global nodal displacement vector and
\begin{equation}
    \boldsymbol B_{\mathrm b}=\nabla^s\boldsymbol N,
    \qquad
    \boldsymbol B_{\mathrm c}=\boldsymbol Q(\boldsymbol N_{\mathrm c}^{+}-\boldsymbol N_{\mathrm c}^{-}),
    \label{eq:bulk_and_cohesive_B_matrices}
\end{equation}
where $\boldsymbol N$ indicates bulk shape functions, $\nabla^s$ is the symmetric gradient operator, and $\boldsymbol N_{\mathrm c}^{\pm}$ are the local shape functions of the cohesive interface.
The element operators, quadrature rules, and assembly procedure are detailed in \cref{app:fe_operators}.

At load step $n$, let $\boldsymbol U_n$ denote the global displacement vector, $\boldsymbol d_n$ the damage states at all cohesive quadrature points, and $\boldsymbol Z$ the unknown constitutive coordinates.
The assembled internal force is
\begin{equation}
    \boldsymbol f_{\mathrm{int}}(\boldsymbol U_n,\boldsymbol d_n;\boldsymbol Z)
    =\boldsymbol K_{\mathrm b}\boldsymbol U_n
    +\sum_{e=1}^{N_{\mathrm c}}\int_{\Gamma_{\mathrm c}^{e}}
    \boldsymbol B_{\mathrm c}^{\top}
    \boldsymbol t(\boldsymbol\delta_n^h,d_n;\boldsymbol z_e)\,\mathrm dS,
    \label{eq:discrete_fint}
\end{equation}
where $\boldsymbol K_{\mathrm b}=\int_{\Omega\setminus\Gamma_{\mathrm c}}\boldsymbol B_{\mathrm b}^{\top}\boldsymbol C\boldsymbol B_{\mathrm b}\,\mathrm dV$ is the bulk stiffness matrix, assembled once for the linear-elastic bulk.
The external force vector is
\begin{equation}
    \boldsymbol f_{\mathrm{ext},n}
    =\int_{\Omega\setminus\Gamma_{\mathrm c}}\boldsymbol N^\top\boldsymbol b_n\,\mathrm dV
    +\int_{\Gamma_t}\boldsymbol N^\top\bar{\boldsymbol t}_n\,\mathrm dS,
    \label{eq:discrete_fext}
\end{equation}
and the equilibrium residual is
\begin{equation}
    \boldsymbol f_n^{\mathrm{res}}(\boldsymbol U_n,\boldsymbol d_n;\boldsymbol Z)
    =\boldsymbol f_{\mathrm{ext},n}-\boldsymbol f_{\mathrm{int}}(\boldsymbol U_n,\boldsymbol d_n;\boldsymbol Z).
    \label{eq:discrete_residual_definition}
\end{equation}
The equilibrium equations and their tangents are restricted to the free displacement degrees of freedom, while reaction forces are obtained from the unreduced nodal force balance.

The loading history is advanced incrementally, retaining the accumulated damage at each cohesive quadrature point.
At step $n$, the rate-independent damage candidate follows from \cref{eq:damage_history} as
\begin{equation}
    d_n^{\mathrm{RI}}
    =\max\{d_{n-1},R(Y_n,\beta_n;\boldsymbol z_e)\}.
    \label{eq:damage_target}
\end{equation}
To regularize abrupt damage evolution during softening, we introduce the incremental update
\begin{equation}
    d_n=d_{n-1}+\rho_n(d_n^{\mathrm{RI}}-d_{n-1}),
    \qquad
    \rho_n=\frac{\Delta\tau_n}{\Delta\tau_n+\tau_{\mathrm v}},
    \label{eq:viscous_damage_main}
\end{equation}
where $\Delta\tau_n=\tau_n-\tau_{n-1}>0$ is the pseudo-time increment and $\tau_{\mathrm v}>0$ controls the numerical damage relaxation.
Collecting the updates at all cohesive quadrature points gives
\begin{equation}
    \boldsymbol d_n=\widehat{\boldsymbol d}_n(\boldsymbol U_n,\boldsymbol d_{n-1};\boldsymbol Z).
    \label{eq:damage_update_vector}
\end{equation}

To improve numerical stability during softening, a pseudo-viscous term is also introduced into the reduced equilibrium residual,
\begin{equation}
    \widetilde{\boldsymbol f}_n^{\mathrm{res}}
    =\boldsymbol f_n^{\mathrm{res}}(\boldsymbol U_n,\widehat{\boldsymbol d}_n;\boldsymbol Z)
    -\gamma_n(\boldsymbol U_n-\boldsymbol U_{n-1}),
    \qquad
    \gamma_n=\frac{c_u}{\Delta\tau_n},
    \label{eq:reduced_residual}
\end{equation}
where $c_u\geq0$ is the displacement-stabilization coefficient and $\widehat{\boldsymbol d}_n$ depends on the current displacement and previous damage state through \cref{eq:damage_update_vector}.
The converged displacement satisfies
\begin{equation}
    \widetilde{\boldsymbol f}_n^{\mathrm{res}}=\boldsymbol 0,
    \label{eq:step_equilibrium}
\end{equation}
after which the converged damage state is obtained from \cref{eq:damage_update_vector}.
Both stabilization mechanisms are retained in the discrete equations used for adjoint differentiation.

The nonlinear equilibrium equations are solved using Newton--Raphson iteration with a consistent tangent that accounts for the updated damage state.
Defining the damage-driving-force gradient and the traction tangent at fixed damage as
\begin{equation}
    \boldsymbol g_n=\nabla_{\boldsymbol\delta}Y_n,
    \qquad
    \boldsymbol D_n^{\mathrm{el}}=\left.\frac{\partial\boldsymbol t}{\partial\boldsymbol\delta}\right|_{d=d_n},
    \label{eq:fixed_damage_tangent}
\end{equation}
the derivative of the damage candidate on the active branch is
\begin{equation}
    \nabla_{\boldsymbol\delta}d_n^{\mathrm{RI}}
    =\frac{\partial R}{\partial Y}\boldsymbol g_n
    +\frac{\partial R}{\partial\beta}\nabla_{\boldsymbol\delta}\beta_n,
    \label{eq:damage_tangent_main}
\end{equation}
and vanishes on the inactive branch.
The viscous update gives
\begin{equation}
    \nabla_{\boldsymbol\delta}d_n
    =\rho_n\nabla_{\boldsymbol\delta}d_n^{\mathrm{RI}},
    \label{eq:viscous_damage_tangent}
\end{equation}
leading to the algorithmic cohesive tangent
\begin{equation}
    \boldsymbol D_n^{\mathrm{alg}}
     \coloneqq \mathrm d\boldsymbol t/\mathrm d\boldsymbol\delta
    =\boldsymbol D_n^{\mathrm{el}}
    -\boldsymbol g_n\otimes\nabla_{\boldsymbol\delta}d_n.
    \label{eq:cohesive_tangent_main}
\end{equation}
The mode-mixity dependence generally renders $\boldsymbol D_n^{\mathrm{alg}}$ nonsymmetric.
The derivatives are evaluated using automatic differentiation, with piecewise derivatives adopted at damage-branch transitions.

At Newton iteration $k$, the displacement increment is obtained from
\begin{equation}
    (\boldsymbol K_n^{(k)}+\eta_n^{(k)}\boldsymbol I)\Delta\boldsymbol U_n^{(k)}
    =\widetilde{\boldsymbol f}_n^{\mathrm{res},(k)},
    \qquad
    \boldsymbol U_n^{(k+1)}=\boldsymbol U_n^{(k)}+\Delta\boldsymbol U_n^{(k)},
    \label{eq:newton_step}
\end{equation}
where
\begin{equation}
    \boldsymbol K_n^{(k)}
    =-\left.\frac{\partial\widetilde{\boldsymbol f}_n^{\mathrm{res}}}{\partial\boldsymbol U_n}\right|_{\boldsymbol U_n=\boldsymbol U_n^{(k)}}.
    \label{eq:consistent_tangent}
\end{equation}
The non-negative shift $\eta_n^{(k)}$ is introduced when needed to improve the conditioning of the Newton system.
Because this shift does not modify the converged residual, it is excluded from the adjoint operator.
Adaptive load-step cutbacks are employed when necessary.

\subsection{Discrete Inverse Problem and Adjoint Sensitivities}
\label{sec:inverse_identification}

With the constitutive prior fixed, inverse identification determines the latent coordinates $\boldsymbol Z$ by minimizing the discrepancy between predicted and measured structural responses.
We formulate the inverse problem using the discrete loading histories and derive the corresponding adjoint sensitivities, including the effects of damage evolution and numerical stabilization.

\subsubsection{Observation Model and Optimization Problem}

For experiment $j$ with loading program $\boldsymbol\ell_j$ and $N_j$ load increments, the converged state history is
\begin{equation}
    \boldsymbol S_j(\boldsymbol Z)
    =(\{\boldsymbol U_{j,n}\}_{n=0}^{N_j},\{\boldsymbol d_{j,n}\}_{n=0}^{N_j}).
    \label{eq:discrete_state_history}
\end{equation}
Let $\mathcal O_{\mathrm G,j}$ and $\mathcal O_{\mathrm F,j}$ denote the global force and full-field observation operators, respectively.
The predicted observations are
\begin{equation}
    \widehat{\boldsymbol y}^{\,c}_j(\boldsymbol Z)
    =\mathcal O_{c,j}(\boldsymbol S_j(\boldsymbol Z),\boldsymbol Z),
    \qquad c\in\{\mathrm G,\mathrm F\},
    \label{eq:observation_operator_general}
\end{equation}
where the explicit dependence on $\boldsymbol Z$ accommodates reaction forces that depend directly on the constitutive response.
The full-field operator extracts the displacement or strain measurements at the prescribed sampling locations.

For $N_{\mathrm{exp}}$ experiments, the observation dataset is
\begin{equation}
    \mathcal D_{\mathrm{obs}}
    =\{(\boldsymbol\ell_j,\boldsymbol y_j^{\,\mathrm G},\boldsymbol y_j^{\,\mathrm F})\}_{j=1}^{N_{\mathrm{exp}}}.
    \label{eq:dataset}
\end{equation}
The latent coordinates are identified by minimizing
\begin{equation}
    \min_{\boldsymbol Z\in\mathcal Z^{N_{\mathrm c}}}\mathcal J(\boldsymbol Z),
    \qquad
    \mathcal J(\boldsymbol Z)
    =\sum_{j=1}^{N_{\mathrm{exp}}}\sum_{c\in\{\mathrm G,\mathrm F\}}w_c\mathcal L_{c,j}(\boldsymbol Z),
    \label{eq:general_min_obj}
\end{equation}
subject to the incremental forward constraints
\begin{equation}
    \widetilde{\boldsymbol f}_{j,n}^{\mathrm{res}}=\boldsymbol 0,
    \qquad
    \boldsymbol d_{j,n}=\widehat{\boldsymbol d}_{j,n},
    \qquad
    n=1,\ldots,N_j,\quad j=1,\ldots,N_{\mathrm{exp}}.
    \label{eq:general_min_res}
\end{equation}
The channel weights $w_{\mathrm G},w_{\mathrm F}\geq0$ control the contributions of the global and full-field observations.
For channel $c$, the normalized misfit is
\begin{equation}
    \mathcal L_{c,j}(\boldsymbol Z)
    =\frac{1}{N_jM_c}\sum_{n=1}^{N_j}\sum_{k=1}^{M_c}
    \left(\frac{\widehat y^{\,c}_{j,n,k}(\boldsymbol Z)-y^{\,c}_{j,n,k}}{s_{c,j}}\right)^2,
    \label{eq:channel_misfit}
\end{equation}
where $M_c$ is the number of scalar observations of channel $c$ recorded at each load increment, indexed by $k$, and $s_{c,j}=\max_{n,k}|y^{\,c}_{j,n,k}|>0$ normalizes the observations of each channel and experiment.
For the global channel, the observations are reaction forces, whereas the full-field channel contains the strain components $(\varepsilon_{xx},\varepsilon_{yy},\gamma_{xy})$ at the selected bulk Gauss points.
Either channel can be used independently by setting the weight of the other to zero.
The normalization and weighting settings are detailed in \cref{app:hyperparams}.

\subsubsection{Discrete Adjoint Sensitivity}

The discrete adjoint evaluates the gradient of $\mathcal J$ with respect to all latent coordinates without computing a separate forward sensitivity for each unknown.
We derive the adjoint for one experiment with $N_{\mathrm{step}}$ increments, omitting the experiment index for clarity.
The initial displacement and damage states are prescribed.

To simplify notation, let $\boldsymbol r_n$ denote the reduced equilibrium residual $\widetilde{\boldsymbol f}_n^{\mathrm{res}}$ and $\boldsymbol h_n$ the damage-update function $\widehat{\boldsymbol d}_n$, with their arguments defined in \cref{eq:damage_update_vector,eq:reduced_residual}.
The forward constraints become
\begin{equation}
    \boldsymbol r_n(\boldsymbol U_n,\boldsymbol U_{n-1},\boldsymbol d_{n-1};\boldsymbol Z)=\boldsymbol 0,
    \label{eq:history_equilibrium}
\end{equation}
\begin{equation}
    \boldsymbol d_n=\boldsymbol h_n(\boldsymbol U_n,\boldsymbol d_{n-1};\boldsymbol Z),
    \qquad n=1,\ldots,N_{\mathrm{step}}.
    \label{eq:history_damage}
\end{equation}
Introducing the equilibrium adjoints $\boldsymbol\lambda_n$ and damage costates $\boldsymbol\mu_n$, the discrete Lagrangian is
\begin{equation}
    \mathscr L
    =\mathcal J+\sum_{n=1}^{N_{\mathrm{step}}}
    \left[\boldsymbol\lambda_n^\top\boldsymbol r_n
    +\boldsymbol\mu_n^\top(\boldsymbol h_n-\boldsymbol d_n)\right].
    \label{eq:adjoint_lagrangian}
\end{equation}
The constraint terms vanish along the converged forward solution, giving $\mathscr L=\mathcal J$.

Stationarity with respect to the displacement state $\boldsymbol U_n$ yields
\begin{equation}
    \boldsymbol K_n^\top\boldsymbol\lambda_n
    =\nabla_{\boldsymbol U_n}\mathcal J
    +(\partial_{\boldsymbol U_n}\boldsymbol h_n)^\top\boldsymbol\mu_n
    +(\partial_{\boldsymbol U_n}\boldsymbol r_{n+1})^\top\boldsymbol\lambda_{n+1},
    \label{eq:adjoint_system}
\end{equation}
where $\boldsymbol K_n=-\partial_{\boldsymbol U_n}\boldsymbol r_n$ is the converged tangent and $\partial_{\boldsymbol U_n}$ denotes the corresponding Jacobian.
The last term accounts for coupling between consecutive displacement states through the stabilization term and vanishes when $c_u=0$.
Stationarity with respect to the damage state gives
\begin{equation}
    \boldsymbol\mu_n
    =\nabla_{\boldsymbol d_n}\mathcal J
    +(\partial_{\boldsymbol d_n}\boldsymbol r_{n+1})^\top\boldsymbol\lambda_{n+1}
    +(\partial_{\boldsymbol d_n}\boldsymbol h_{n+1})^\top\boldsymbol\mu_{n+1}.
    \label{eq:damage_costate}
\end{equation}
Terms indexed by $N_{\mathrm{step}}+1$ are omitted.
For purely kinematic observations, $\nabla_{\boldsymbol d_n}\mathcal J=\boldsymbol 0$, whereas reaction-force observations may depend explicitly on the current damage state.
The adjoint variables are evaluated backward through the loading history.

The total gradient with respect to the latent coordinates is
\begin{equation}
    \frac{\mathrm d\mathcal J}{\mathrm d\boldsymbol Z}
    =\nabla_{\boldsymbol Z}\mathcal J
    +\sum_{n=1}^{N_{\mathrm{step}}}
    \left[(\partial_{\boldsymbol Z}\boldsymbol r_n)^\top\boldsymbol\lambda_n
    +(\partial_{\boldsymbol Z}\boldsymbol h_n)^\top\boldsymbol\mu_n\right].
    \label{eq:total_gradient}
\end{equation}
The explicit derivative $\nabla_{\boldsymbol Z}\mathcal J$ accounts for observables that depend directly on the constitutive parameters at fixed state.
For multiple experiments, the corresponding gradient contributions are summed.

All derivatives are evaluated using automatic differentiation.
The backward adjoint sweep requires one linear solve with the transpose of the converged tangent per load step, with the number of such solves independent of the number of latent coordinates.
The resulting gradient is used in Adam optimization followed by L-BFGS refinement.
The same adjoint formulation also applies to direct identification of neural constitutive parameters by replacing $\boldsymbol Z$ with the corresponding network parameters.

\subsection{Offline--Online Identification Framework}
\label{sec:offline_online_algorithm}

\Cref{alg:prior_inverse_framework} summarizes the separation between offline constitutive-prior training and online inverse identification.
The reference constitutive histories are used only to learn the admissible family, while the subsequent identification uses structural observations of the target specimen with the pretrained networks fixed.
No direct measurements of interfacial tractions or separations are required.

\begin{algorithm}[tbp]
    \caption{Offline constitutive-prior training and online inverse identification}
    \label{alg:prior_inverse_framework}
    \small
    \algrenewcommand\algorithmicindent{1.2em}
    \begin{algorithmic}[1]
        \Statex \textbf{Offline input:} Reference constitutive library $\mathcal P$.
        \Statex \textbf{Online input:} Specimen model and observations $\mathcal D_{\mathrm{obs}}$.
        \Statex \textbf{Output:} Identified latent field $\boldsymbol Z^\star$ and cohesive laws $\{\mathcal T_{\boldsymbol z_e^\star}\}_{e=1}^{N_{\mathrm c}}$.
        \Statex \textbf{Offline training}
        \State Jointly optimize $\boldsymbol\Theta$ and $\{\boldsymbol z_m\}$ using \cref{eq:meta_loss_gauge}.
        \State Fix the trained constitutive parameters $\boldsymbol\Theta^\star$.
        \Statex \textbf{Online identification}
        \State Initialize $\boldsymbol Z\in\mathcal Z^{N_{\mathrm c}}$.
        \Repeat
            \State Solve the incremental forward problem for $\boldsymbol Z$.
            \State Evaluate $\mathcal J(\boldsymbol Z)$ and its adjoint gradient $\mathrm d\mathcal J/\mathrm d\boldsymbol Z$.
            \State Update $\boldsymbol Z$ to reduce $\mathcal J$ subject to $\boldsymbol Z\in\mathcal Z^{N_{\mathrm c}}$.
        \Until{the stopping criterion or iteration budget is reached}
        \State \Return $\boldsymbol Z^\star$ and $\{\mathcal T_{\boldsymbol z_e^\star}\}_{e=1}^{N_{\mathrm c}}$.
    \end{algorithmic}
\end{algorithm}

\section{Numerical Studies}
\label{sec:examples}

We investigate the identification of cohesive laws from synthetic measurements of double-cantilever-beam (DCB) and fiber--matrix composite specimens.
Three measurement configurations are considered, consisting of force--displacement data alone (``force-only''), bulk strain fields representative of digital image correlation (DIC) measurements alone (``DIC-only''), and their combination (``force+DIC'').
These comparisons assess whether either data source alone is sufficient for accurate identification and whether, and to what extent, combining them improves the recovered cohesive laws.

For the DIC-only configuration, force measurements are withheld during identification, allowing the predicted force--displacement response to serve as an independent validation.
In all configurations, the cohesive laws are identified without direct measurements of interfacial tractions or separations.

The separable active-energy representation in \cref{eq:psi_decoupled} is used in all examples except the homogeneous composite specimen in \cref{sec:composite}, for which the coupled representation is adopted.
All examples employ the full-MNN architecture described in \cref{sec:nn_representation}, which enforces monotonicity with respect to the damage-driving force and incorporates an inductive bias favoring higher fracture toughness with increasing mode mixity.

The data-driven constitutive prior is trained offline using Adam \citep{diederik2014adam} to jointly optimize the network parameters and the latent coordinates associated with the reference cohesive laws.
During heterogeneous inverse identification, the trained network parameters are held fixed.
The inverse problems are solved using adjoint-based gradients, with Adam employed for initial optimization followed by L-BFGS \citep{liu1989limited} for refinement.

We first examine homogeneous interfaces using the neural cohesive model described in \cref{sec:examples:homo}.
A Mode~I DCB test and a composite specimen subjected to two loading paths assess cohesive law recovery under opening and mixed-mode loading, respectively.
The DCB study compares two configurations to investigate how the relative size of the cohesive process zone affects identifiability across small- and large-scale bridging regimes, and whether strain-field measurements improve identification when the force response is weakly sensitive to the cohesive law.

We next consider spatially heterogeneous interfaces using the data-driven constitutive prior described in \cref{sec:meta_data_driven}.
With the pretrained network parameters held fixed, inverse identification is performed by optimizing the latent variables.
A DCB specimen and a composite specimen containing three embedded fibers are used to assess the recovery of spatially varying cohesive laws.
Comparisons with a bilinear cohesive model with spatially varying parameters and Bayesian optimization evaluate the roles of spatial heterogeneity, the learned constitutive prior, and adjoint-based optimization, respectively.

After inverse identification, the accuracy of the recovered cohesive laws is assessed by comparing the identified and reference traction--separation responses.
Although the cohesive law generally depends on both normal and tangential separations, we simplify the comparison by evaluating the traction error along the pure-opening and pure-sliding paths.
For pure opening, the normal traction is evaluated at $(\delta_n,\delta_t)=(s,0)$, whereas for pure sliding, the tangential traction is evaluated at $(\delta_n,\delta_t)=(0,s)$.
The relative traction error at each cohesive element is then defined as
\begin{equation}
    \varepsilon^2 =
    \frac{\|\widehat{\boldsymbol{t}}_n-\boldsymbol{t}_n\|_2^2}
    {2N_n\|\boldsymbol{t}_n\|_\infty^2}
    +
    \frac{\|\widehat{\boldsymbol{t}}_t-\boldsymbol{t}_t\|_2^2}
    {2N_t\|\boldsymbol{t}_t\|_\infty^2},
    \label{eq:rel_traction_error}
\end{equation}
where $\boldsymbol{t}_n$ and $\boldsymbol{t}_t$ are the reference traction vectors sampled at $N_n$ and $N_t$ separation points along the respective paths, and the corresponding identified tractions are denoted by $\widehat{\boldsymbol{t}}_n$ and $\widehat{\boldsymbol{t}}_t$.
The errors are normalized by the reference peak tractions and reported as percentages.
For the homogeneous DCB specimens, only the pure-opening contribution is considered scaled by 2.

Hyperparameter settings and implementation details are provided in \cref{app:hyperparams}.

\subsection{Direct Identification of Homogeneous Cohesive Laws}
\label{sec:examples:homo}

We first investigate the identification of a single cohesive law shared along the interface by directly optimizing the neural constitutive parameters, without a pretrained prior.
The DCB examples examine how cohesive process-zone size influences identification under Mode~I loading, while the subsequent composite example extends the investigation to mixed-mode loading along a curved fiber--matrix interface.

\subsubsection{Double Cantilever Beam with Different Process-Zone Sizes}
\label{sec:precrack}

\begin{figure}[htbp]
    \centering
    \begin{subfigure}[b]{0.57\textwidth}
        \centering
        \includegraphics[width=\linewidth]{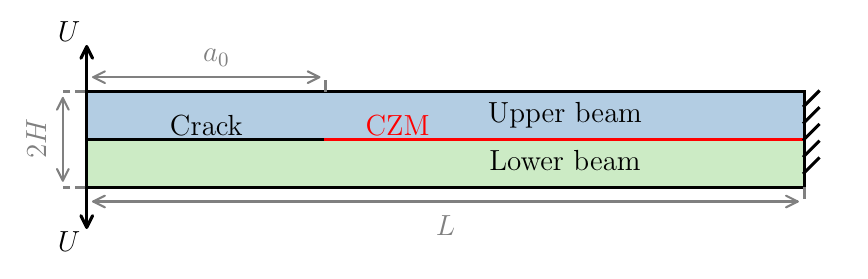}
        \caption{}
        \label{fig:precrack_structure}
    \end{subfigure}
    \begin{subfigure}[b]{0.25\textwidth}
        \centering
        \includegraphics[width=\linewidth]{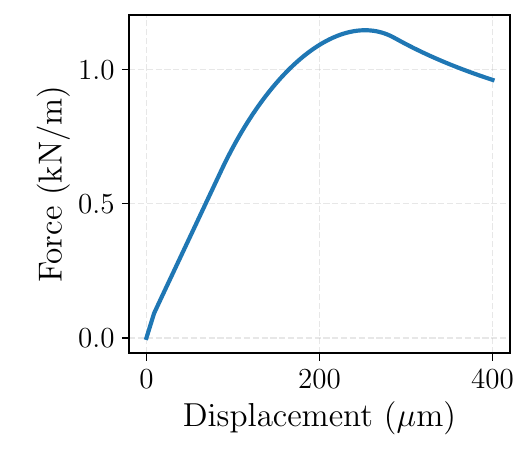}
        \caption{}
        \label{fig:precrack_forward_fd}
    \end{subfigure}
    \caption{DCB specimen: (a) geometry and boundary conditions and (b) reference force--displacement response for the longer process-zone configuration.
    Force is reported per unit out-of-plane thickness.}
    \label{fig:precrack_setup}
\end{figure}

We consider a double cantilever beam (DCB) consisting of two identical beams of length $L=30$~mm and height $H=1$~mm, joined along a cohesive interface with an initial crack of length $a_0=10$~mm, as illustrated in \cref{fig:precrack_structure}.
The right ends are clamped, while equal and opposite vertical displacements $U$ are applied at the left ends.
The bulk material is isotropic and linear elastic, with $E=100$~GPa and $\nu=0.3$, and is discretized using a $240\times8$ structured mesh of plane-strain quadrilateral elements and 160 cohesive elements along the bonded interface.

To investigate the influence of cohesive process-zone size on inverse identification, we consider two bilinear reference laws.
The first has a fracture energy $G_{Ic}=45$~J/m$^2$ and peak traction $\sigma_n^0=0.5$~MPa, with an applied displacement reaching $U=400$~\textmu m.
The second has $G_{Ic}=15$~J/m$^2$ and $\sigma_n^0=10$~MPa, with $U$ reaching $150$~\textmu m.
Both laws have an initiation-to-failure separation ratio $\delta_n^0/\delta_n^f=0.2$.
The reference force--displacement response for the first configuration is shown in \cref{fig:precrack_forward_fd}, exhibiting an initial elastic branch followed by a force reduction as debonding progresses.

\begin{figure}[htbp]
    \centering
    \begin{subfigure}[b]{0.35\textwidth}
        \centering
        \includegraphics[width=\linewidth]{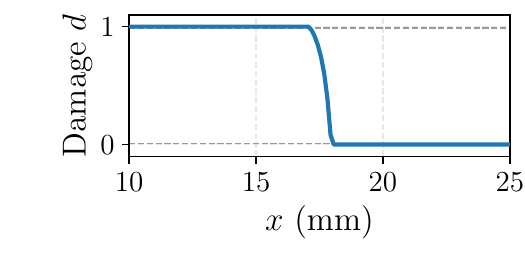}
        \caption{SSB}
        \label{fig:precrack_damage_ssb}
    \end{subfigure}
    \begin{subfigure}[b]{0.35\textwidth}
        \centering
        \includegraphics[width=\linewidth]{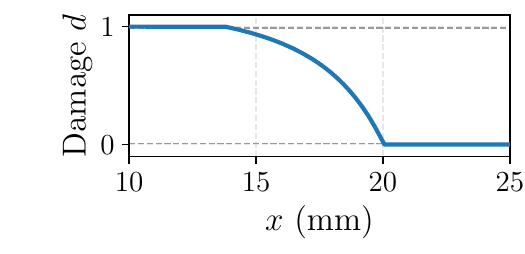}
        \caption{LSB}
        \label{fig:precrack_damage_lsb}
    \end{subfigure}
    \caption{Final interfacial damage distributions for the (a) SSB and (b) LSB configurations.
    The interface coordinate $x$ is measured from the left end, with the profiles beginning at the initial crack tip $x=a_0$.}
    \label{fig:precrack_damage}
\end{figure}

The final damage distributions in \cref{fig:precrack_damage} indicate cohesive process-zone lengths of approximately $l_{\mathrm{cz}}=0.88$ for the second law and $6.00$~mm for the first law, corresponding to $l_{\mathrm{cz}}/a_0=0.088$ and $0.60$, respectively.
We refer to these configurations as small-scale bridging (SSB) and large-scale bridging (LSB), using these terms comparatively rather than implying a strict asymptotic limit.
Their process-zone lengths relative to the beam height are $l_{\mathrm{cz}}/H=0.88$ and $6.00$.
As further examined in \cref{app:precrack_sensitivity}, the global force response is considerably less sensitive to variations in cohesive strength in the SSB configuration.

\begin{remark}
    We emphasize that the terms SSB and LSB are used here to distinguish configurations with relatively localized and extended cohesive process zones, rather than to identify rigorously defined asymptotic regimes. The terminology adopted here is intended to characterize the contrasting cohesive-zone scales and their influence on the global mechanical response and inverse identification, without implying that either configuration necessarily satisfies a strict asymptotic limit.
\end{remark}

\begin{figure}[htbp]
    \centering
    \includegraphics[width=0.55\textwidth]{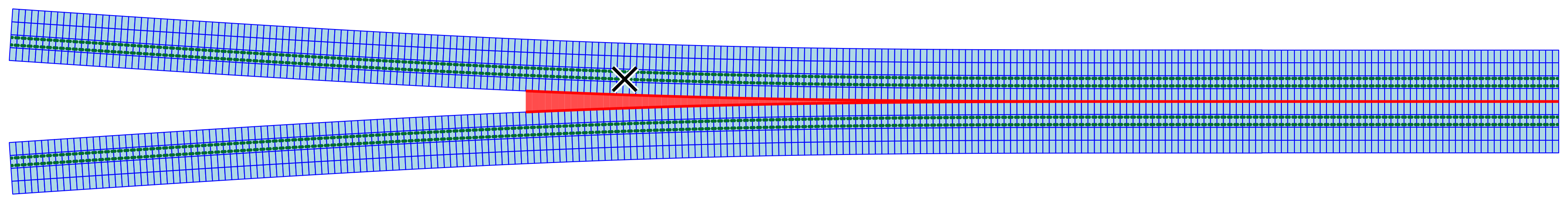}
    \caption{Deformed LSB specimen at $U=200$~\textmu m, with displacements magnified by a factor of 4.
    Red shading indicates cohesive elements, green markers denote DIC strain-sampling points, and the cross identifies the point used for the LSB strain-history comparison in \cref{fig:precrack_dic_lsb}.}
    \label{fig:precrack_deformed}
\end{figure}

Synthetic DIC observations are obtained from bulk Gauss points one element row away from the interface, avoiding direct measurements of the displacement discontinuity.
\Cref{fig:precrack_deformed} illustrates the sampling locations and the deformed configuration for the LSB specimen at $U=200$~\textmu m.

For inverse identification, the cohesive law is represented by separate normal and tangential ICNN branches, each with two hidden layers of two units, together with a full-MNN containing two hidden layers of 16 units.
The reference law is withheld during identification, and the network parameters are optimized directly using force-only, DIC-only, or combined force--DIC observations.
Each strategy is tested with five random initializations, using the same initial laws across observation types.
The optimization employs 500 Adam iterations followed by 25 L-BFGS iterations, with additional settings provided in \cref{app:hyperparams}.

\begin{figure}[htbp]
    \centering
    \begin{subfigure}[b]{0.32\textwidth}
        \centering
        \includegraphics[width=\linewidth]{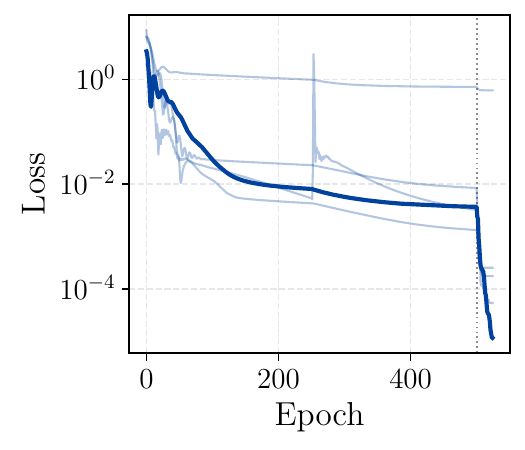}
        \caption{Force-only}
        \label{fig:precrack_loss_force}
    \end{subfigure}
    \hfill
    \begin{subfigure}[b]{0.32\textwidth}
        \centering
        \includegraphics[width=\linewidth]{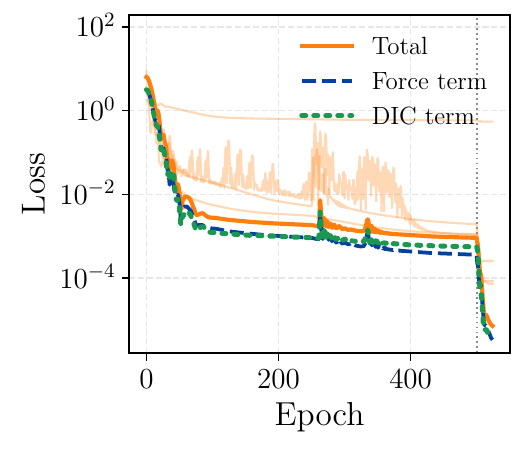}
        \caption{Force+DIC}
        \label{fig:precrack_loss_both}
    \end{subfigure}
    \hfill
    \begin{subfigure}[b]{0.32\textwidth}
        \centering
        \includegraphics[width=\linewidth]{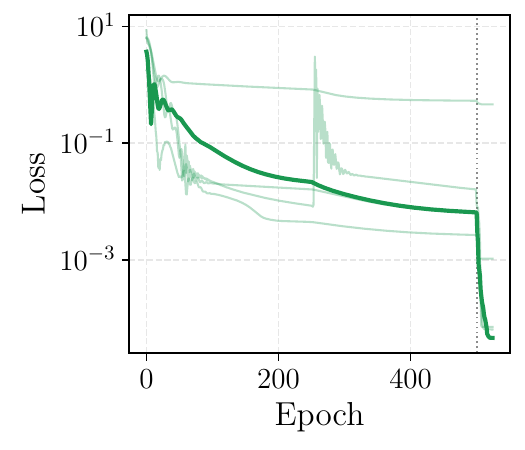}
        \caption{DIC-only}
        \label{fig:precrack_loss_dic}
    \end{subfigure}
    \caption{Normalized loss histories for LSB identification using (a) force-only, (b) combined force--DIC, and (c) DIC-only observations.
    Each panel includes five initializations, with the lowest-loss run emphasized.
    Vertical dotted lines mark the transition from Adam to L-BFGS after 500 iterations.}
    \label{fig:precrack_loss}
\end{figure}

The optimization histories for the LSB configuration are shown in \cref{fig:precrack_loss}.
Four of the five initializations achieve substantial loss reductions, particularly after switching to L-BFGS, whereas one initialization stagnates in an essentially undamaged state because the damage resistance remains above the attained driving force.
This behavior occurs for all three observation choices.
The loss magnitudes across panels are not directly comparable because the observation objectives differ.

\begin{figure}[htbp]
    \centering
    \begin{subfigure}[b]{0.33\textwidth}
        \centering
        \includegraphics[width=\linewidth]{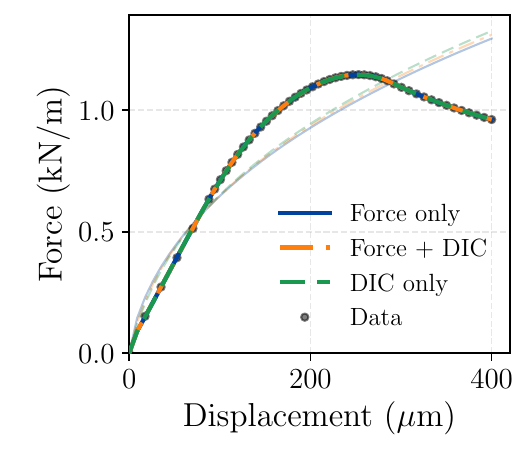}
        \caption{LSB: force}
        \label{fig:precrack_fd_lsb}
    \end{subfigure}
    \begin{subfigure}[b]{0.33\textwidth}
        \centering
        \includegraphics[width=\linewidth]{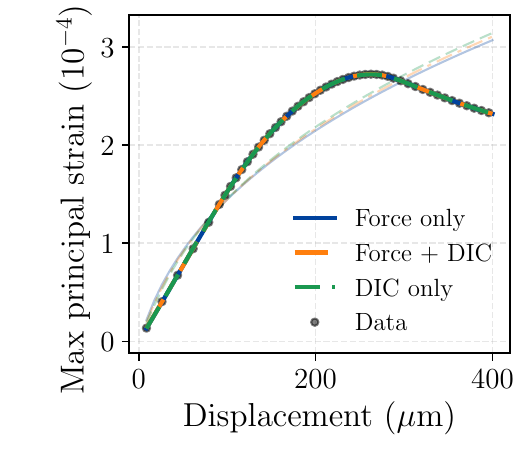}
        \caption{LSB: strain}
        \label{fig:precrack_dic_lsb}
    \end{subfigure}

    \medskip

    \begin{subfigure}[b]{0.33\textwidth}
        \centering
        \includegraphics[width=\linewidth]{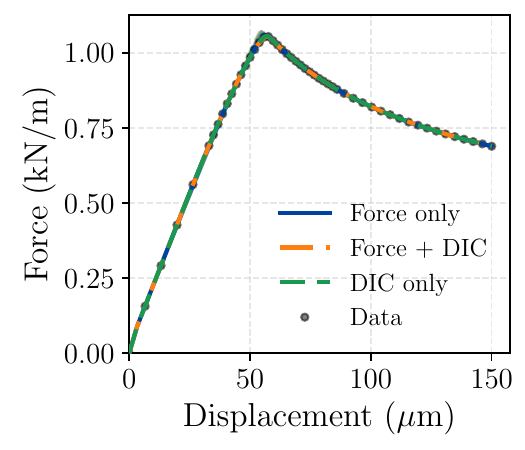}
        \caption{SSB: force}
        \label{fig:precrack_fd_ssb}
    \end{subfigure}
    \begin{subfigure}[b]{0.33\textwidth}
        \centering
        \includegraphics[width=\linewidth]{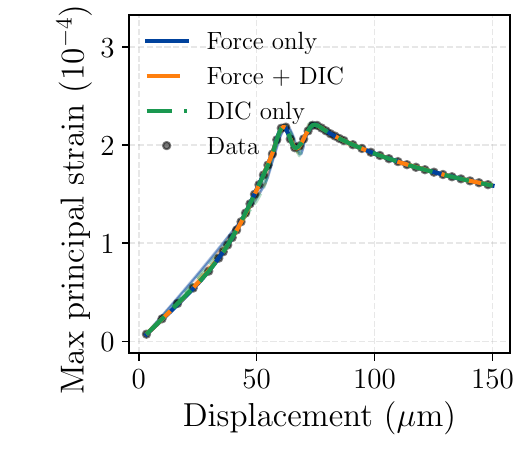}
        \caption{SSB: strain}
        \label{fig:precrack_dic_ssb}
    \end{subfigure}
    \caption{Structural responses predicted by the identified cohesive laws for (a,b) LSB and (c,d) SSB.
    The maximum principal strain is evaluated at DIC points located 11.90 and 11.53~mm from the left end for LSB and SSB, respectively, both 0.30~mm above the interface.
    Black dots denote reference data.
    Each panel shows five initializations per observation type, with the lowest-loss result emphasized.}
    \label{fig:precrack_fit_comparison}
\end{figure}

\Cref{fig:precrack_fit_comparison} compares the force--displacement and maximum principal strain histories predicted by the identified models for both configurations.
For LSB, four of the five initializations accurately reproduce the reference force and strain responses, while the remaining initialization fails to capture the structural response.
For SSB, all initializations closely reproduce both the global force response and the strain history at the selected DIC point.
Notably, the DIC-only identifications also reproduce the force histories despite these measurements being excluded from their optimization objectives.

\begin{figure}[htbp]
    \centering
    \begin{subfigure}[b]{0.32\textwidth}
        \centering
        \includegraphics[width=\linewidth]{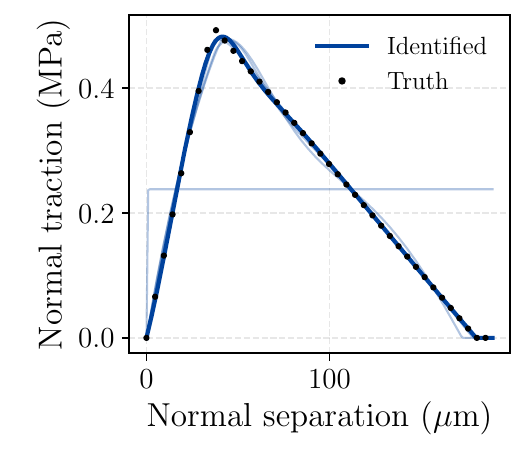}
        \caption{LSB: force-only}
        \label{fig:precrack_ts_lsb_force}
    \end{subfigure}
    \hfill
    \begin{subfigure}[b]{0.32\textwidth}
        \centering
        \includegraphics[width=\linewidth]{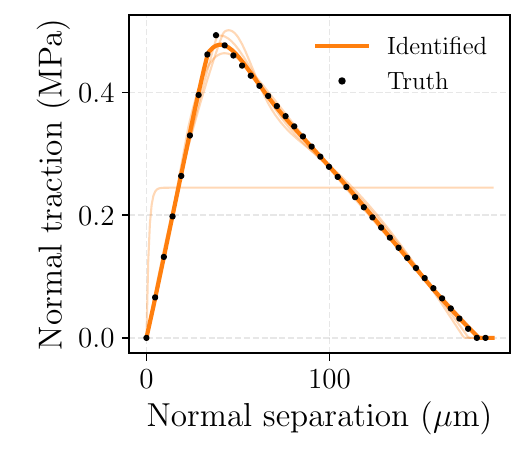}
        \caption{LSB: force+DIC}
        \label{fig:precrack_ts_lsb_both}
    \end{subfigure}
    \hfill
    \begin{subfigure}[b]{0.32\textwidth}
        \centering
        \includegraphics[width=\linewidth]{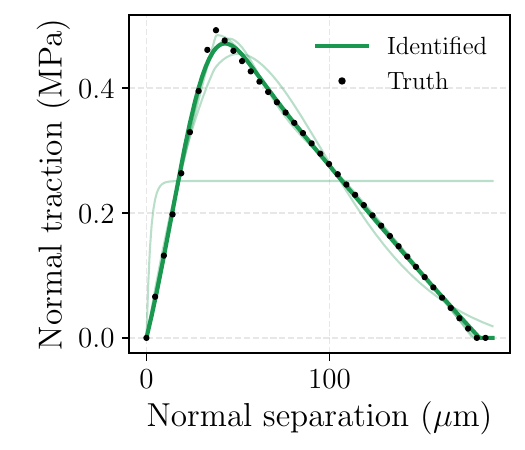}
        \caption{LSB: DIC-only}
        \label{fig:precrack_ts_lsb_dic}
    \end{subfigure}

    \medskip

    \begin{subfigure}[b]{0.32\textwidth}
        \centering
        \includegraphics[width=\linewidth]{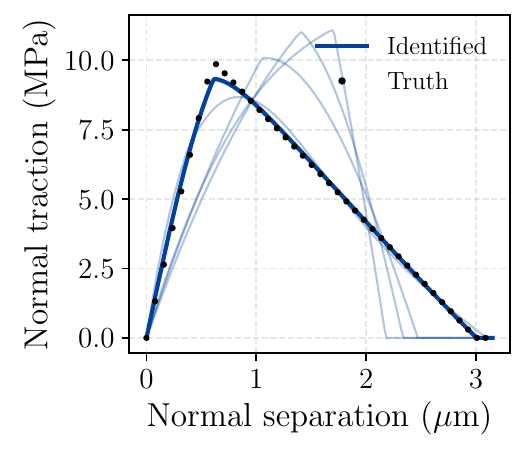}
        \caption{SSB: force-only}
        \label{fig:precrack_ts_ssb_force}
    \end{subfigure}
    \hfill
    \begin{subfigure}[b]{0.32\textwidth}
        \centering
        \includegraphics[width=\linewidth]{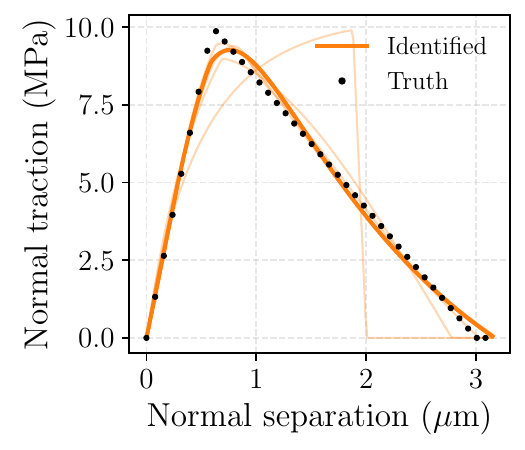}
        \caption{SSB: force+DIC}
        \label{fig:precrack_ts_ssb_both}
    \end{subfigure}
    \hfill
    \begin{subfigure}[b]{0.32\textwidth}
        \centering
        \includegraphics[width=\linewidth]{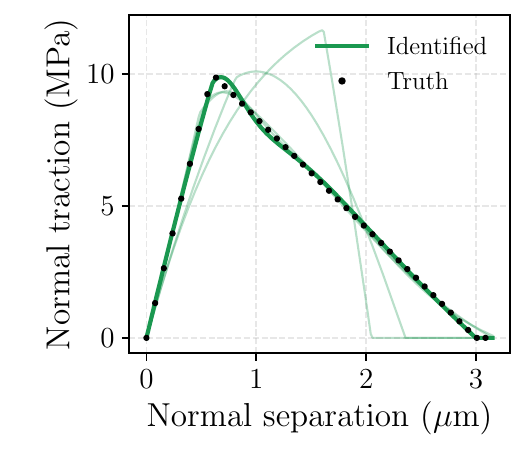}
        \caption{SSB: DIC-only}
        \label{fig:precrack_ts_ssb_dic}
    \end{subfigure}
    \caption{Recovered Mode~I traction--separation laws for (a--c) LSB and (d--f) SSB using the three observation types.
    Black dots indicate the reference law.
    Each panel includes five initializations, with the lowest-loss result emphasized and the others shown in lighter shades.}
    \label{fig:precrack_ts_comparison}
\end{figure}

Despite the similar agreement with the structural observations, the recovered cohesive laws in \cref{fig:precrack_ts_comparison} exhibit substantially different accuracy between the two configurations.
For LSB, four of the five initializations closely recover the reference traction--separation law for each observation choice, consistent with the stronger sensitivity of the global response to cohesive law shape.
For SSB, however, force-only identification produces substantial deviations in three of the five initializations, particularly near the traction peak and along the softening branch, despite accurately reproducing the force--displacement response.
Including DIC measurements improves the consistency of constitutive recovery, although the lowest-loss run for each observation strategy also closely matches the reference law.

\begin{figure}[htbp]
    \centering
    \begin{subfigure}[b]{0.32\textwidth}
        \centering
        \includegraphics[width=\linewidth]{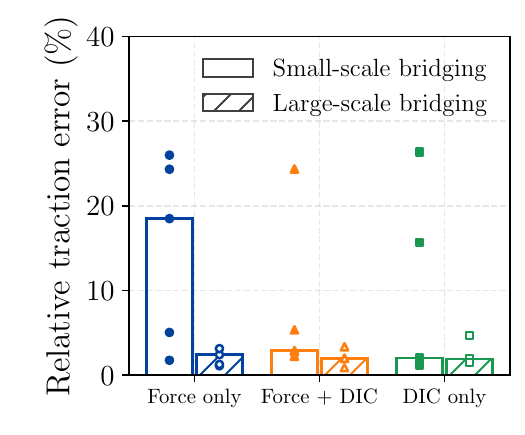}
        \caption{}
        \label{fig:precrack_arms_traction}
    \end{subfigure}
    \begin{subfigure}[b]{0.32\textwidth}
        \centering
        \includegraphics[width=\linewidth]{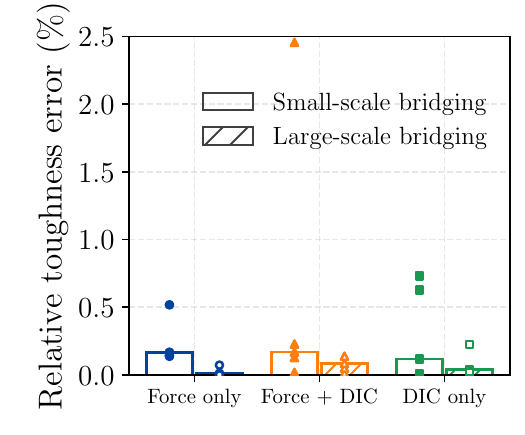}
        \caption{}
        \label{fig:precrack_arms_gc}
    \end{subfigure}
    \caption{Constitutive recovery errors across five initializations for each observation type: (a) relative traction error and (b) relative fracture-energy error $|\widehat G_{Ic}-G_{Ic}|/G_{Ic}$, expressed as percentages.
    Markers denote individual runs, with filled and open symbols for SSB and LSB, respectively; bars indicate the corresponding medians.
    The failed LSB initialization lies outside the plotted traction-error range but is included in the medians.}
    \label{fig:precrack_arms}
\end{figure}

The constitutive recovery errors are quantified in \cref{fig:precrack_arms}, which compares the relative traction error defined in \cref{eq:rel_traction_error} with the relative error in fracture energy.
Across both configurations, the median fracture-energy error remains below $0.2\%$ for every observation choice, whereas the median traction error ranges from $2\%$ to $18\%$.
For SSB, incorporating strain observations reduces the median traction error from approximately $18\%$ for force-only identification to below $3\%$.
These results demonstrate that accurately reproducing the global force response and fracture energy does not necessarily imply accurate identification of the cohesive law shape.
The combined force--DIC objective does not consistently outperform DIC-only identification under the present observation weights and optimization settings.

These observations are consistent with the distinction between small- and large-scale bridging discussed by \citet{bao1992remarks}.
As the cohesive process zone becomes small relative to the structural dimensions, crack propagation approaches an energy-controlled regime in which different traction--separation laws can produce similar global responses.
A larger process zone increases the influence of the local traction distribution on structural compliance and load transfer, providing greater sensitivity to constitutive shape without guaranteeing unique identification.
Spatially resolved strain measurements supply complementary information that improves practical identifiability, particularly in the shorter process-zone configuration.

\subsubsection{Composite Specimen Analysis}
\label{sec:composite}

We extend the investigation to mixed-mode cohesive law identification using a fiber--matrix composite specimen.
Unlike pure Mode~I opening, mixed-mode identification requires recovering the constitutive response under different combinations of normal and tangential separations.
Since a single loading condition may explore only a limited range of these combinations, we consider two distinct loading scenarios and jointly identify a single cohesive law from their structural responses.
The curved fiber--matrix interface and the contrast in elastic properties produce spatially varying interfacial deformation, while the different loading conditions provide additional variation in the relative contributions of opening and sliding.
Furthermore, synthetic DIC strain measurements are considered in both the fiber and matrix to incorporate information about the deformation of the two constituents.

\begin{figure}[htbp]
    \centering
    \begin{subfigure}[b]{0.25\textwidth}
        \centering
        \includegraphics[width=\textwidth]{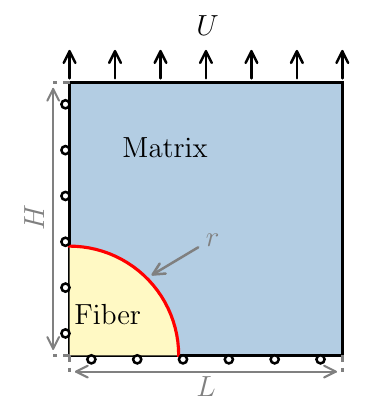}
        \caption{}
    \end{subfigure}%
    \hspace{0.03\textwidth}%
    \begin{subfigure}[b]{0.285\textwidth}
        \centering
        \includegraphics[width=\textwidth]{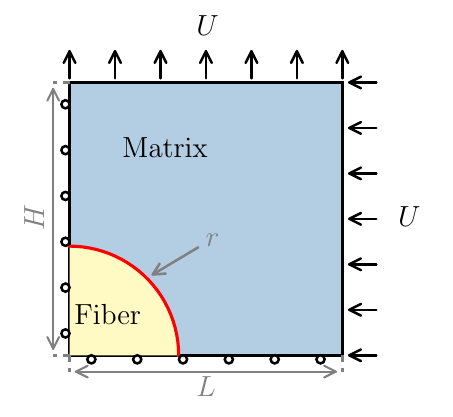}
        \caption{}
    \end{subfigure}
    \caption{Quarter-specimen geometry and boundary conditions under (a) constrained vertical tension and (b) combined vertical tension and horizontal compression.
    The red arc represents the cohesive fiber--matrix interface, while the left and bottom edges satisfy symmetry conditions.}
    \label{fig:composite_structure}
\end{figure}

The specimen is modeled under plane-strain conditions using a square quarter-domain with dimensions $L=H=10$~mm and a quarter-circular fiber of radius $r=4$~mm centered at the origin, as shown in \cref{fig:composite_structure}.
Both constituents are isotropic and linearly elastic, with Young's modulus $E_f=100$~GPa and Poisson's ratio $\nu_f=0.35$ for the fiber, and $E_m=10$~GPa and $\nu_m=0.3$ for the matrix.
The finite-element mesh consists of 464 bulk elements and 16 cohesive elements, with an approximate element length of $0.4$~mm along the interface.

The reference cohesive response is generated using the mixed-mode damage law with exponential softening described in \cref{app:ref_exponential}.
The mixed-mode failure envelope follows a Benzeggagh--Kenane-type interpolation \citep{benzeggagh1996measurement} with $\eta=2.0$.
The pure-mode properties are defined by peak tractions $\sigma_n^0=10$~MPa and $\sigma_t^0=12$~MPa, initial stiffnesses $k_n=k_t=10^{13}$~Pa/m, separation parameters $\delta_n^f=10$~\textmu m and $\delta_t^f=12$~\textmu m, and an exponential softening parameter $\alpha=4.0$.

Two displacement-controlled loading scenarios are considered, as illustrated in \cref{fig:composite_structure}.
In both cases, symmetry conditions are imposed along $x=0$ and $y=0$.
The first scenario applies constrained vertical tension, with $u_y=U$ and $u_x=0$ prescribed along the top edge, while the right edge remains traction-free.
The second scenario combines vertical tension and horizontal compression, with $u_y=U$ imposed along the top edge and $u_x=-U$ along the right edge, while the lateral constraint on the top edge is released.
The prescribed displacement reaches $U=15$~\textmu m in both scenarios.

\begin{figure}[htbp]
    \centering
    \begin{subfigure}[b]{0.20\textwidth}
        \centering
        \includegraphics[height=1.2in]{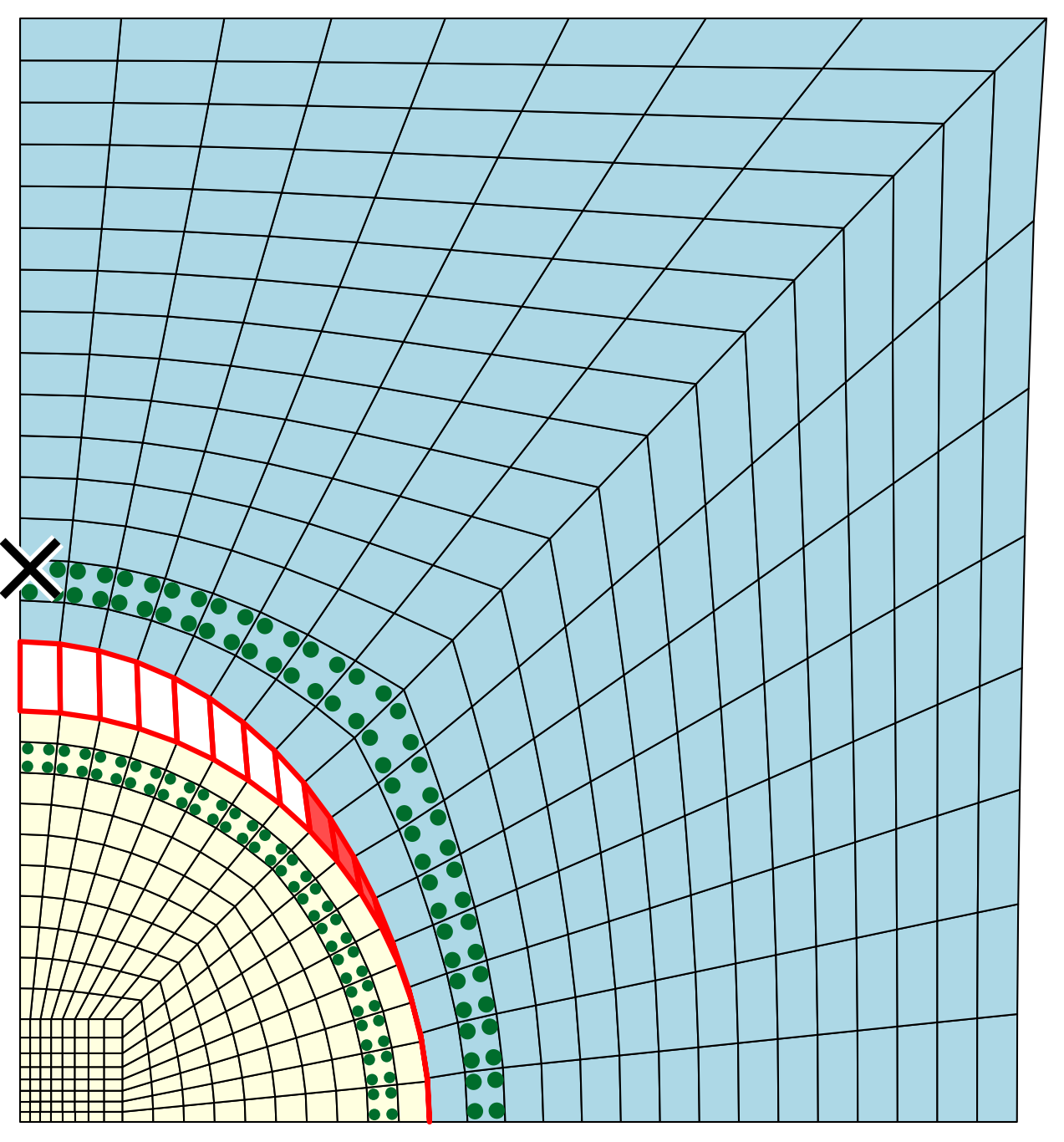}
        \caption{}
        \label{fig:composite_forward_deformed_mesh_uniaxial}
    \end{subfigure}%
    \begin{subfigure}[b]{0.20\textwidth}
        \centering
        \includegraphics[height=1.2in]{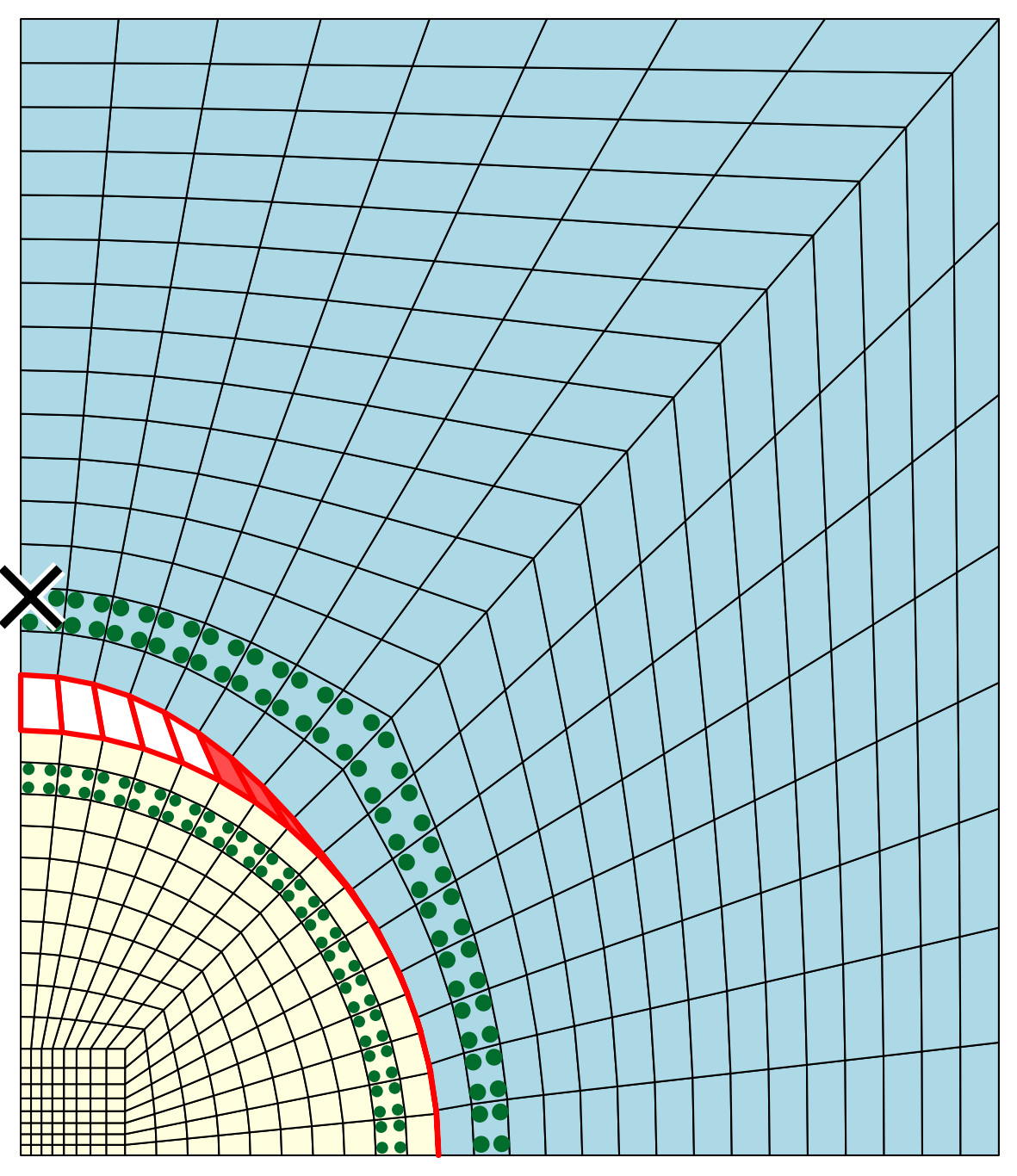}
        \caption{}
        \label{fig:composite_forward_deformed_mesh_shear}
    \end{subfigure}
    \caption{Final deformed configurations under (a) constrained vertical tension and (b) combined vertical tension and horizontal compression, with displacements magnified by a factor of 50.
    Red shading indicates cohesive elements that are not fully damaged.
    Green markers indicate the locations of synthetic DIC strain measurements in the fiber and matrix.
    The cross identifies the sampling point used for the strain-history comparison in \cref{fig:composite_nn_gpb_strain_vs_truth}.}
    \label{fig:composite_forward_deformed_mesh}
\end{figure}

The final deformed configurations for both loading scenarios are shown in \cref{fig:composite_forward_deformed_mesh}.
In both scenarios, the force observation is defined as the vertical reaction along the top boundary.
The horizontal reaction along the right boundary in the second scenario is not used as an additional observation for inverse identification.
Synthetic DIC observations are obtained by sampling the bulk strain fields at selected Gauss points in both the fiber and matrix.
The sampling locations are indicated in \cref{fig:composite_forward_deformed_mesh}.

The inverse problem is solved using observations from both loading scenarios to identify a single cohesive law shared across the entire interface.
The neural cohesive model employs the coupled ICNN model described in \cref{sec:nn_representation}, with two hidden layers of three neurons, together with a full-MNN containing two hidden layers of eight exponential units.
As in the DCB example, each measurement configuration is evaluated using five random initializations.

\begin{figure}[htbp]
     \centering
     \begin{subfigure}[b]{0.22\textwidth}
         \centering
         \includegraphics[width=1\textwidth]{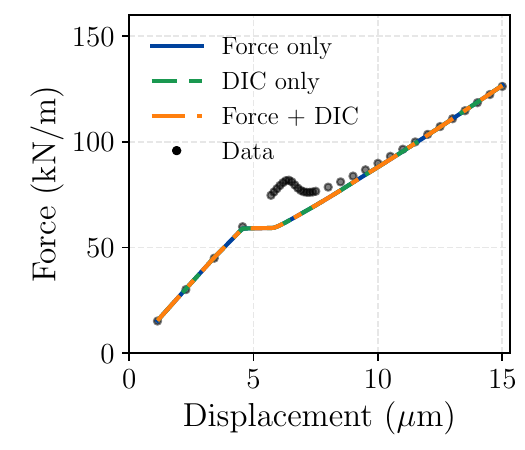}
         \caption*{Initial guess}
     \end{subfigure}
     \begin{subfigure}[b]{0.22\textwidth}
         \centering
         \includegraphics[width=1\textwidth]{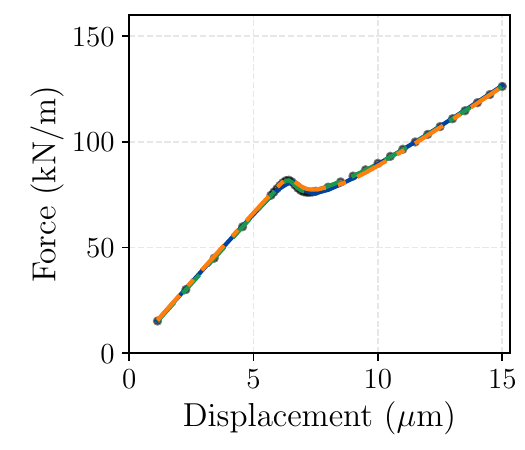}
         \caption*{End of optimization}
     \end{subfigure}
     \\[1.5ex]
     \begin{subfigure}[b]{0.22\textwidth}
         \centering
         \includegraphics[width=1\textwidth]{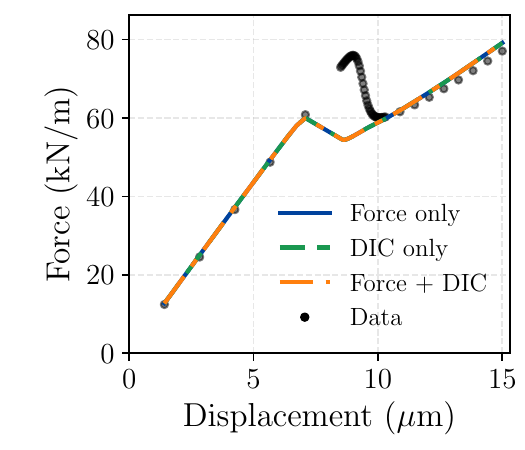}
         \caption*{Initial guess}
     \end{subfigure}
     \begin{subfigure}[b]{0.22\textwidth}
         \centering
         \includegraphics[width=1\textwidth]{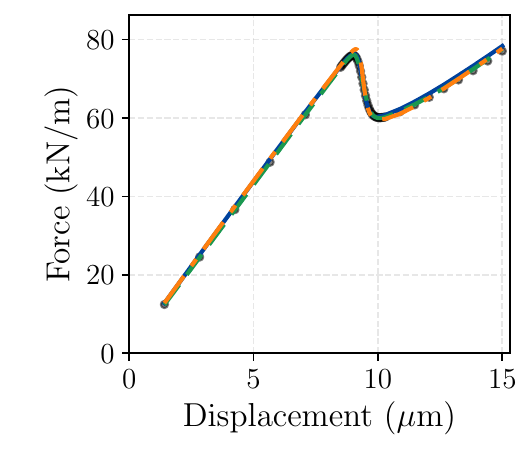}
         \caption*{End of optimization}
     \end{subfigure}
\caption{Predicted top-edge force--displacement responses during joint identification using both loading scenarios.
The top and bottom rows correspond to constrained vertical tension and combined vertical tension and horizontal compression, respectively.
The left and right columns show the responses before and at the end of optimization, respectively.
The run with the lowest final objective value among five random initializations is shown.}
    \label{fig:composite_nn_gpb_fit_vs_truth}
\end{figure}

\begin{figure}[htbp]
     \centering
     \begin{subfigure}[b]{0.22\textwidth}
         \centering
         \includegraphics[width=1\textwidth]{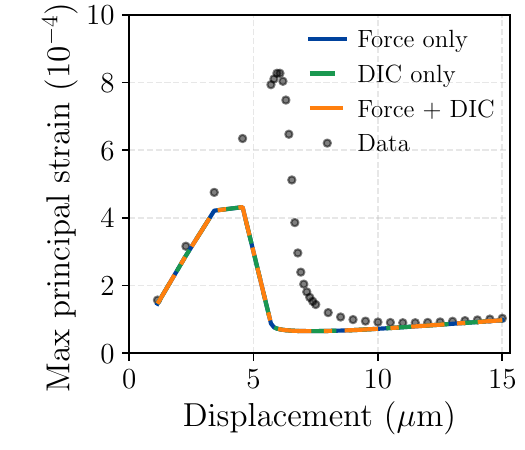}
         \caption*{Initial guess}
     \end{subfigure}
     \begin{subfigure}[b]{0.22\textwidth}
         \centering
         \includegraphics[width=1\textwidth]{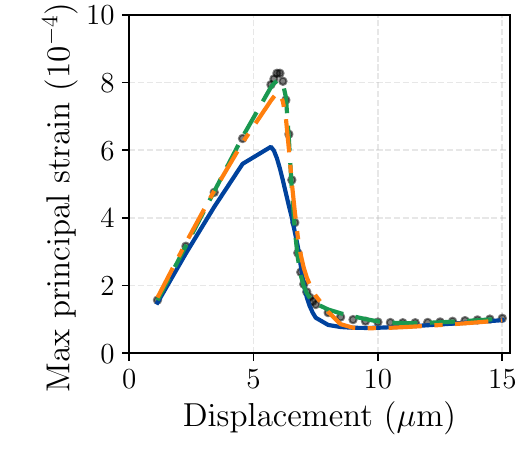}
         \caption*{End of optimization}
     \end{subfigure}
     \\[1.5ex]
     \begin{subfigure}[b]{0.22\textwidth}
         \centering
         \includegraphics[width=1\textwidth]{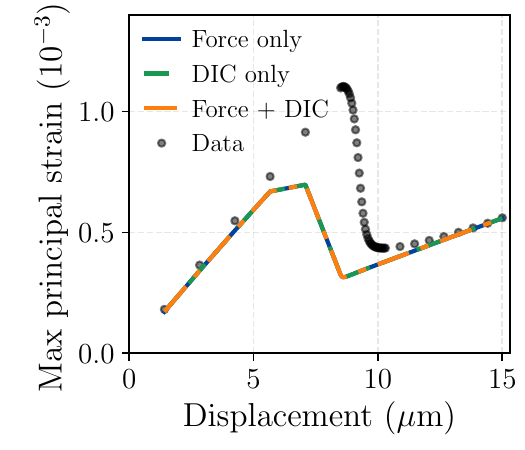}
         \caption*{Initial guess}
     \end{subfigure}
     \begin{subfigure}[b]{0.22\textwidth}
         \centering
         \includegraphics[width=1\textwidth]{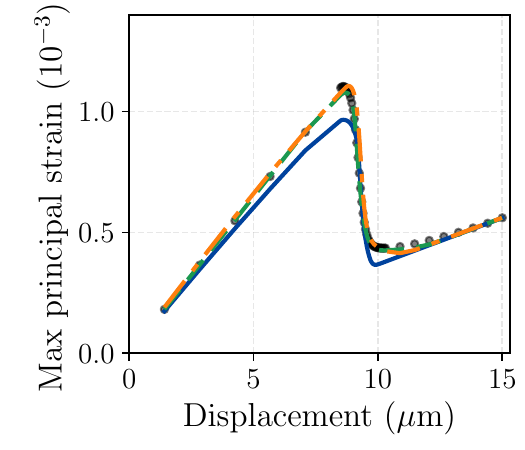}
         \caption*{End of optimization}
     \end{subfigure}
\caption{Predicted maximum principal strain histories at a selected DIC point during joint identification using both loading scenarios, with the sampling location indicated in \cref{fig:composite_forward_deformed_mesh}.
The top row corresponds to constrained vertical tension, and the bottom row corresponds to combined vertical tension and horizontal compression.
The left and right columns show the responses before and at the end of optimization, respectively.
The run with the lowest final objective value among five random initializations is shown.}
    \label{fig:composite_nn_gpb_strain_vs_truth}
\end{figure}

The predicted force--displacement responses shown in \cref{fig:composite_nn_gpb_fit_vs_truth} closely match the reference responses for all three measurement configurations under both loading scenarios, including the load reductions associated with interfacial debonding.
However, differences emerge in the local strain predictions at the DIC point marked in \cref{fig:composite_forward_deformed_mesh}.
As shown in \cref{fig:composite_nn_gpb_strain_vs_truth}, the force-only identification underestimates the peak maximum principal strain under both loading scenarios, despite accurately reproducing the global force responses.
In contrast, the DIC-only and force+DIC configurations reproduce the strain histories more accurately.
These results demonstrate that agreement with global force measurements does not necessarily ensure accurate local deformation predictions and highlight the additional information provided by strain-field measurements in this mixed-mode composite example.

\begin{figure}[htbp]
     \centering
     \begin{subfigure}[b]{0.22\textwidth}
         \centering
         \includegraphics[width=1\textwidth]{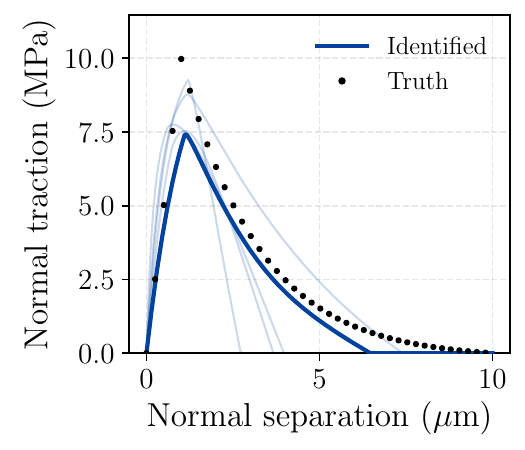}
         \caption{Mode~I, force-only}
     \end{subfigure}
     \begin{subfigure}[b]{0.22\textwidth}
         \centering
         \includegraphics[width=1\textwidth]{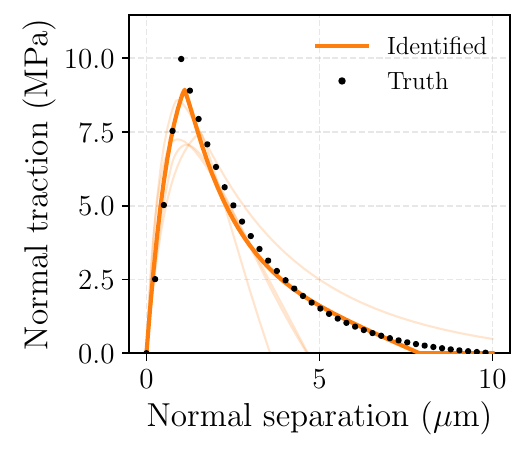}
         \caption{Mode~I, force+DIC}
     \end{subfigure}
     \begin{subfigure}[b]{0.22\textwidth}
         \centering
         \includegraphics[width=1\textwidth]{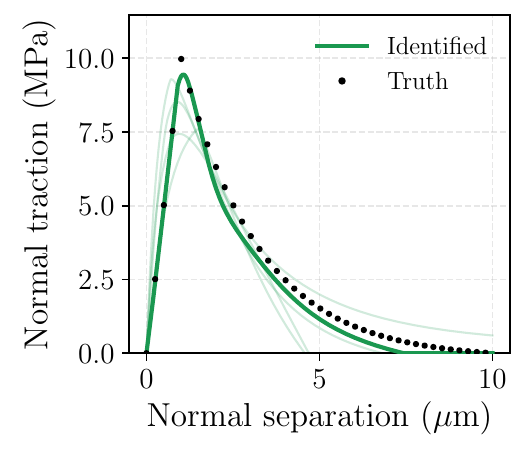}
         \caption{Mode~I, DIC-only}
     \end{subfigure}

     \centering
     \begin{subfigure}[b]{0.22\textwidth}
         \centering
         \includegraphics[width=1\textwidth]{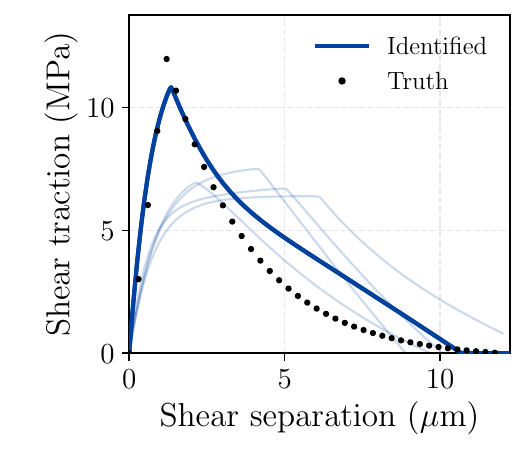}
         \caption{Mode~II, force-only}
     \end{subfigure}
     \begin{subfigure}[b]{0.22\textwidth}
         \centering
         \includegraphics[width=1\textwidth]{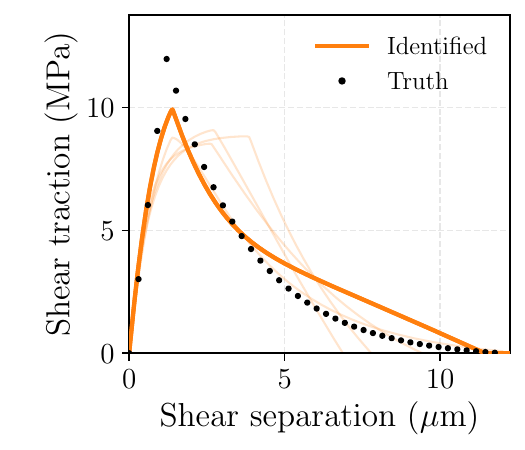}
         \caption{Mode~II, force+DIC}
     \end{subfigure}
     \begin{subfigure}[b]{0.22\textwidth}
         \centering
         \includegraphics[width=1\textwidth]{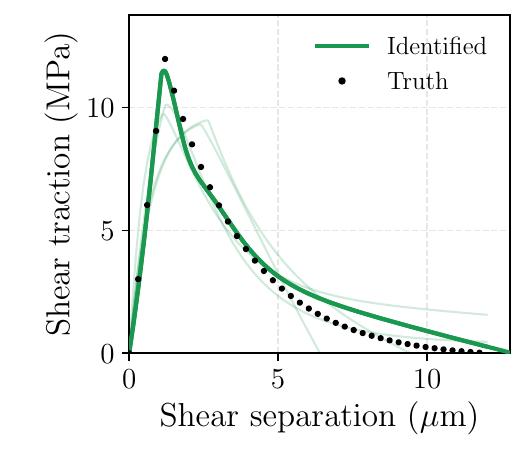}
         \caption{Mode~II, DIC-only}
     \end{subfigure}
\caption{Recovered cohesive laws along pure Mode~I and Mode~II separation paths following joint identification from the two loading scenarios.
The top and bottom rows show the Mode~I opening and Mode~II sliding responses, respectively, while the columns correspond to force-only, force+DIC, and DIC-only measurement configurations.
Each panel shows five random initializations, with the run achieving the lowest final objective value highlighted and the remaining runs shown in lighter colors.
}
    \label{fig:composite_nn_gpb_traction_separation}
\end{figure}

\Cref{fig:composite_nn_gpb_traction_separation} compares the recovered cohesive laws with the reference response along pure Mode~I and Mode~II separation paths.
Despite the close agreement in the global force--displacement responses, noticeable differences remain in the identified cohesive laws.
The force-only configuration underestimates the Mode~I peak traction and overpredicts the traction along much of the Mode~II softening branch, with considerable variability across random initializations.
Incorporating strain measurements improves the overall agreement with the reference laws, with the DIC-only configuration providing the closest recovery in both modes among the best-performing runs.

% \alert{Unlike the DCB example}, where force-only identification provided the most accurate and consistent recovery, the composite example benefits more from strain-field measurements.
The composite example benefits more from strain-field measurements, similar to the SSB DCB case.
The ability of different cohesive laws to reproduce nearly identical force--displacement responses suggests that global force measurements alone may not sufficiently constrain all features of the constitutive response.
Spatially resolved strain measurements provide complementary information about local deformation and load transfer, although their effectiveness depends on the specimen geometry, loading conditions, and measurement locations.
The better performance of DIC-only compared with force+DIC in this example does not necessarily imply that combining observations is disadvantageous, since the outcome also depends on the relative weighting of the observation terms and optimization convergence.

These findings suggest that inverse cohesive law identification may suffer from limited practical identifiability, motivating strategies that introduce additional information or restrict the admissible constitutive space.
One such strategy is to regularize the inverse problem by restricting the constitutive response to a learned generative family, an approach previously explored in inverse problems using deep generative models \citep{bora2017compressed}.
This motivates the data-driven constitutive prior introduced next, which constrains the admissible cohesive laws through a pretrained generative model and enables inverse identification by optimizing its latent variables rather than the network parameters.

\subsection{Data-Driven Constitutive Prior}
\label{sec:meta_data_driven}

We consider a setting in which a collection of representative traction--separation responses is available from different interfacial material systems or interface conditions, providing prior knowledge of the range of cohesive behaviors that may arise in a structure.
Rather than prescribing a particular analytical traction--separation law, we use these responses to learn a low-dimensional family of constitutive models that captures variations in interfacial strength, separation scales, softening behavior, and mode dependence.
This approach builds on the constitutive-prior framework introduced in our previous work \cite{han2027constitutive}, where a family of nonlinear elastic material responses was learned from data and subsequently used in inverse design.
Here, we extend this concept to interfacial damage by constructing a family of cohesive laws within the thermodynamically consistent formulation developed in \cref{sec:priors}.
The resulting constitutive prior provides a restricted but flexible space of admissible cohesive responses within which subsequent inverse identification can be performed.

To demonstrate this approach, we construct a synthetic constitutive library $\mathcal P$ containing $N_m=9$ paired Mode~I and Mode~II traction--separation histories, corresponding to 18 reference curves.
These responses represent different cohesive behaviors with variations in peak traction, characteristic separation, and post-peak softening.
Following \cref{eq:prior_family_train}, each pair is assigned a shared scalar latent coordinate $z_m$, which distinguishes its constitutive response from the other members of the library.
The normal and tangential responses belonging to the same pair share this coordinate, allowing their variations to be represented jointly rather than through independently fitted constitutive parameters.
We specialize the latent representation to $n_z=1$ and define the admissible interval as
\begin{equation*}
    \mathcal Z
    =
    \left[\min_m z_m^\star,\,\max_m z_m^\star\right],
\end{equation*}
where $z_m^\star$ denotes the optimized latent coordinate associated with reference pair $m$ after offline training using \cref{eq:meta_loss_gauge}.
Varying $z$ within $\mathcal Z$ then generates a continuous family of cohesive responses, allowing the model to interpolate between the reference behaviors without prescribing a conventional parametric traction--separation law.

\begin{figure}[htbp]
    \centering
    \includegraphics[width=0.28\textwidth]{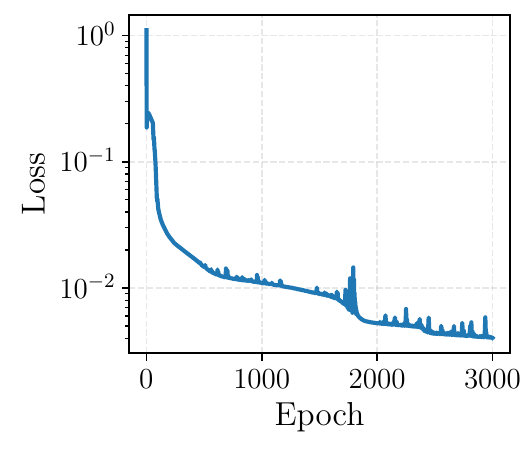}
    \caption{
    Training history of the data-driven cohesive constitutive prior.
    }
    \label{fig:train_prior_loss}
\end{figure}

The constitutive family is represented using the separable energy formulation in \cref{eq:psi_decoupled}.
The normal and tangential active-energy branches are parameterized by separate PICNNs, each containing three hidden layers with 18 units in the convex stream and 12 units in the unconstrained stream.
The inverse-resistance map is represented by a full-MNN with three hidden layers and 12 units in each stream.
The latent coordinate enters the unconstrained streams, enabling variations in the constitutive response while preserving the convexity and monotonicity properties required by the underlying formulation.
The shared network parameters and the nine reference latent coordinates are optimized jointly using Adam as shown in \cref{fig:train_prior_loss}.

\begin{figure}[htbp]
    \centering
    \begin{subfigure}[b]{0.23\textwidth}
        \centering
        \includegraphics[width=\textwidth]{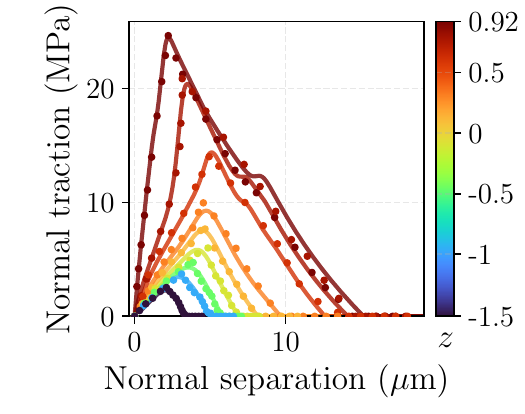}
        \caption{Mode~I\protect\label{fig:meta2_ts_mode_I}}
    \end{subfigure}
    \begin{subfigure}[b]{0.23\textwidth}
        \centering
        \includegraphics[width=\textwidth]{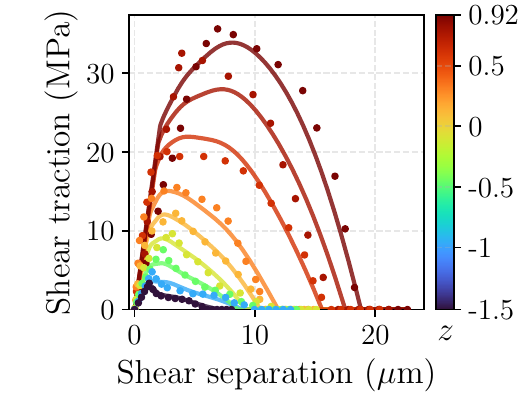}
        \caption{Mode~II\protect\label{fig:meta2_ts_mode_II}}
    \end{subfigure}
    \begin{subfigure}[b]{0.23\textwidth}
        \centering
        \includegraphics[width=\textwidth]{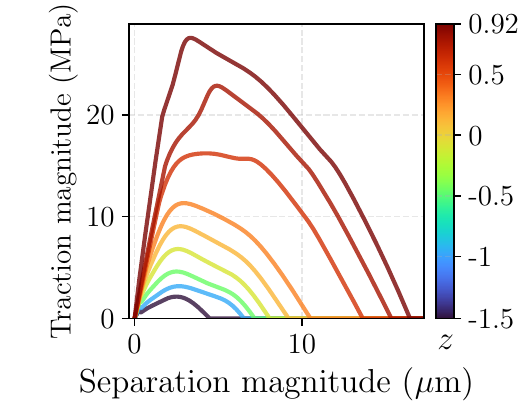}
        \caption{Mixed mode\protect\label{fig:meta2_ts_mixed}}
    \end{subfigure}
    \caption{
    Learned family of cohesive traction--separation responses.
    Panels (a) and (b) compare the predicted pure Mode~I and Mode~II responses (solid curves) with the nine paired reference histories (dots).
    Colors indicate the shared scalar latent coordinate $z_m^\star$ associated with each material pair.
    Panel (c) shows the predicted traction magnitude as a function of separation magnitude under proportional mixed-mode loading with $\delta_t=\delta_n\geq 0$.
    }
    \label{fig:meta2_ts}
\end{figure}

The learned constitutive family is illustrated in \cref{fig:meta2_ts}.
The fitted responses reproduce the principal variations in peak traction, separation scales, and softening behavior exhibited by the reference data in both loading modes.
The mixed-mode curves in \cref{fig:meta2_ts_mixed} illustrate the responses generated by the constitutive model along the proportional loading path $\delta_t=\delta_n$.

\begin{remark}
Because only pure-mode data are used for training, these mixed-mode predictions follow from the assumed constitutive structure and should not be interpreted as independently validated mixed-mode behavior.    
\end{remark}

Having constructed the constitutive prior, we next consider two inverse identification problems involving a DCB and a three-fiber composite.
In both examples, we assume that the unknown local cohesive laws belong to the learned constitutive family, although their spatial distributions are unknown.
Under this assumption, identifying a heterogeneous interface no longer requires independently determining the complete traction--separation law at every location.
Instead, each cohesive element $e$ is assigned a scalar coordinate $z_e\in\mathcal Z$, which selects its local constitutive response from the learned family.
The shared network parameters $\boldsymbol\Theta^\star$ remain fixed during this stage, restricting the inverse solution to the constitutive variations represented by the offline prior.

For each specimen, synthetic structural observations are generated using a prescribed spatial distribution of latent coordinates that is withheld during inverse identification.
The objective is to recover this distribution from global force--displacement measurements, full-field DIC observations, or their combination.
These examples therefore examine whether spatially heterogeneous cohesive behavior can be identified from remote structural measurements when the admissible constitutive family is known in advance.
Because the reference responses are generated within the same learned family, the experiments assess identification accuracy under a correctly specified constitutive prior rather than robustness to constitutive model mismatch.
The DCB and three-fiber composite examples further examine the relationship between matching the observed structural response and recovering the underlying local cohesive laws.
Additional comparisons in \cref{sec:meta_dd_comparison} investigate the effects of constitutive parameterization and optimization strategy.

\subsubsection{Double Cantilever Beam with Heterogeneous Interface}
\label{sec:meta_dd_dcb}

\begin{figure}[htbp]
    \centering
    \sbox0{\includegraphics[width=0.25\textwidth]{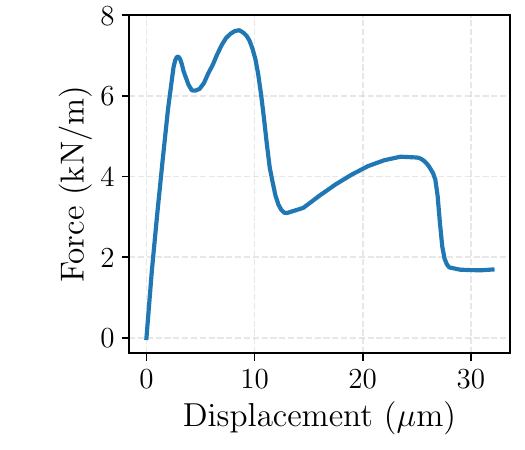}}
    \begin{subfigure}[b]{0.48\textwidth}
        \centering
        \parbox[b][\ht0][c]{\linewidth}{\centering\includegraphics[width=\linewidth]{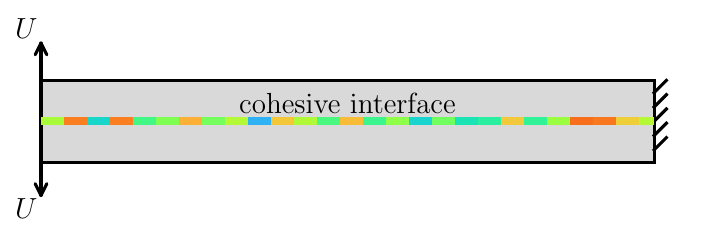}}
        \caption{Reference interface}
        \label{fig:meta2_dcb_forward_xi_profile}
    \end{subfigure}
    \begin{subfigure}[b]{0.25\textwidth}
        \centering
        \usebox0
        \caption{Synthetic force record}
        \label{fig:meta2_dcb_forward_fd}
    \end{subfigure}
    \caption{Heterogeneous DCB used to generate the inverse data.
    Panel (a) shows the prescribed spatial distribution of cohesive laws, with consecutive three-element blocks sharing the same latent coordinate.
    The colors correspond to the constitutive responses in \cref{fig:meta2_ts}.
    Panel (b) shows the resulting force--displacement response, with $U$ denoting the single-arm displacement and force reported per unit out-of-plane thickness.}
    \label{fig:meta2_dcb_setup}
\end{figure}

We consider an initially intact DCB with the same beam dimensions, bulk material, and Mode~I loading configuration as in \cref{sec:precrack}, but without a pre-existing crack.
The cohesive interface spans the entire specimen length and is allowed to vary spatially within the learned constitutive family.
The bulk is discretized using an $80\times8$ structured mesh of plane-strain quadrilateral elements, with 80 cohesive elements along the interface.
To generate synthetic observations, we prescribe a piecewise-constant latent field in which consecutive blocks of three cohesive elements share the same constitutive law, as illustrated in \cref{fig:meta2_dcb_forward_xi_profile}.
The colors identify the local cohesive laws and correspond to those used for the learned constitutive family in \cref{fig:meta2_ts}.
Evaluating the trained prior at these prescribed coordinates and solving the forward problem produces the synthetic force--displacement and strain histories.
As shown in \cref{fig:meta2_dcb_forward_fd}, the force response exhibits multiple local peaks and load drops as debonding propagates through regions with different cohesive properties.

\begin{figure}[htbp]
    \centering
    \begin{subfigure}[b]{0.22\textwidth}
        \centering
        \includegraphics[width=\linewidth]{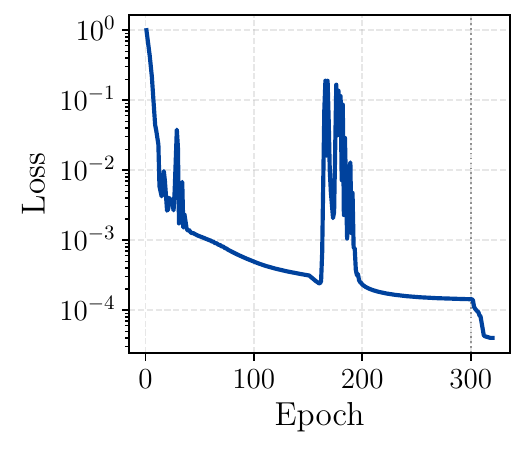}
        \caption{Force-only}
        \label{fig:meta2_dcb_inverse_loss_force}
    \end{subfigure}
    \begin{subfigure}[b]{0.22\textwidth}
        \centering
        \includegraphics[width=\linewidth]{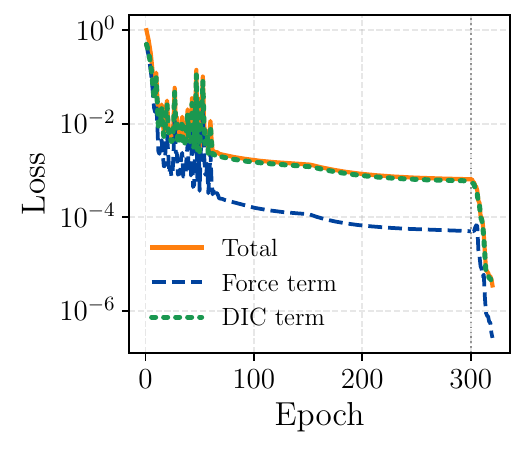}
        \caption{Force+DIC}
        \label{fig:meta2_dcb_inverse_loss_both}
    \end{subfigure}
    \begin{subfigure}[b]{0.22\textwidth}
        \centering
        \includegraphics[width=\linewidth]{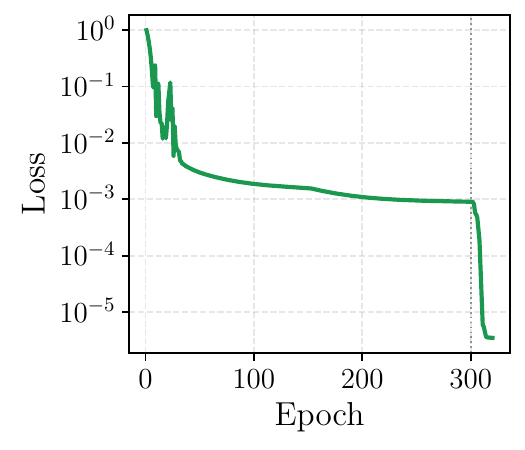}
        \caption{DIC-only}
        \label{fig:meta2_dcb_inverse_loss_dic}
    \end{subfigure}
    \caption{Normalized loss histories for inverse identification of the heterogeneous DCB, with
    the vertical dotted lines marking the transition from 300 Adam iterations to 20 L-BFGS iterations.}
    \label{fig:meta2_dcb_inverse_loss}
\end{figure}

We investigate the recovery of the heterogeneous interface using force-only, DIC-only, and combined force--DIC observations, while keeping the pretrained constitutive network fixed.
The corresponding optimization histories are shown in \cref{fig:meta2_dcb_inverse_loss}.
Consistent with the preceding examples, the loss reduction under Adam eventually slows, whereas switching to L-BFGS yields a substantial further decrease for all three identification strategies.

\begin{figure}[htbp]
    \centering
    \begin{subfigure}[b]{0.30\textwidth}
        \centering
        \includegraphics[width=\linewidth]{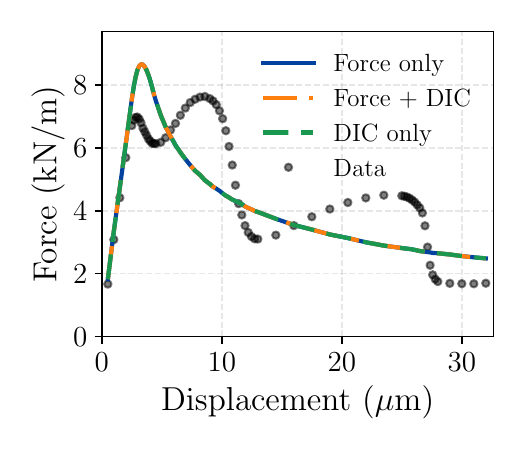}
        \caption{Initial guess}
        \label{fig:meta2_dcb_inverse_fd_epoch1}
    \end{subfigure}
    \begin{subfigure}[b]{0.30\textwidth}
        \centering
        \includegraphics[width=\linewidth]{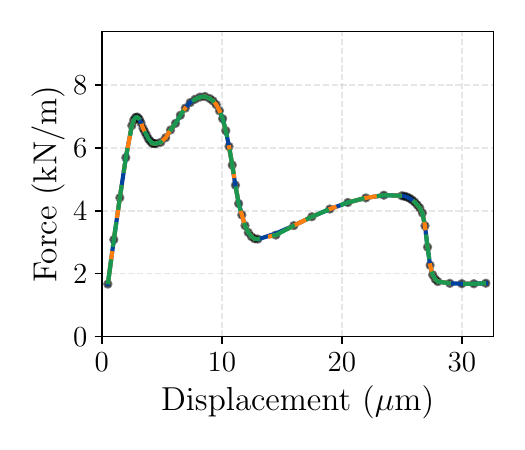}
        \caption{After optimization}
        \label{fig:meta2_dcb_inverse_fd_epoch320}
    \end{subfigure}
    \caption{Comparison of predicted and reference force--displacement responses for the heterogeneous DCB at (a) the initial guess and (b) after inverse identification using different observation types.}
    \label{fig:meta2_dcb_inverse_fit_vs_truth}
\end{figure}

\begin{figure}[htbp]
    \centering
    \begin{subfigure}[b]{0.30\textwidth}
        \centering
        \includegraphics[width=\linewidth]{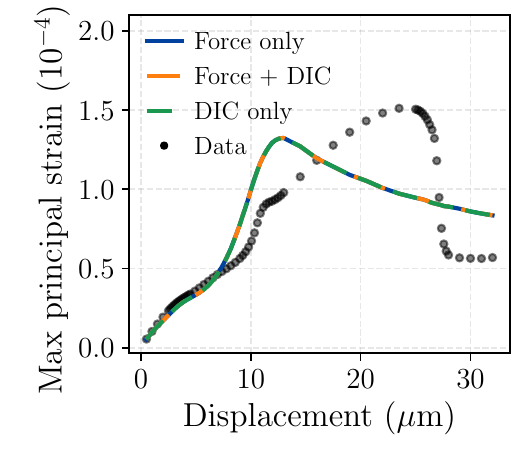}
        \caption{Initial guess}
        \label{fig:meta2_dcb_inverse_field_epoch1}
    \end{subfigure}
    \begin{subfigure}[b]{0.30\textwidth}
        \centering
        \includegraphics[width=\linewidth]{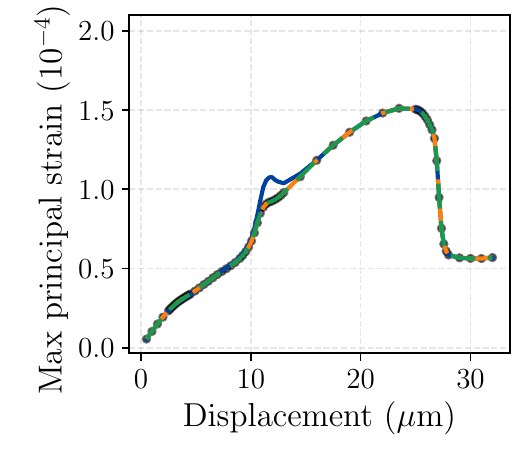}
        \caption{After optimization}
        \label{fig:meta2_dcb_inverse_field_epoch320}
    \end{subfigure}
    \caption{Comparison of predicted and reference maximum principal strain histories at a selected DIC point in the heterogeneous DCB.
    The sampling point is located 2.55~mm from the left end along the interface and 0.45~mm above it.}
    \label{fig:meta2_dcb_inverse_field_fit}
\end{figure}

\Cref{fig:meta2_dcb_inverse_fit_vs_truth} compares the force--displacement responses predicted by the initial and identified models against the reference data.
After optimization, all three identification strategies closely reproduce the global force response, demonstrating that even the DIC-only identification can predict the overall structural behavior.
However, agreement with the global force response does not necessarily imply an equally accurate representation of the local deformation field.
To examine this distinction, \cref{fig:meta2_dcb_inverse_field_fit} compares the predicted maximum principal strain histories at a selected DIC point with the reference response.
Both the DIC-only and combined force--DIC identifications closely reproduce the strain evolution, whereas the force-only identification exhibits a discrepancy near the first abrupt change in the strain history, despite matching the global force response.
These results suggest that global force measurements alone may be insufficient to distinguish between heterogeneous cohesive properties that produce similar structural responses, while local strain measurements provide additional constraints on the inverse solution.

\begin{figure}[htbp]
    \centering
    \begin{subfigure}[b]{0.25\textwidth}
        \centering
        \includegraphics[width=\linewidth]{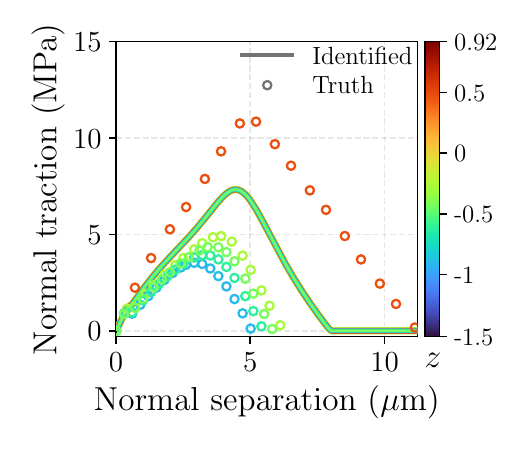}
        \caption*{Initial guess}
    \end{subfigure}
    \begin{subfigure}[b]{0.25\textwidth}
        \centering
        \includegraphics[width=\linewidth]{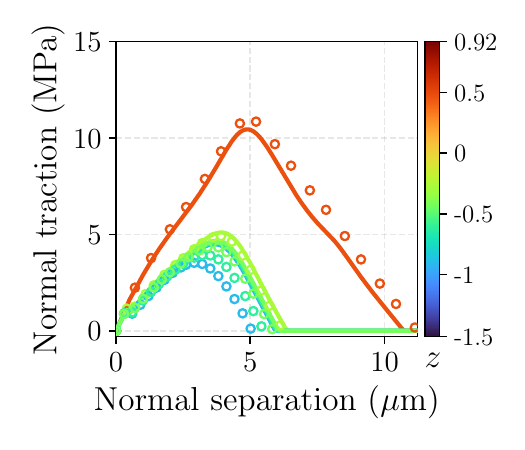}
        \caption*{Epoch 10}
    \end{subfigure}
    \begin{subfigure}[b]{0.25\textwidth}
        \centering
        \includegraphics[width=\linewidth]{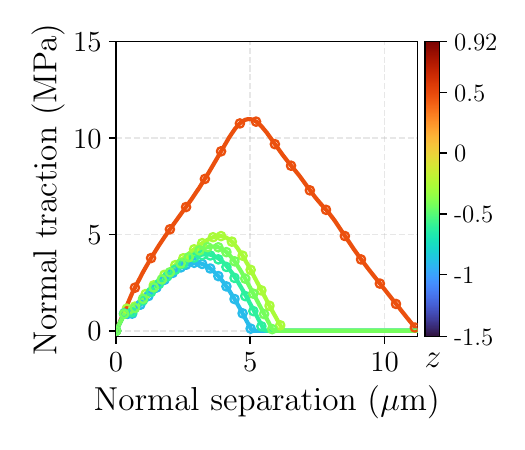}
        \caption*{Epoch 320}
    \end{subfigure}
    \caption{Evolution of the recovered Mode~I traction--separation laws at five representative cohesive elements using combined force--DIC observations.
    Solid curves represent the identified laws, while open circles denote the reference responses.
    Colors indicate the reference latent coordinate assigned to each element.}
    \label{fig:meta2_dcb_inverse_ts_three_epochs}
\end{figure}

\begin{figure}[htbp]
    \centering
    \begin{subfigure}[b]{0.25\textwidth}
        \centering
        \includegraphics[width=\linewidth]{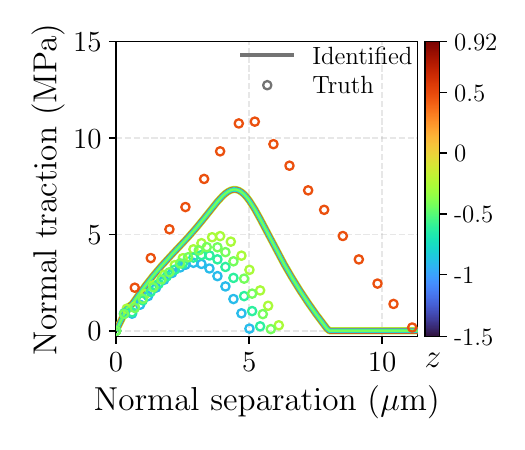}
        \caption*{Initial guess}
    \end{subfigure}
    \begin{subfigure}[b]{0.25\textwidth}
        \centering
        \includegraphics[width=\linewidth]{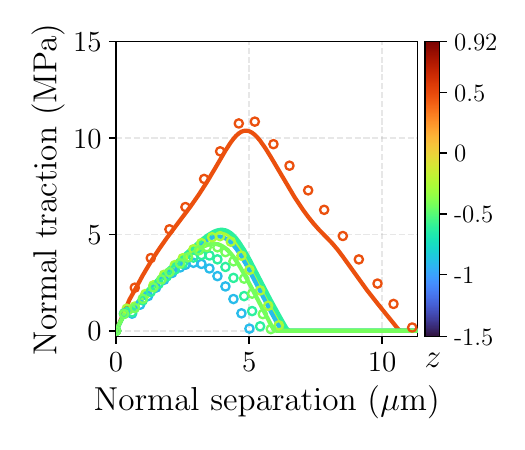}
        \caption*{Epoch 10}
    \end{subfigure}
    \begin{subfigure}[b]{0.25\textwidth}
        \centering
        \includegraphics[width=\linewidth]{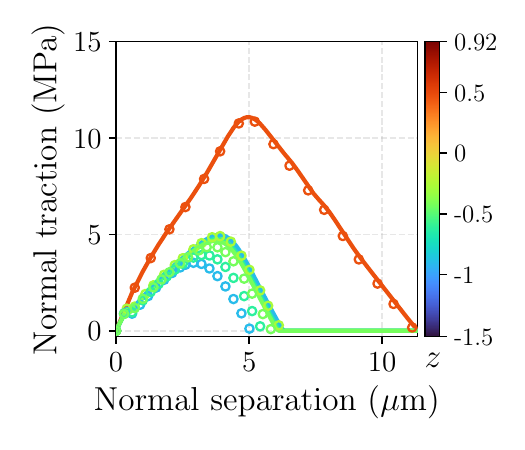}
        \caption*{Epoch 320}
    \end{subfigure}
    \caption{Evolution of the recovered Mode~I traction--separation laws using force-only observations at the same five elements as in \cref{fig:meta2_dcb_inverse_ts_three_epochs}.
    Solid curves represent the identified laws, while open circles denote the reference responses.
    Colors indicate the reference latent coordinate assigned to each element.}
    \label{fig:meta2_dcb_inverse_ts_force_only}
\end{figure}

We next examine whether the agreement with the structural observations translates into accurate recovery of the local cohesive laws.
\Cref{fig:meta2_dcb_inverse_ts_three_epochs} illustrates the evolution of the identified Mode~I traction--separation responses at five representative cohesive elements using combined force--DIC observations.
Starting from the initial estimates, the recovered laws progressively approach their reference responses, with the final predictions closely capturing differences in peak traction and post-peak softening.
The corresponding results using force-only observations are shown in \cref{fig:meta2_dcb_inverse_ts_force_only}.
Although the force-only identification recovers the overall trends, larger discrepancies remain for some of the local cohesive laws, despite the close agreement with the global force response discussed above.

\begin{figure}[htbp]
    \centering
    \begin{subfigure}[b]{0.22\textwidth}
        \centering
        \includegraphics[width=\linewidth]{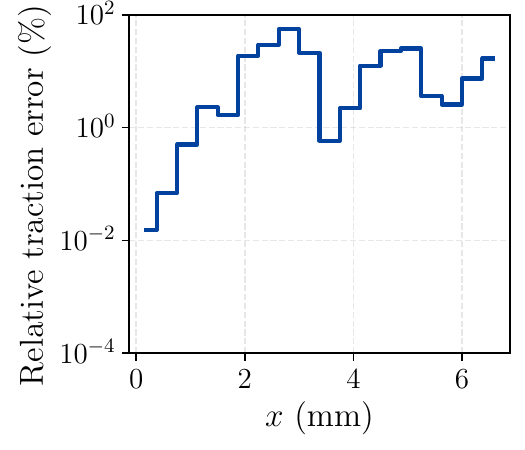}
        \caption*{Force-only}
    \end{subfigure}
    \begin{subfigure}[b]{0.22\textwidth}
        \centering
        \includegraphics[width=\linewidth]{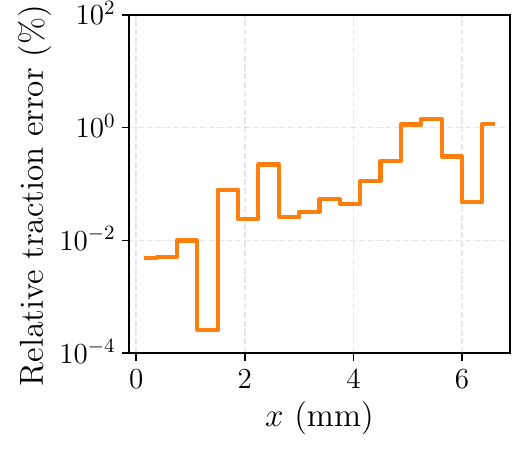}
        \caption*{Force+DIC}
    \end{subfigure}
    \begin{subfigure}[b]{0.22\textwidth}
        \centering
        \includegraphics[width=\linewidth]{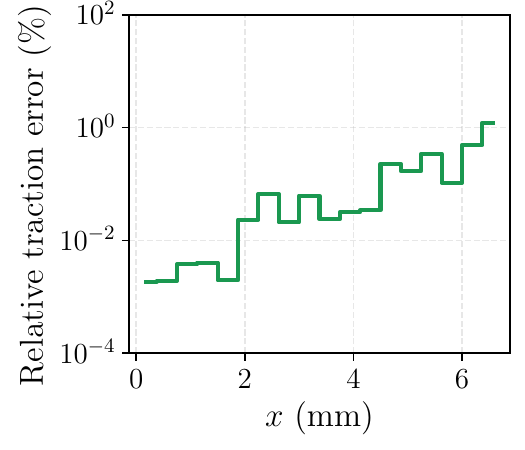}
        \caption*{DIC-only}
    \end{subfigure}
    \caption{Relative error between the identified and reference traction--separation laws along the heterogeneous DCB interface for the three observation types.
    Results are shown for the first 18 cohesive elements, corresponding to six three-element blocks traversed by the crack front.
    All panels use the same logarithmic error scale.}
    \label{fig:meta2_dcb_ts_error}
\end{figure}

To assess recovery accuracy across the interface, \cref{fig:meta2_dcb_ts_error} reports the relative traction error defined in \cref{eq:rel_traction_error} for the first 18 cohesive elements traversed by the crack front.
The traction--separation responses are evaluated using a separation increment of $0.1$~\textmu m, which is also adopted in the subsequent examples involving the learned prior.
Both identification strategies incorporating DIC measurements yield substantially smaller errors than the force-only approach across the evaluated region.
Together with the structural-response comparisons, these results demonstrate that different distributions of cohesive properties can produce similar global force responses, while spatially resolved strain measurements provide additional information for distinguishing the underlying local constitutive behavior.

Although recovering spatially heterogeneous cohesive laws from global force--displacement measurements alone would be particularly attractive, the present results suggest that such identification may remain unreliable even when the inverse search space is constrained by a learned constitutive prior.
One possible explanation is that the learned family, while substantially restricting the admissible constitutive responses, remains sufficiently expressive to allow different spatial distributions of cohesive properties to produce similar global force responses.
A more restrictive constitutive prior could potentially improve identifiability, provided that it retains sufficient flexibility to represent the underlying material behavior.
This observation motivates further investigation into the balance between constitutive expressiveness and inverse identifiability, including strategies for adaptively refining the admissible constitutive family based on available information.
We leave these directions for future work.

\subsubsection{Composite Specimen with Three Fibers}
\label{sec:meta_dd_comp3fiber}

\begin{figure}[htbp]
    \centering
    \includegraphics[width=0.3\textwidth]{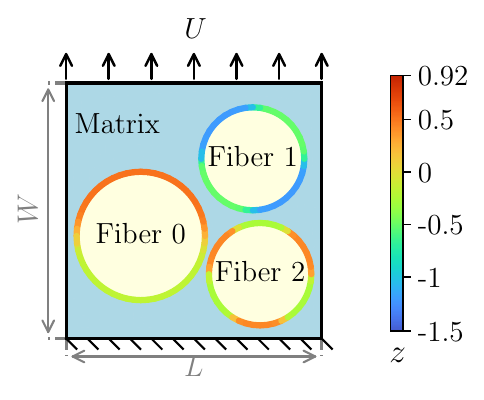}
\caption{Geometry, boundary conditions, and spatial distribution of cohesive laws for the three-fiber composite specimen.
The bottom edge is clamped, and a vertical displacement $U$ is prescribed along the top edge.
Colors along the fiber--matrix interfaces distinguish the local traction--separation laws selected from the learned constitutive family.}
    \label{fig:c3fdd_structure}
\end{figure}

We next consider a three-fiber composite specimen to investigate the identification of spatially heterogeneous cohesive laws along multiple curved interfaces.
The plane-strain model uses the same linear-elastic fiber and matrix properties as in \cref{sec:composite}.
As illustrated in \cref{fig:c3fdd_structure}, the specimen is a square of dimensions $L\times W=10\times10$~mm containing three fibers with radii of $2.51$, $2.02$, and $2.00$~mm, centered at $(2.92,\,4.02)$, $(7.30,\,7.03)$, and $(7.59,\,2.52)$~mm, respectively, measured from the bottom-left corner.
The bottom edge is clamped, while the top edge is subjected to a prescribed vertical displacement reaching $U=40$~\textmu m, with its horizontal displacement constrained and the lateral edges traction free.
The mesh consists of 1353 nodes, 1152 bulk elements, and 144 cohesive elements distributed along the three fiber--matrix interfaces.
As shown in \cref{fig:c3fdd_structure}, the cohesive laws vary spatially along the fiber--matrix interfaces, with colors indicating the local constitutive responses selected from the learned family.

\begin{figure}[htbp]
    \centering
    \begin{subfigure}[b]{0.25\textwidth}
        \centering
        \includegraphics[width=\linewidth]{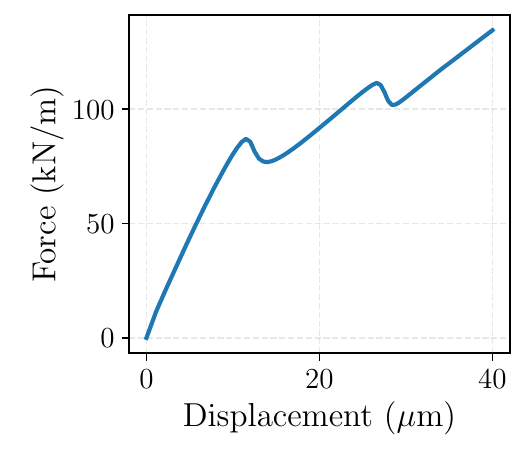}
        \caption{}
        \label{fig:c3fdd_forward_fd}
    \end{subfigure}
    \begin{subfigure}[b]{0.21\textwidth}
        \centering
        \includegraphics[width=\linewidth]{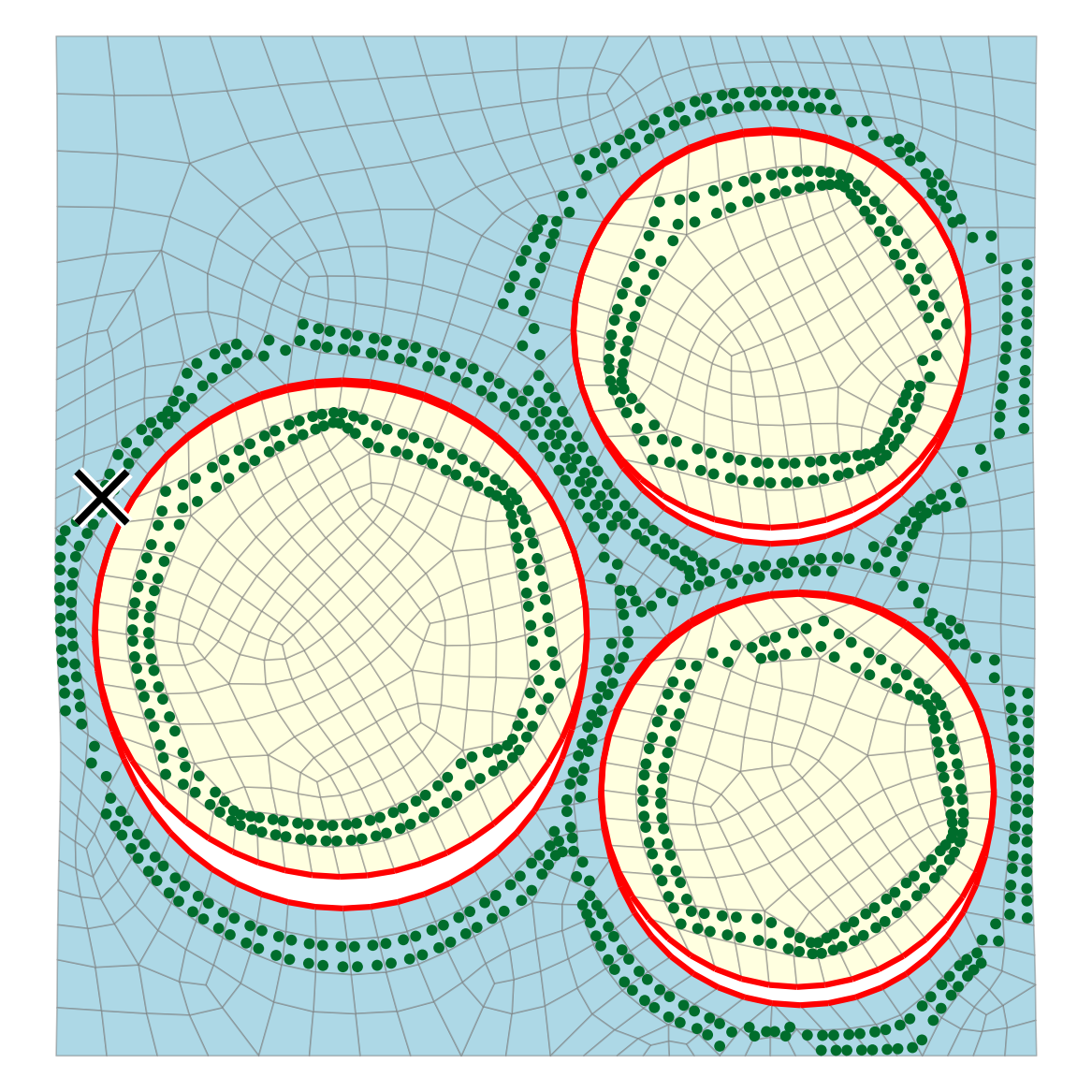}
        \caption{}
        \label{fig:c3fdd_mesh}
    \end{subfigure}
    \caption{Synthetic reference observations for the three-fiber composite specimen.
    (a) Force--displacement response and (b) deformed configuration with displacements magnified by a factor of 10.
    Green markers indicate DIC strain-sampling points, and the cross identifies the point used for the strain-history comparison in \cref{fig:c3fdd_inverse_field}.}
    \label{fig:c3fdd_forward}
\end{figure}

The prescribed constitutive field is used to generate the synthetic structural observations for inverse identification.
The resulting force--displacement response in \cref{fig:c3fdd_forward_fd} exhibits two load drops associated with progressive interfacial debonding and redistribution of load within the composite.
\Cref{fig:c3fdd_mesh} shows the deformed configuration and the bulk Gauss points used to sample the synthetic DIC strain measurements.
Unlike the preceding DCB example, the present specimen involves multiple interacting interfaces with different orientations and local cohesive properties, providing a more complex setting for heterogeneous constitutive identification.

\begin{figure}[htbp]
    \centering
    \begin{subfigure}[b]{0.22\textwidth}
        \centering
        \includegraphics[width=\linewidth]{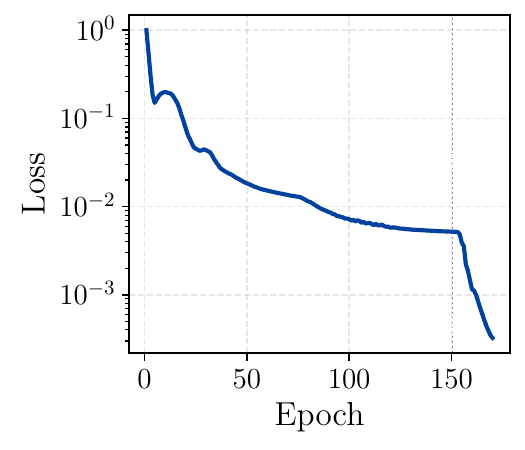}
        \caption{Force-only}
        \label{fig:c3fdd_inverse_loss_force}
    \end{subfigure}
    \begin{subfigure}[b]{0.22\textwidth}
        \centering
        \includegraphics[width=\linewidth]{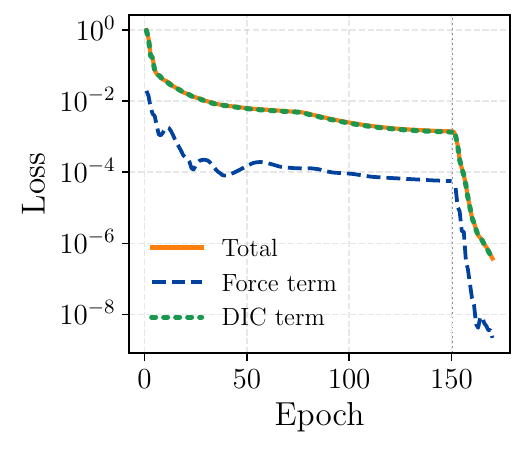}
        \caption{Force+DIC}
        \label{fig:c3fdd_inverse_loss_both}
    \end{subfigure}
    \begin{subfigure}[b]{0.22\textwidth}
        \centering
        \includegraphics[width=\linewidth]{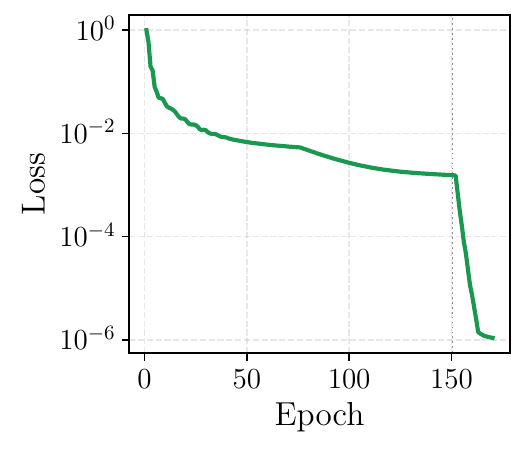}
        \caption{DIC-only}
        \label{fig:c3fdd_inverse_loss_dic}
    \end{subfigure}
    \caption{Normalized loss histories for inverse identification of the three-fiber composite.
    The vertical dotted lines denote the transition from 150 Adam iterations to 20 L-BFGS iterations.
    }
    \label{fig:c3fdd_inverse_loss}
\end{figure}

As in the DCB example, inverse identification is performed using force-only, DIC-only, and combined force--DIC observations, while the pretrained constitutive network remains fixed.
An independent scalar coordinate $z_e$ is assigned to each cohesive element, resulting in 144 unknown latent variables that characterize the spatial distribution of interfacial behavior.
The optimization histories for the three identification strategies are shown in \cref{fig:c3fdd_inverse_loss}.
Consistent with the preceding examples, the loss decreases during the initial Adam iterations, while switching to L-BFGS produces a further substantial reduction.

\begin{figure}[htbp]
    \centering
    \begin{subfigure}[b]{0.25\textwidth}
        \centering
        \includegraphics[width=\linewidth]{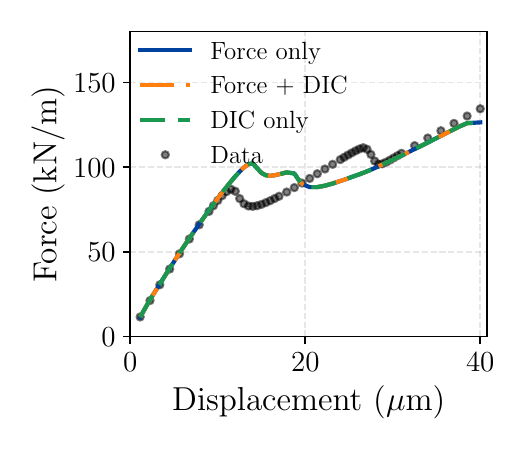}
        \caption{Initial guess}
    \end{subfigure}
    \begin{subfigure}[b]{0.25\textwidth}
        \centering
        \includegraphics[width=\linewidth]{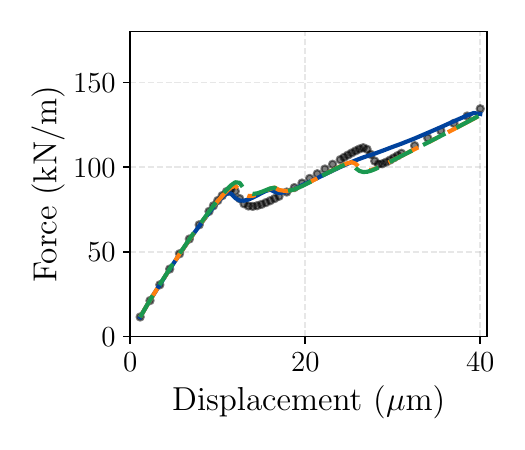}
        \caption{Epoch 4}
    \end{subfigure}
    \begin{subfigure}[b]{0.25\textwidth}
        \centering
        \includegraphics[width=\linewidth]{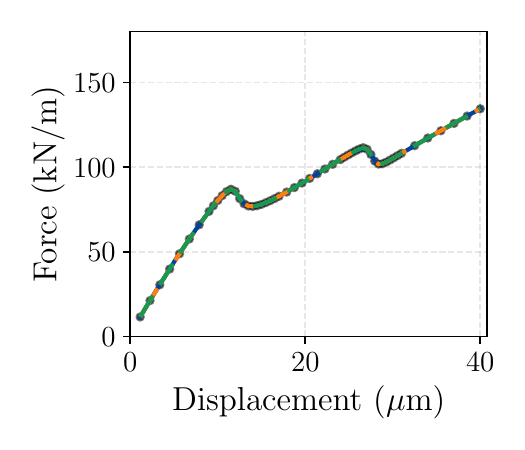}
        \caption{Epoch 170}
    \end{subfigure}
    \caption{Evolution of the predicted force--displacement responses for the three-fiber composite during inverse identification.
    Dots denote the synthetic reference data, and solid curves represent predictions obtained using the three observation types.}
    \label{fig:c3fdd_inverse_fit}
\end{figure}

\begin{figure}[htbp]
    \centering
    \begin{subfigure}[b]{0.25\textwidth}
        \centering
        \includegraphics[width=\linewidth]{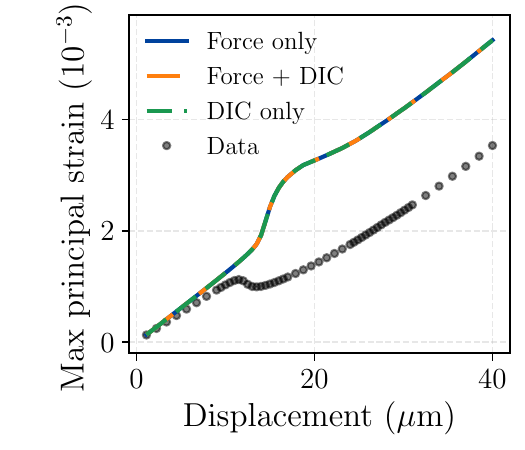}
        \caption{Initial guess}
    \end{subfigure}
    \begin{subfigure}[b]{0.25\textwidth}
        \centering
        \includegraphics[width=\linewidth]{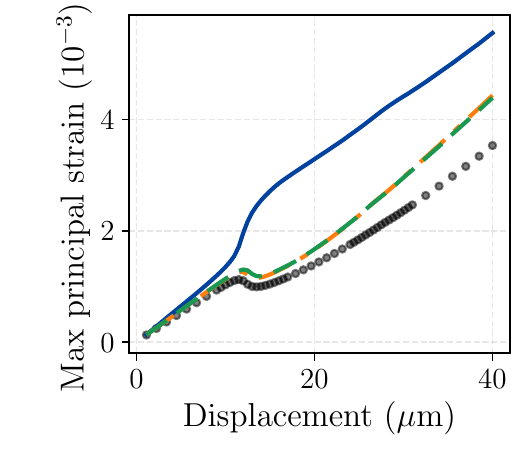}
        \caption{Epoch 4}
    \end{subfigure}
    \begin{subfigure}[b]{0.25\textwidth}
        \centering
        \includegraphics[width=\linewidth]{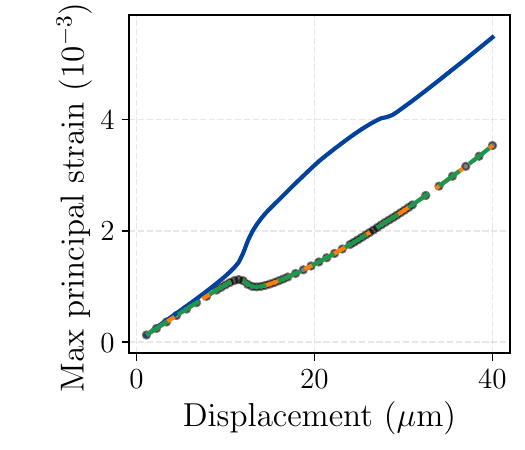}
        \caption{Epoch 170}
    \end{subfigure}
    \caption{Evolution of the predicted maximum principal strain histories at the selected DIC point indicated in \cref{fig:c3fdd_mesh}.
    }
    \label{fig:c3fdd_inverse_field}
\end{figure}

\Cref{fig:c3fdd_inverse_fit} compares the predicted force--displacement responses with the reference data at different stages of optimization. All three identification strategies closely reproduce the global force response. However, their predictions of the local deformation response differ considerably, as shown in \cref{fig:c3fdd_inverse_field}, which compares the maximum principal strain histories at a selected DIC point.
Both the DIC-only and combined force--DIC identifications closely reproduce the reference strain evolution, whereas the force-only identification exhibits substantial deviations after the first load drop.
Consistent with the preceding DCB example, these results suggest that global force measurements alone may be insufficient to reliably recover local deformation and interfacial damage behavior, whereas spatially resolved strain measurements provide additional information about load transfer and damage evolution around the fibers.

\begin{figure}[htbp]
    \centering
    \begin{subfigure}[b]{0.25\textwidth}
        \centering
        \includegraphics[width=\linewidth]{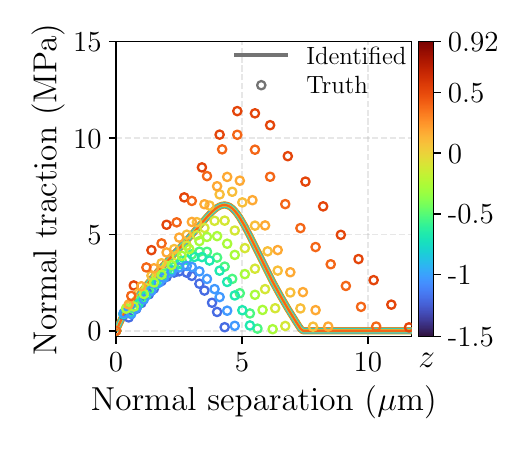}
        \caption*{Initial guess}
    \end{subfigure}
    \begin{subfigure}[b]{0.25\textwidth}
        \centering
        \includegraphics[width=\linewidth]{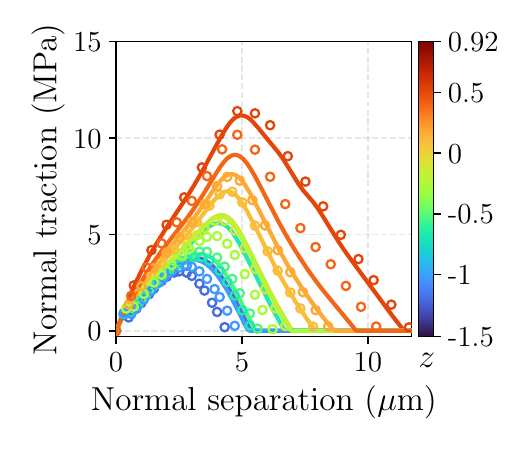}
        \caption*{Epoch 86}
    \end{subfigure}
    \begin{subfigure}[b]{0.25\textwidth}
        \centering
        \includegraphics[width=\linewidth]{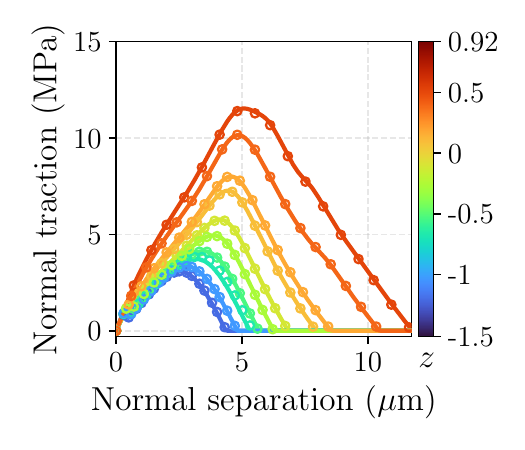}
        \caption*{Epoch 170}
    \end{subfigure}

    \medskip

    \begin{subfigure}[b]{0.25\textwidth}
        \centering
        \includegraphics[width=\linewidth]{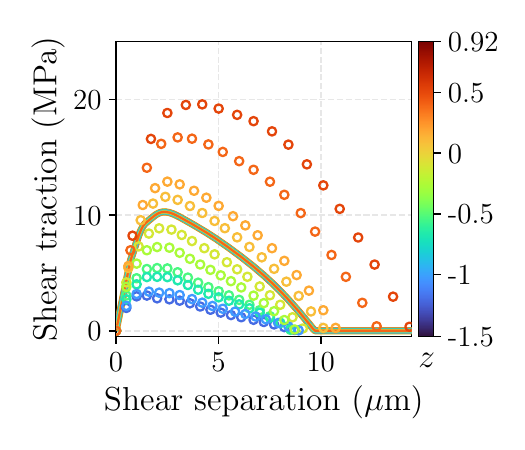}
        \caption*{Initial guess}
    \end{subfigure}
    \begin{subfigure}[b]{0.25\textwidth}
        \centering
        \includegraphics[width=\linewidth]{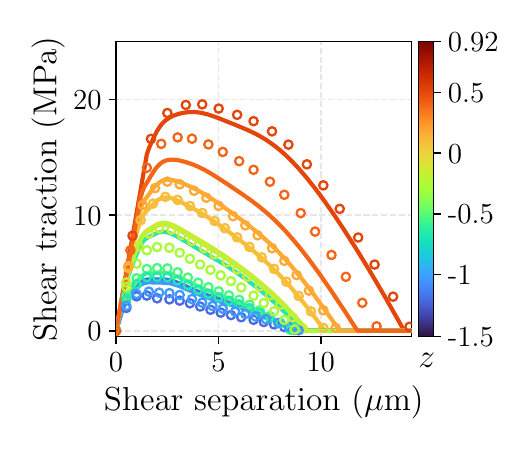}
        \caption*{Epoch 86}
    \end{subfigure}
    \begin{subfigure}[b]{0.25\textwidth}
        \centering
        \includegraphics[width=\linewidth]{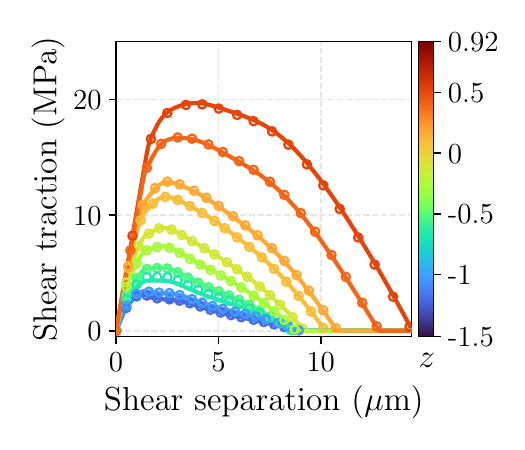}
        \caption*{Epoch 170}
    \end{subfigure}
    \caption{Evolution of the recovered local traction--separation laws at ten representative cohesive elements using combined force--DIC observations.
    The top and bottom rows show Mode~I and Mode~II responses, respectively, at the initial guess and two subsequent optimization stages.
    Solid curves represent the identified laws, while open circles denote the reference responses.
    Colors indicate the reference latent coordinate of each element.}
    \label{fig:c3fdd_inverse_ts_recovery}
\end{figure}

To determine whether these differences in structural-response predictions are reflected in the recovered constitutive behavior, we next examine the local traction--separation laws.
\Cref{fig:c3fdd_inverse_ts_recovery} illustrates the evolution of the identified Mode~I and Mode~II responses at ten representative cohesive elements using combined force--DIC observations.
Starting from the initial estimates, the identified laws progressively approach their reference responses, with the final predictions closely capturing the variations in peak traction, characteristic separation, and post-peak softening across both loading modes.
Notably, these responses are recovered through the element-wise latent coordinates of the shared constitutive prior, without independently fitting the network parameters at each interface location.

\begin{figure}[htbp]
    \centering
    \begin{subfigure}[b]{0.25\textwidth}
        \centering
        \includegraphics[width=\linewidth]{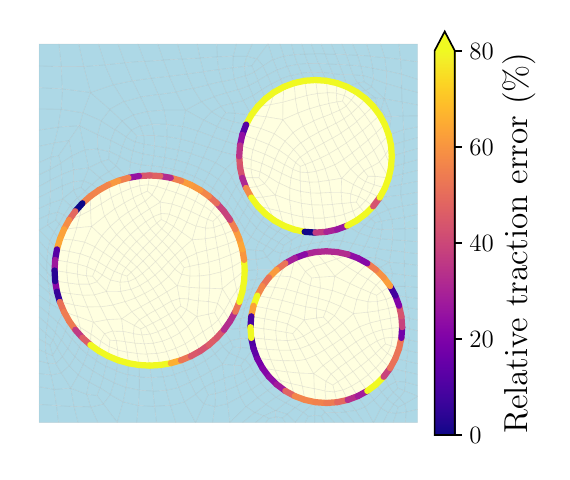}
        \caption{Force-only}
    \end{subfigure}
    \begin{subfigure}[b]{0.25\textwidth}
        \centering
        \includegraphics[width=\linewidth]{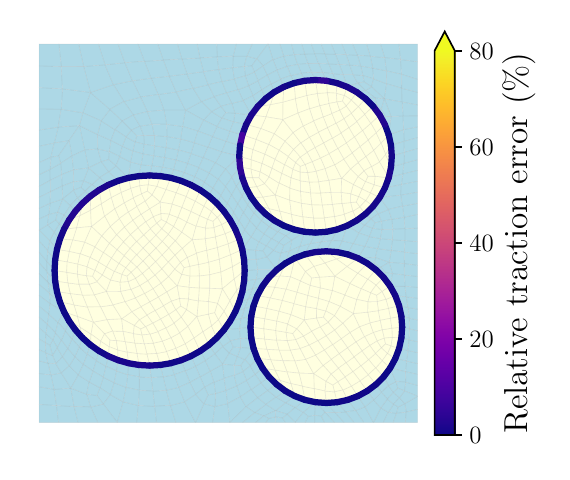}
        \caption{Force+DIC}
    \end{subfigure}
    \begin{subfigure}[b]{0.25\textwidth}
        \centering
        \includegraphics[width=\linewidth]{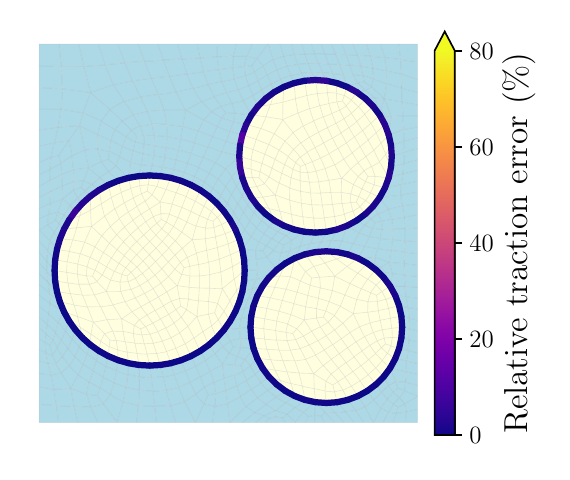}
        \caption{DIC-only}
    \end{subfigure}
    \caption{Element-wise relative traction--separation error along the three fiber--matrix interfaces after inverse identification.
    All panels share the same percentage color scale, with the upper extension indicating errors exceeding $80\%$.}
    \label{fig:c3fdd_ts_error_mesh}
\end{figure}

The accuracy of constitutive recovery over the entire specimen is quantified in \cref{fig:c3fdd_ts_error_mesh}, which maps the relative traction--separation error defined in \cref{eq:rel_traction_error} along all three fiber--matrix interfaces.
The force-only identification exhibits substantial errors across large portions of the interfaces, with a mean error of $89.8\%$ and a maximum of $407\%$ over the 144 cohesive elements.
In contrast, the combined force--DIC identification reduces the mean and maximum errors to $0.75\%$ and $8.8\%$, respectively, while the DIC-only identification achieves corresponding errors of $1.06\%$ and $11.8\%$.
These differences are particularly significant given that all three identification strategies reproduce the global force--displacement response with comparable accuracy.
The results demonstrate that incorporating spatial strain measurements substantially improves the recovery of heterogeneous cohesive properties in this benchmark, even when the admissible constitutive responses are already constrained by a learned prior.

Together with the heterogeneous DCB results, this example further illustrates the limitations of identifying spatially varying cohesive properties from global force measurements alone.
Although the learned constitutive prior substantially reduces the admissible space of local material responses, different spatial distributions can still produce similar structural force histories.
Spatially resolved strain observations provide additional constraints that help distinguish these distributions and improve constitutive recovery.
These findings highlight the importance of combining physically informed constitutive representations with sufficiently informative structural measurements for inverse identification of heterogeneous interfaces.

\subsubsection{Comparison with Baseline Methods}
\label{sec:meta_dd_comparison}

We next compare the proposed approach with baseline methods to investigate the roles of constitutive representation and optimization strategy in identifying heterogeneous cohesive behavior.
While the preceding DCB example established the importance of accounting for spatial variations in cohesive properties, we now focus on two complementary questions.
First, we examine whether a conventional bilinear cohesive model with spatially varying parameters can achieve comparable identification accuracy to the learned constitutive prior.
Second, we assess the performance and computational cost of the proposed adjoint-based identification method against derivative-free Bayesian optimization (BO).
All comparisons use the same three-fiber composite specimen and combined force--DIC observations.

\paragraph{Spatially heterogeneous bilinear cohesive laws.}

We investigate whether allowing the parameters of a conventional bilinear cohesive model to vary spatially is sufficient to recover the underlying heterogeneous interfacial behavior from structural observations.
For this comparison, each cohesive element is assigned five independent parameters comprising the Mode~I and Mode~II fracture energies $G_{Ic}$ and $G_{IIc}$, the corresponding peak tractions $\sigma_n^0$ and $\sigma_t^0$, and the initiation-to-failure separation ratio $\lambda_c$, as described in \cref{app:ref_bilinear}.
This results in $144\times5=720$ unknown parameters, compared with 144 latent coordinates for the learned constitutive prior.
Both models are identified using the same combined force--DIC observations and adjoint-based optimization strategy.

\begin{figure}[htbp]
    \centering
    \includegraphics[width=0.28\textwidth]{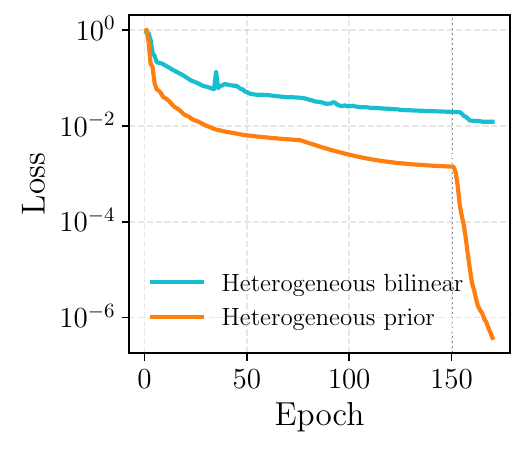}
    \caption{Identification loss histories for the spatially heterogeneous bilinear model and learned constitutive prior under the same combined force--DIC objective.
    The vertical dotted line marks the transition from Adam to L-BFGS after 150 iterations.}
    \label{fig:c3fdd_bilinear_dic_loss}
\end{figure}

As shown by the optimization histories in \cref{fig:c3fdd_bilinear_dic_loss}, the heterogeneous bilinear model cannot reproduce the observed data as accurately as the learned constitutive prior, despite having more adjustable parameters.

\begin{figure}[htbp]
    \centering
    \begin{subfigure}[b]{0.25\textwidth}
        \centering
        \includegraphics[width=\linewidth]{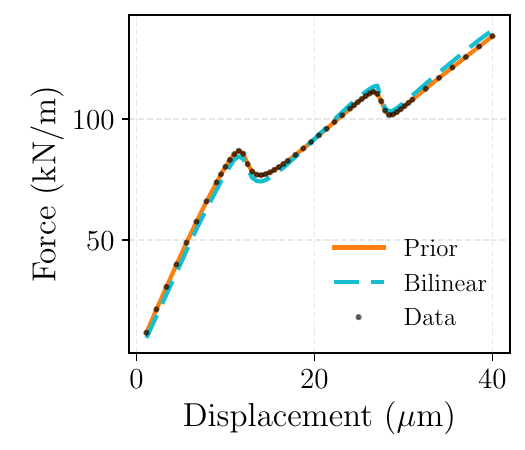}
        \caption{}
        \label{fig:c3fdd_bilinear_dic_fd_hetero}
    \end{subfigure}
    \begin{subfigure}[b]{0.25\textwidth}
        \centering
        \includegraphics[width=\linewidth]{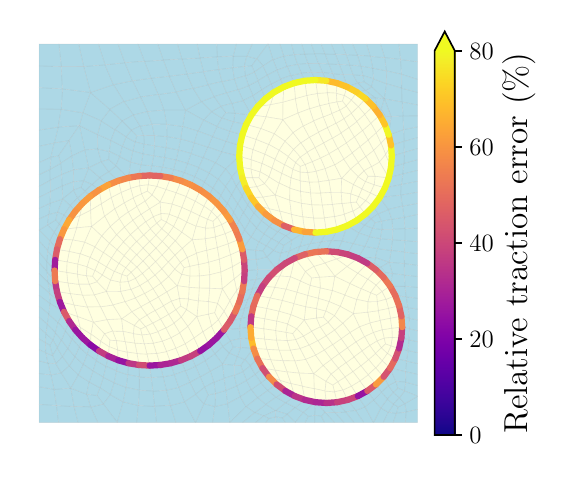}
        \caption{}
        \label{fig:c3fdd_ts_error_bilinear}
    \end{subfigure}
    \begin{subfigure}[b]{0.25\textwidth}
        \centering
        \includegraphics[width=\linewidth]{pics/meta_comp3fiber_data_driven/meta_comp3fiber_data_driven_inverse_ts_error_mesh_force_dic.pdf}
        \caption{}
    \end{subfigure}
    \caption{Effect of constitutive parameterization on the identification of heterogeneous cohesive behavior.
    (a) Force--displacement predictions for the heterogeneous bilinear model and learned prior compared with the reference data.
    (b) Relative cohesive law errors for the bilinear model with 720 unknown parameters and (c) the learned prior with 144 latent coordinates.
    Both error maps use the same color scale as \cref{fig:c3fdd_ts_error_mesh}.}
    \label{fig:c3fdd_bilinear}
\end{figure}

\Cref{fig:c3fdd_bilinear} compares the corresponding force--displacement predictions and spatial distributions of cohesive law errors.
The heterogeneous bilinear model closely reproduces the global force response, including both load drops, but exhibits substantial errors in the recovered local constitutive behavior.
The mean and maximum relative traction errors reach $59.5\%$ and $135\%$, respectively, compared with $0.75\%$ and $8.8\%$ for the learned prior.
The reference cohesive responses exhibit variations in softening shape that cannot be represented exactly by conventional bilinear laws.
Consequently, allowing the bilinear parameters to vary independently across the interfaces does not eliminate the limitations of the underlying constitutive representation.

\paragraph{Comparison with Bayesian optimization.}
\label{sec:meta_dd_bo}

Finally, we assess the role of the optimization strategy by comparing the proposed adjoint-based method with derivative-free Bayesian optimization (BO).
Both heterogeneous constitutive representations are considered, allowing the effect of the optimizer to be examined while holding the constitutive parameterization fixed.
For each model, BO uses the same forward problem, observations, and identification objective as its adjoint-based counterpart.

The BO surrogate is a single-task Gaussian process with an automatic-relevance-determination radial-basis-function kernel.
Its hyperparameters are fitted by maximizing the exact marginal likelihood, with inputs normalized to the unit cube and the objective standardized.
The log expected improvement acquisition function is optimized using eight restarts and 256 raw samples.
Each run begins with 20 scrambled Sobol points, followed by 300 sequential evaluations, giving 320 forward-model evaluations in total.

The learned prior reduces the BO search dimension from 720 bilinear parameters to 144 latent coordinates, although both remain high-dimensional optimization problems.
Unlike the proposed differentiable approach, BO does not use the available adjoint sensitivities of the structural objective.

\begin{figure}[htbp]
    \centering
    \begin{subfigure}[b]{0.3\textwidth}
        \centering
        \includegraphics[width=\linewidth]{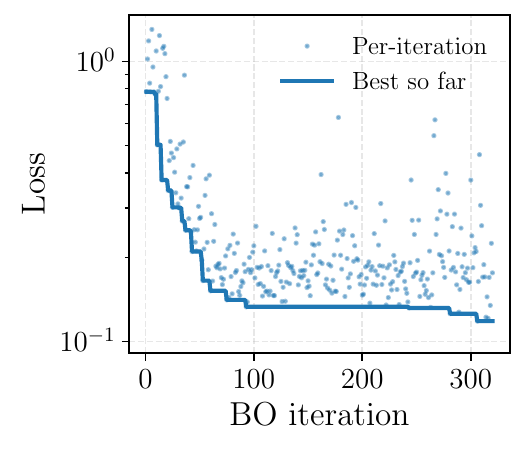}
        \caption{Learned prior}
        \label{fig:c3fdd_bo_loss}
    \end{subfigure}
    \begin{subfigure}[b]{0.3\textwidth}
        \centering
        \includegraphics[width=\linewidth]{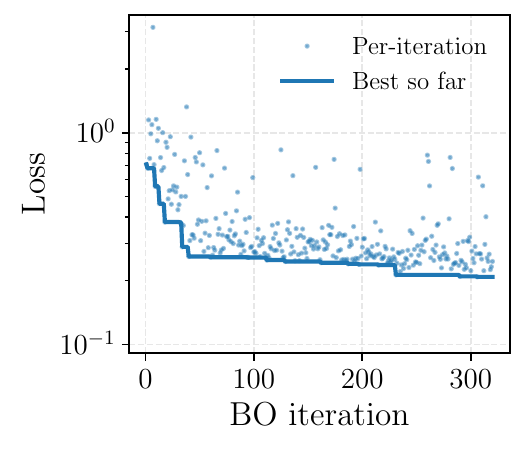}
        \caption{Bilinear model}
        \label{fig:c3fdd_bo_bilinear_loss}
    \end{subfigure}
    \caption{Bayesian optimization loss histories for the three-fiber composite using (a) the learned constitutive prior with 144 latent coordinates and (b) the heterogeneous bilinear model with 720 parameters.
    Points denote individual candidate evaluations, and solid curves indicate the lowest loss attained up to each evaluation.
    Each run includes 20 initial Sobol points and 300 subsequent BO evaluations.}
    \label{fig:c3fdd_bo_losses}
\end{figure}

The optimization histories in \cref{fig:c3fdd_bo_loss,fig:c3fdd_bo_bilinear_loss} show that most improvements occur during the early evaluations, followed by extended intervals of limited progress.
The individual candidate losses fluctuate considerably, while the best-found losses decrease only gradually after approximately the first 100 evaluations.
For the learned constitutive prior, BO attains a final loss of order $10^{-1}$, whereas adjoint-based identification reduces the same objective below $10^{-6}$.
This difference demonstrates that restricting the search to an appropriate constitutive family is not sufficient by itself, and that the optimization strategy also plays a critical role in effectively identifying the unknown spatial distribution.

\begin{figure}[htbp]
    \centering
    \begin{subfigure}[b]{0.3\textwidth}
        \centering
        \includegraphics[width=\linewidth]{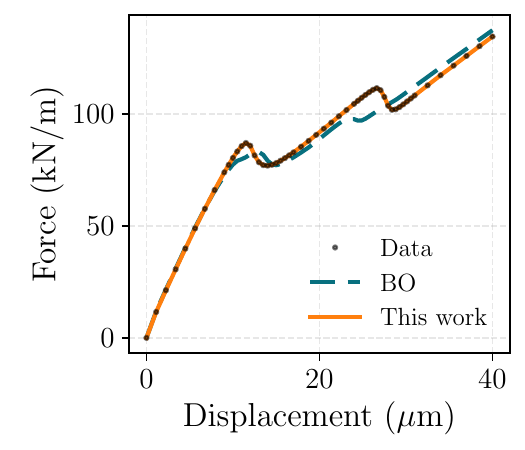}
        \caption{Learned prior}
    \end{subfigure}
    \begin{subfigure}[b]{0.3\textwidth}
        \centering
        \includegraphics[width=\linewidth]{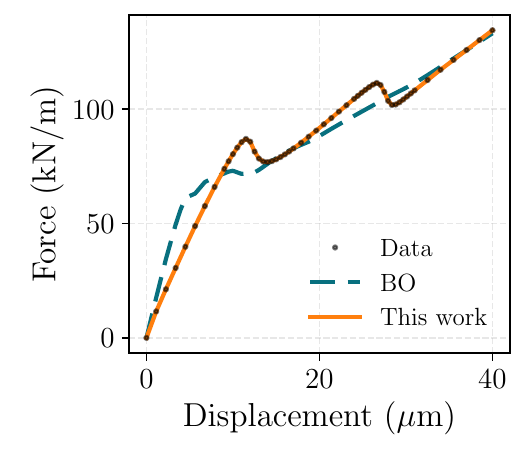}
        \caption{Bilinear model}
    \end{subfigure}
    \caption{Best-found force--displacement predictions after 320 BO evaluations using (a) the learned prior and (b) the heterogeneous bilinear model.
    Dots denote the reference data, dashed curves represent BO predictions, and solid curves show the adjoint-based result with the learned prior for comparison.}
    \label{fig:c3fdd_bo_fit}
\end{figure}

\begin{figure}[htbp]
    \centering
    \begin{subfigure}[b]{0.25\textwidth}
        \centering
        \includegraphics[width=\linewidth]{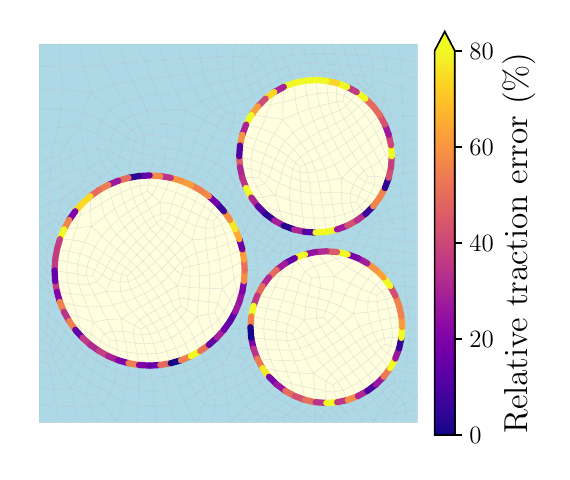}
        \caption{BO with prior}
    \end{subfigure}
    \begin{subfigure}[b]{0.25\textwidth}
        \centering
        \includegraphics[width=\linewidth]{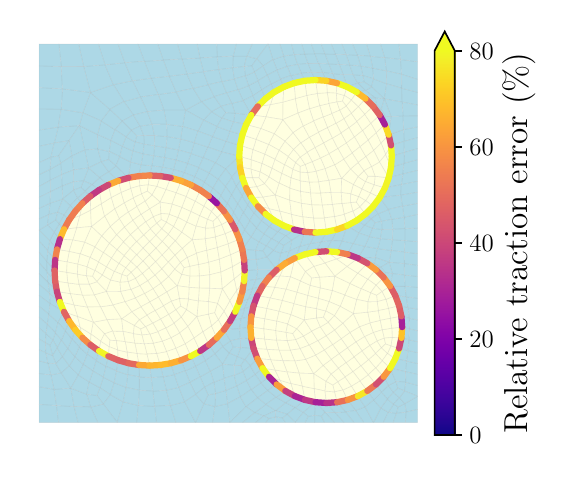}
        \caption{BO with bilinear law}
    \end{subfigure}
    \begin{subfigure}[b]{0.25\textwidth}
        \centering
        \includegraphics[width=\linewidth]{pics/meta_comp3fiber_data_driven/meta_comp3fiber_data_driven_inverse_ts_error_mesh_force_dic.pdf}
        \caption{Adjoint with prior}
    \end{subfigure}
    \caption{Relative cohesive law errors after identification using (a) BO with the learned prior, (b) BO with the heterogeneous bilinear model, and (c) adjoint-based identification with the learned prior.
    All panels use the same color scale as \cref{fig:c3fdd_ts_error_mesh}.}
    \label{fig:c3fdd_ts_error_bo}
\end{figure}

The best-found force--displacement predictions after 320 BO evaluations are compared with the reference data in \cref{fig:c3fdd_bo_fit}.
Both parameterizations capture the overall force trend but retain discrepancies around the load drops.
The corresponding local cohesive law errors in \cref{fig:c3fdd_ts_error_bo} are also substantial, with mean and maximum errors of $59.4\%$ and $432\%$ for the learned prior and $71.2\%$ and $257\%$ for the bilinear model.
Although the learned prior yields a smaller mean error in these BO runs, this comparison changes both the constitutive family and the number of unknown parameters.
More importantly, comparison with adjoint-based identification using the same learned prior reveals markedly better constitutive recovery by the differentiable approach under the tested optimization settings.

Beyond recovery accuracy, computational efficiency is another important consideration.
Direct comparison of iteration counts is insufficient because each adjoint-based iteration involves forward and sensitivity computations, with additional objective evaluations potentially required during L-BFGS refinement, whereas each BO evaluation involves a forward simulation, surrogate updating, and acquisition optimization.
We therefore compare the elapsed computational times and achieved losses, excluding offline constitutive-prior training.

For this comparison, all four identifications were run on the same laptop, equipped with an AMD Ryzen AI 9 365 processor.
For the 144-dimensional learned prior, adjoint-based identification required $5524$~s for 150 Adam and 20 L-BFGS iterations, whereas BO required $6961$~s for 320 evaluations.
For the 720-parameter bilinear model, the corresponding times were $1963$~s and $10\,272$~s.
BO was therefore approximately $1.26$ times slower for the learned prior and $5.23$ times slower for the bilinear model, while attaining final objective values approximately $3.5\times10^5$ and $17$ times larger, respectively.
In both prior and bilinear cases, Gaussian-process fitting becomes increasingly costly as more observations are accumulated and accounts for approximately $59$--$64\%$ of the total BO runtime.

In the bilinear BO run, 18 of the 320 trial forward simulations terminated before reaching the prescribed final displacement, between $48\%$ and $99\%$ of the loading history.
For these trials, repeated load-step reductions either exhausted the permitted number of steps or failed to achieve convergence.
Their objective values were evaluated over the attained portion of the loading history and normalized by the fraction of displacement reached.

The limited progress observed in the BO runs is consistent with the difficulty of constructing and optimizing a surrogate of a high-dimensional structural objective using a limited number of evaluations.
However, these results should not be interpreted as a general limitation of Bayesian optimization, since alternative surrogate models, acquisition strategies, and high-dimensional BO methods may yield different outcomes \citep{Eriksson2021,papenmeier2025}.

Overall, the baseline comparisons highlight the complementary roles of constitutive representation and differentiable optimization in the proposed framework.
The bilinear comparison demonstrates that allowing cohesive parameters to vary independently across the interfaces does not necessarily ensure accurate recovery when the assumed constitutive family cannot adequately represent the underlying behavior.
The BO comparison further demonstrates the advantages of exploiting adjoint sensitivities for improving identification accuracy and computational efficiency in high-dimensional inverse problems.
These findings are established using synthetic data generated within the learned constitutive family, and assessing the proposed approach under constitutive model mismatch and experimental uncertainty remains an important direction for future work.

\section{Conclusions}
\label{sec:conclusions}

This work develops a differentiable finite element framework based on an intrinsic cohesive-zone formulation for identifying traction--separation laws from indirect structural observations, including global force--displacement responses and spatially resolved strain fields.
Building on a thermodynamically structured neural cohesive formulation, the framework combines finite element simulations with discrete adjoint sensitivity analysis that accounts for the history-dependent evolution of interfacial damage.
Two complementary identification settings were investigated, including direct identification of neural constitutive parameters for homogeneous interfaces and identification of spatially heterogeneous cohesive laws within a pretrained constitutive family.

The homogeneous examples demonstrated that flexible cohesive laws can be recovered from structural observations without prescribing a particular analytical traction--separation form.
However, the accuracy of constitutive recovery depends strongly on the information contained in the measurements.
The DCB comparisons revealed that, under small-scale bridging, substantially different cohesive laws can reproduce nearly identical global force--displacement responses, even when their fracture energies are accurately recovered.
In contrast, the larger process zone provides greater sensitivity to the cohesive law shape, enabling more reliable identification from global force measurements.
The mixed-mode composite example further demonstrated that accurate reproduction of the global structural response does not necessarily imply accurate recovery of the underlying constitutive behavior.
Across these examples, incorporating spatial strain measurements generally improved the identification of cohesive law features that were insufficiently constrained by global force observations alone.

For spatially heterogeneous interfaces, we extended the constitutive-prior concept to irreversible cohesive damage by learning a shared family of traction--separation laws from representative constitutive histories.
The pretrained family restricts the admissible constitutive responses, allowing the inverse problem to be formulated in terms of low-dimensional, element-wise latent coordinates rather than independently identifying complete constitutive models at each location.
The heterogeneous DCB and three-fiber composite examples demonstrated the effectiveness of this approach for recovering spatial variations in cohesive behavior.
Nevertheless, the learned prior alone did not eliminate the limited identifiability associated with global force measurements.
In the three-fiber composite, all three observation strategies reproduced the global force response closely, while the mean local traction error decreased from $89.8\%$ for force-only identification to $0.75\%$ when force and DIC observations were combined.
These findings highlight the complementary roles of constitutive prior knowledge and spatially informative measurements in identifying heterogeneous interfacial behavior.

The baseline comparisons further demonstrated the importance of constitutive representation and optimization strategy.
A spatially heterogeneous bilinear cohesive model, despite having more adjustable parameters, exhibited substantially larger constitutive recovery errors than the learned prior, illustrating the limitations of an assumed constitutive family that cannot adequately represent the underlying responses.
Comparisons with Bayesian optimization also showed that exploiting adjoint sensitivities substantially improved identification accuracy and computational efficiency under the investigated settings.
Together, these results indicate that restricting the admissible constitutive space and efficiently optimizing within that space are complementary requirements for accurate high-dimensional inverse identification.

The present numerical studies are limited to two-dimensional, small-strain, quasi-static problems with linear-elastic bulk materials and scalar interfacial damage.
The inverse identification experiments use synthetic, noise-free observations generated on the same discretizations used for identification, while the heterogeneous reference laws belong to the learned constitutive family.
Consequently, the results do not establish robustness to experimental uncertainty, discretization errors, or constitutive behavior outside the learned prior.
Future work will investigate these effects, including the influence of prior expressiveness on identifiability and the potential for adaptive refinement of learned constitutive families.
Extensions to finite deformations, rate- and temperature-dependent behavior, cyclic loading, and three-dimensional interfaces will further broaden the applicability of the framework.

\section*{Data Availability}
All data used in can be reproduced from the information provided and are available from the corresponding author upon reasonable request.

\section*{Acknowledgment}
B.B. acknowledges support from the startup fund provided by Northwestern University.
The authors acknowledge the use of the open-source machine learning libraries PyTorch~\citep{paszke2019pytorch}, SciPy~\citep{2020SciPy-NMeth}, and BoTorch~\citep{balandat2020botorch}. This research was supported in part through the computational resources and staff contributions provided for the Quest high-performance computing facility at Northwestern University.

\section*{Declaration of generative AI and AI-assisted technologies in the manuscript preparation process}
During the preparation of this work, the authors used ChatGPT and Claude to assist with language refinement and improve the clarity of the manuscript. Additionally, the first and second authors used LLM-based coding assistants to assist with the partial implementation of certain code components and improve code modularity. The authors reviewed and revised all AI-assisted content and code as necessary and take full responsibility for the content of the published article.

\bibliographystyle{plainnat}
\bibliography{bibliography}

@article{hillerborg1976analysis,
  title={Analysis of crack formation and crack growth in concrete by means of fracture mechanics and finite elements},
  author={Hillerborg, A. and Mod{\'e}er, M. and Petersson, P.-E.},
  journal={Cement and Concrete Research},
  volume={6},
  number={6},
  pages={773--781},
  year={1976},
  publisher={Elsevier}
}

@article{park2011,
  title={Cohesive zone models: a critical review of traction-separation relationships across fracture surfaces},
  author={Park, Kyoungsoo and Paulino, Glaucio H},
  journal={Applied Mechanics Reviews},
  volume={64},
  number={6},
  pages={060802},
  year={2011},
  publisher={American Society of Mechanical Engineers}
}

@article{benzeggagh1996measurement,
  title={Measurement of mixed-mode delamination fracture toughness of unidirectional glass/epoxy composites with mixed-mode bending apparatus},
  author={Benzeggagh, ML and Kenane, MJCS},
  journal={Composites Science and Technology},
  volume={56},
  number={4},
  pages={439--449},
  year={1996},
  publisher={Elsevier}
}

@article{sills2015cohesive,
  title={Cohesive-length scales for damage and toughening mechanisms},
  author={Sills, RB and Thouless, MD},
  journal={International Journal of Solids and Structures},
  volume={55},
  pages={32--43},
  year={2015},
  publisher={Elsevier}
}

@article{Shen2010,
  title={Inverse computation of cohesive fracture properties from displacement fields},
  author={Shen, B. and Stanciulescu, I. and Paulino, G. H.},
  journal={Inverse Problems in Science and Engineering},
  volume={18},
  number={8},
  pages={1103--1128},
  year={2010}
}

@article{becker2024extracting,
  title={Extracting fracture properties from digital image and volume correlation displacement data: A review},
  author={Becker, Thorsten H},
  journal={Strain},
  volume={60},
  number={4},
  pages={e12469},
  year={2024},
  publisher={Wiley Online Library}
}

@article{Wei2026,
  title={Extracting interfacial interactions using elastic field decomposition and a deep complex neural network},
  author={Wei, Congjie and Zhang, Jiaxin and Liechti, Kenneth M and Wu, Chenglin},
  journal={International Journal of Solids and Structures},
  volume={334},
  pages={113980},
  year={2026},
  publisher={Elsevier}
}

@article{Jensen2019,
  title={Inverse parameter identification of n-segmented multilinear cohesive laws using parametric finite element modeling},
  author={Jensen, S. M. and Martos, M. J. and Lindgaard, E. and Bak, B. L. V.},
  journal={Composite Structures},
  volume={225},
  pages={111074},
  year={2019}
}

@article{tang2025bayesian,
  title={Bayesian inference for the calibration of cohesive zone models of dovetail specimens from laminated composite fan blade},
  author={Tang, Xu and Chen, Yong},
  journal={Journal of Engineering for Gas Turbines and Power},
  volume={147},
  number={5},
  pages={051021},
  year={2025},
  publisher={American Society of Mechanical Engineers}
}

@article{albuquerque2018bayesian,
  title={{A Bayesian framework for the calibration of cohesive zone models}},
  author={Albuquerque, Emerson B and Guzman, Cynthia and Borges, Lavinia A and Castello, Daniel A},
  journal={The Journal of Adhesion},
  volume={94},
  number={4},
  pages={255--277},
  year={2018},
  publisher={Taylor \& Francis}
}

@inproceedings{Eriksson2021,
  title={High-dimensional {Bayesian} optimization with sparse axis-aligned subspaces},
  author={Eriksson, David and Jankowiak, Martin},
  booktitle={Uncertainty in Artificial Intelligence},
  pages={493--503},
  year={2021},
  organization={PMLR}
}

@inproceedings{papenmeier2025,
  title={Understanding High-Dimensional {Bayesian} Optimization},
  author={Papenmeier, Leonard and Poloczek, Matthias and Nardi, Luigi},
  booktitle={International Conference on Machine Learning},
  pages={47902--47923},
  year={2025},
  organization={PMLR}
}

@article{han2024topology,
  title={Topology optimization with a finite strain nonlocal damage model using the continuous adjoint method},
  author={Han, Jike and Furuta, Kozo and Kondoh, Tsuguo and Izui, Kazuhiro and Nishiwaki, Shinji and Terada, Kenjiro},
  journal={Computer Methods in Applied Mechanics and Engineering},
  volume={432},
  pages={117333},
  year={2024},
  publisher={Elsevier}
}

@article{oberai2003,
  title={Solution of inverse problems in elasticity imaging using the adjoint method},
  author={Oberai, Assad A and Gokhale, Nachiket H and Feij{\'o}o, Gonzalo R},
  journal={Inverse Problems},
  volume={19},
  number={2},
  pages={297--313},
  year={2003}
}

@article{Dugdale1960,
  title={Yielding of steel sheets containing slits},
  author={Dugdale, Donald S.},
  journal={Journal of the Mechanics and Physics of Solids},
  volume={8},
  number={2},
  pages={100--104},
  year={1960},
  publisher={Elsevier}
}

@article{Barenblatt1962,
  title={The mathematical theory of equilibrium cracks in brittle fracture},
  author={Barenblatt, Grigory I.},
  journal={Advances in Applied Mechanics},
  volume={7},
  pages={55--129},
  year={1962},
  publisher={Elsevier}
}

@article{xu2014inverse,
  title={Inverse parameter identification of cohesive zone model for simulating mixed-mode crack propagation},
  author={Xu, Yangjian and Li, Xiangyu and Wang, Xiaogui and Liang, Lihua},
  journal={International Journal of Solids and Structures},
  volume={51},
  number={13},
  pages={2400--2410},
  year={2014},
  publisher={Elsevier}
}

@article{li2020spatial,
  title={How the spatial correlation in adhesion properties influences the performance of secondary bonding of laminated composites},
  author={Li, Xiaole and Tao, Ran and Yudhanto, Arief and Lubineau, Gilles},
  journal={International Journal of Solids and Structures},
  volume={196},
  pages={41--52},
  year={2020},
  publisher={Elsevier}
}

@article{bao1992remarks,
  title={Remarks on crack-bridging concepts},
  author={Bao, G and Suo, Z},
  journal={Applied Mechanics Reviews},
  volume={45},
  number={8},
  pages={355--366},
  year={1992}
}

@inproceedings{kim2024scalable,
  title={Scalable monotonic neural networks},
  author={Kim, Hyunho and Lee, Jong-Seok},
  booktitle={The Twelfth International Conference on Learning Representations},
  year={2024}
}

@article{turon2006damage,
  title={A damage model for the simulation of delamination in advanced composites under variable-mode loading},
  author={Turon, A. and Camanho, P. P. and Costa, J. and D{\'a}vila, C. G.},
  journal={Mechanics of Materials},
  volume={38},
  number={11},
  pages={1072--1089},
  year={2006},
  publisher={Elsevier}
}

@article{thakolkaran2022nn,
  title={{NN-EUCLID}: Deep-learning hyperelasticity without stress data},
  author={Thakolkaran, Prakash and Joshi, Akshay and Zheng, Yiwen and Flaschel, Moritz and De Lorenzis, Laura and Kumar, Siddhant},
  journal={Journal of the Mechanics and Physics of Solids},
  volume={169},
  pages={105076},
  year={2022},
  publisher={Elsevier}
}

@article{paszke2019pytorch,
  title={Pytorch: An imperative style, high-performance deep learning library},
  author={Paszke, Adam and Gross, Sam and Massa, Francisco and Lerer, Adam and Bradbury, James and Chanan, Gregory and Killeen, Trevor and Lin, Zeming and Gimelshein, Natalia and Antiga, Luca and others},
  journal={Advances in neural information processing systems},
  volume={32},
  year={2019}
}

@article{diederik2014adam,
  title={Adam: A method for stochastic optimization},
  author={Kingma, Diederik P and Ba, Jimmy},
  journal={International Conference on Learning Representations},
  year={2015}
}

@article{loshchilov2016sgdr,
  title={{SGDR}: Stochastic gradient descent with warm restarts},
  author={Loshchilov, Ilya and Hutter, Frank},
  journal={arXiv preprint arXiv:1608.03983},
  year={2016}
}

@inproceedings{pascanu2013difficulty,
  title={On the difficulty of training recurrent neural networks},
  author={Pascanu, Razvan and Mikolov, Tomas and Bengio, Yoshua},
  booktitle={International conference on machine learning},
  pages={1310--1318},
  year={2013},
  organization={{PMLR}}
}

@article{liu1989limited,
  title={On the limited memory {BFGS} method for large scale optimization},
  author={Liu, Dong C and Nocedal, Jorge},
  journal={Mathematical Programming},
  volume={45},
  number={1},
  pages={503--528},
  year={1989},
  publisher={Springer}
}

@article{sorensen1998large,
  title={Large-scale bridging in composites: R-curves and bridging laws},
  author={S{\o}rensen, Bent F and Jacobsen, Torben K},
  journal={Composites Part A: Applied Science and Manufacturing},
  volume={29},
  number={11},
  pages={1443--1451},
  year={1998},
  publisher={Elsevier}
}

@article{sorensen2003determination,
  title={{Determination of cohesive laws by the J integral approach}},
  author={S{\o}rensen, Bent F and Jacobsen, Torben K},
  journal={Engineering Fracture Mechanics},
  volume={70},
  number={14},
  pages={1841--1858},
  year={2003},
  publisher={Elsevier}
}

@article{wei2023data,
  title={Data driven modeling of interfacial traction--separation relations using a thermodynamically consistent neural network},
  author={Wei, Congjie and Zhang, Jiaxin and Liechti, Kenneth M and Wu, Chenglin},
  journal={Computer Methods in Applied Mechanics and Engineering},
  volume={404},
  pages={115826},
  year={2023},
  publisher={Elsevier}
}

@article{wang2019meta,
  title={Meta-modeling game for deriving theory-consistent, microstructure-based traction--separation laws via deep reinforcement learning},
  author={Wang, Kun and Sun, WaiChing},
  journal={Computer Methods in Applied Mechanics and Engineering},
  volume={346},
  pages={216--241},
  year={2019},
  publisher={Elsevier}
}

@article{camanho2003numerical,
  title={Numerical simulation of mixed-mode progressive delamination in composite materials},
  author={Camanho, Pedro Ponces and D{\'a}vila, Carlos G and de Moura, M{\'a}rio F},
  journal={Journal of Composite Materials},
  volume={37},
  number={16},
  pages={1415--1438},
  year={2003},
  publisher={Sage Publications}
}

@article{tao2024reconstruction,
  title={Reconstruction and prediction of Mode-I cohesive law using artificial neural network},
  author={Tao, Chongcong and Zhang, Chao and Ji, Hongli and Qiu, Jinhao},
  journal={Composites Science and Technology},
  volume={256},
  pages={110755},
  year={2024},
  publisher={Elsevier}
}

@inproceedings{park2019deepsdf,
  title={{DeepSDF}: Learning continuous signed distance functions for shape representation},
  author={Park, Jeong Joon and Florence, Peter and Straub, Julian and Newcombe, Richard and Lovegrove, Steven},
  booktitle={2019 IEEE/CVF conference on computer vision and pattern recognition (CVPR)},
  pages={165--174},
  year={2019},
  organization={IEEE}
}

@article{alfano2006influence,
  title={On the influence of the shape of the interface law on the application of cohesive-zone models},
  author={Alfano, G.},
  journal={Composites Science and Technology},
  volume={66},
  number={6},
  pages={723--730},
  year={2006},
  publisher={Elsevier},
}

@article{fedele2009identification,
  title={Identification of adhesive properties in {GLARE} assemblies using digital image correlation},
  author={Fedele, Roberto and Raka, B. and Hild, Fran{\c{c}}ois and Roux, St{\'e}phane},
  journal={Journal of the Mechanics and Physics of Solids},
  volume={57},
  number={7},
  pages={1003--1016},
  year={2009},
  publisher={Elsevier},
}

@article{ruybalid2018mixed,
  title={Mixed-mode cohesive zone parameters from integrated digital image correlation on micrographs only},
  author={Ruybalid, A. P. and Hoefnagels, J. P. M. and van der Sluis, O. and van Maris, M.P.F.H.L. and Geers, M. G. D.},
  journal={International Journal of Solids and Structures},
  volume={156--157},
  pages={179--193},
  year={2019},
  publisher={Elsevier},
}

@article{richefeu2012toward,
  title={Toward local identification of cohesive zone models using digital image correlation},
  author={Richefeu, Vincent and Chrysochoos, Andr{\'e} and Huon, Vincent and Monerie, Yann and Peyroux, Robert and Wattrisse, Bertrand},
  journal={European Journal of Mechanics -- A/Solids},
  volume={34},
  pages={38--51},
  year={2012},
  publisher={Elsevier},
}

@article{wei2022deep,
  title={Deep-green inversion to extract traction-separation relations at material interfaces},
  author={Wei, Congjie and Zhang, Jiaxin and Liechti, Kenneth M. and Wu, Chenglin},
  journal={International Journal of Solids and Structures},
  volume={250},
  pages={111698},
  year={2022},
  publisher={Elsevier},
}

@article{xue2023jax,
  title={{JAX-FEM}: A differentiable {GPU}-accelerated {3D} finite element solver for automatic inverse design and mechanistic data science},
  author={Xue, Tianju and Liao, Shuheng and Gan, Zhengtao and Park, Chanwook and Xie, Xiaoyu and Liu, Wing Kam and Cao, Jian},
  journal={Computer Physics Communications},
  volume={291},
  pages={108802},
  year={2023},
  publisher={Elsevier},
}

@article{hernandez2017toughness,
  title={Toughness amplification in copper/epoxy joints through pulsed laser micro-machined interface heterogeneities},
  author={Hernandez, Edwin and Alfano, Marco and Pulungan, Ditho and Lubineau, Gilles},
  journal={Scientific reports},
  volume={7},
  number={1},
  pages={16344},
  year={2017},
  publisher={Nature Publishing Group UK London}
}

@article{fulco2026tailoring,
  title={Tailoring toughness and adhesion through architected interfaces},
  author={Fulco, Sage and Budzik, Michal K and Turner, Kevin T},
  journal={npj Metamaterials},
  volume={2},
  number={1},
  pages={5},
  year={2026},
  publisher={Nature Publishing Group UK London}
}

@article{hao2026data,
  title={Data-Driven Cohesive Zone Modeling within the Generalized Standard Materials Framework},
  author={Hao, Sida and Han, Jinkyo and Bahmani, Bahador},
  journal={arXiv preprint arXiv:2609.13629},
  year={2026}
}

@inproceedings{bora2017compressed,
  title={Compressed sensing using generative models},
  author={Bora, Ashish and Jalal, Ajil and Price, Eric and Dimakis, Alexandros G},
  booktitle={International conference on machine learning},
  pages={537--546},
  year={2017},
  organization={PMLR}
}

@article{han2027constitutive,
  title={Constitutive priors for inverse design},
  author={Han, Jinkyo and Bahmani, Bahador},
  journal={Computer Methods in Applied Mechanics and Engineering},
  volume={463},
  pages={119345},
  year={2027},
  publisher={Elsevier}
}

@article{pundir2026versatile,
  title={A versatile FEM framework with native GPU scalability via globally-applied AD},
  author={Pundir, Mohit and Lorez, Flavio and Kammer, David S},
  journal={Computer Methods in Applied Mechanics and Engineering},
  volume={460},
  pages={119118},
  year={2026},
  publisher={Elsevier}
}

@article{chandrasekhar2021auto,
  title   = {{AuTO}: A framework for automatic differentiation in topology optimization},
  author  = {Chandrasekhar, Aaditya and Sridhara, Saketh and Suresh, Krishnan},
  journal = {Structural and Multidisciplinary Optimization},
  volume  = {64},
  pages   = {4355--4365},
  year    = {2021},
  doi     = {10.1007/s00158-021-03025-8}
}

@article{caruana1997multitask,
  title={Multitask learning},
  author={Caruana, Rich},
  journal={Machine learning},
  volume={28},
  number={1},
  pages={41--75},
  year={1997},
  publisher={Springer}
}

@article{arora1977design,
  title={Design sensitivity analysis with substructuring},
  author={Arora, Jasbir S and Govil, Ashok K},
  journal={Journal of the Engineering Mechanics Division},
  volume={103},
  number={4},
  pages={537--548},
  year={1977},
  publisher={American Society of Civil Engineers}
}

@article{bendsoe1988generating,
  title={Generating optimal topologies in structural design using a homogenization method},
  author={Bends{\o}e, Martin Philip and Kikuchi, Noboru},
  journal={Computer methods in applied mechanics and engineering},
  volume={71},
  number={2},
  pages={197--224},
  year={1988},
  publisher={Elsevier}
}

@inproceedings{balandat2020botorch,
    title = {{BoTorch: A Framework for Efficient Monte-Carlo Bayesian Optimization}},
    author = {Balandat, Maximilian and Karrer, Brian and Jiang, Daniel R. and Daulton, Samuel and Letham, Benjamin and Wilson, Andrew Gordon and Bakshy, Eytan},
    booktitle = {Advances in Neural Information Processing Systems 33},
    year = 2020,
    url = {https://proceedings.neurips.cc/paper/2020/hash/f5b1b89d98b7286673128a5fb112cb9a-Abstract.html}
}

@ARTICLE{2020SciPy-NMeth,
  author  = {Virtanen, Pauli and Gommers, Ralf and Oliphant, Travis E. and
            Haberland, Matt and Reddy, Tyler and Cournapeau, David and
            Burovski, Evgeni and Peterson, Pearu and Weckesser, Warren and
            Bright, Jonathan and {van der Walt}, St{\'e}fan J. and
            Brett, Matthew and Wilson, Joshua and Millman, K. Jarrod and
            Mayorov, Nikolay and Nelson, Andrew R. J. and Jones, Eric and
            Kern, Robert and Larson, Eric and Carey, C J and
            Polat, {\.I}lhan and Feng, Yu and Moore, Eric W. and
            {VanderPlas}, Jake and Laxalde, Denis and Perktold, Josef and
            Cimrman, Robert and Henriksen, Ian and Quintero, E. A. and
            Harris, Charles R. and Archibald, Anne M. and
            Ribeiro, Ant{\^o}nio H. and Pedregosa, Fabian and
            {van Mulbregt}, Paul and {SciPy 1.0 Contributors}},
  title   = {{{SciPy} 1.0: Fundamental Algorithms for Scientific
            Computing in Python}},
  journal = {Nature Methods},
  year    = {2020},
  volume  = {17},
  pages   = {261--272},
  adsurl  = {https://rdcu.be/b08Wh},
  doi     = {10.1038/s41592-019-0686-2},
}

@article{gavris2026discovering,
  title={Discovering neural cohesive zone laws from displacement fields},
  author={Gavris, Georgios Barkoulis and Sun, WaiChing},
  journal={Computer Methods in Applied Mechanics and Engineering},
  volume={452},
  pages={118733},
  year={2026},
  publisher={Elsevier}
}

\appendix

\section*{Appendix}

\section{Hyperparameter Settings and Implementation Details}
\label[appendix]{app:hyperparams}

This appendix summarizes the numerical settings for network scaling, constitutive regularization, observation weighting, offline prior training, inverse optimization, and forward-solver stabilization.

\paragraph{Network scaling.}
The ICNN/PICNN and MNN are evaluated using nondimensionalized inputs and outputs to improve numerical conditioning.
Separations are normalized by a reference separation $\delta_{\mathrm{ref}}$, while the active energy and damage-driving force are scaled by reference quantities with units of energy per interface area.
The damage variable remains dimensionless.

For homogeneous identification, these scales are numerical choices determined from prescribed bilinear initialization laws rather than the withheld reference constitutive responses.
The separation and driving-force scales are
\begin{equation}
    \delta_{\mathrm{ref}}=\frac{\delta_n^0}{2},
    \qquad
    Y_{\mathrm{ref}}=\frac{\sigma_n^0\delta_n^0}{4},
\end{equation}
with the energy scale chosen such that the initialized normal stiffness is $2\sigma_n^0/\delta_n^0$.
For the pre-cracked DCB, initial laws are sampled independently for each initialization, excluding candidates within $20\%$ of the reference law to avoid near-truth starting points.
This screening is performed only when constructing the synthetic benchmark initializations.
The observation-loss normalizers, introduced below, are instead evaluated using a fixed off-truth reference law shared across all initializations of a given configuration.

For heterogeneous identification, the scales are fixed during offline prior training and subsequently retained during inversion.
The learned prior uses $\delta_{\mathrm{ref}}=3$~\textmu m and $Y_{\mathrm{ref}}=2.03$~J/m$^2$, with the energy scaling determined during training.

\paragraph{Smoothing of the active separation measures.}
\label{app:fe_smoothing}
The active separation measures in \cref{eq:active_separation} are nonsmooth at $\delta_n=0$ and $\delta_t=0$.
For the coupled PICNN representation, this may introduce discontinuities in the cohesive traction and hinder Newton convergence.
We therefore employ the Huber-type regularization
\begin{equation}
    \boldsymbol\delta_{\mathrm a}^{\kappa}
    =\bigl(h_\kappa(\langle\delta_n\rangle_+),h_\kappa(|\delta_t|)\bigr)^\top,
    \qquad
    h_\kappa(x)=
    \begin{cases}
        x^2/(2\kappa_\delta), & 0\leq x\leq\kappa_\delta,\\
        x-\kappa_\delta/2, & x>\kappa_\delta,
    \end{cases}
    \label{eq:psi_blend}
\end{equation}
with $\kappa_\delta=10^{-3}\delta_{\mathrm{ref}}$.
The regularized measures are continuously differentiable and preserve convexity, while their derivatives vanish at zero.
Outside the smoothing interval, they differ from the original measures only by the constant offset $\kappa_\delta/2$.
The separable energy representation in \cref{eq:psi_decoupled} does not require this regularization because its individual branches have zero slope at the origin.

\paragraph{Damage-onset regularization.}
Let $\widetilde R(Y,\beta;\boldsymbol z)$ denote the raw MNN output before enforcing the admissible damage interval.
Hard clipping to $[0,1]$ produces zero derivatives when $\widetilde R<0$, which can hinder gradient-based identification near damage initiation.
To retain a positive derivative in this regime, we introduce the smooth lower-clamp approximation
\begin{equation}
    R_\kappa(Y,\beta;\boldsymbol z)=
    \begin{cases}
        \dfrac{\kappa_d^2}{2\kappa_d-\widetilde R},
        & \widetilde R<\kappa_d,\\[2mm]
        \min\{\widetilde R,1\},
        & \widetilde R\geq\kappa_d,
    \end{cases}
    \qquad \kappa_d=0.01.
    \label{eq:onset_floor}
\end{equation}
The branches are continuously differentiable at $\widetilde R=\kappa_d$, and $\partial R_\kappa/\partial\widetilde R>0$ below this threshold.
The upper clamp remains unchanged, and $R_\kappa$ approaches the hard-clipped map as $\kappa_d\to0$.
When used, this approximation replaces $R$ in \cref{eq:damage_target}, while the irreversible history update and numerical damage relaxation remain unchanged.

The regularization is applied during offline prior training and homogeneous identification, whereas heterogeneous identification uses the frozen prior with hard clipping.
For the reported networks, $\widetilde R(0,\beta;\boldsymbol z)<0$, so the smooth clamp produces a small positive damage value at zero driving force,
$R_\kappa(0,\beta;\boldsymbol z)<\kappa_d/2$.
Consequently, the initial stiffness is reduced by less than $0.5\%$, representing a numerical approximation of the exact condition in \cref{eq:mnn_constraints}.
Switching from the smooth clamp during training to hard clipping during heterogeneous identification changes the damage prediction near onset by at most $\kappa_d/2$.

\paragraph{Observation normalization and weighting.}
Following \cref{eq:general_min_obj}, $\mathrm G$ denotes global force measurements and $\mathrm F$ denotes full-field strain measurements.
Let $\mathcal E_c$ be the total misfit of channel $c$ over the loading histories, and $\mathcal E_c^0>0$ its value at a prescribed reference state.
The normalized objective is
\begin{equation}
    \mathcal J
    =a_{\mathrm G}\frac{\mathcal E_{\mathrm G}}{\mathcal E_{\mathrm G}^0}
    +a_{\mathrm F}\frac{\mathcal E_{\mathrm F}}{\mathcal E_{\mathrm F}^0},
    \qquad
    a_{\mathrm G},a_{\mathrm F}\geq0,
    \quad a_{\mathrm G}+a_{\mathrm F}=1.
    \label{eq:normalized_channel_objective}
\end{equation}
The corresponding weights in \cref{eq:general_min_obj} are
\begin{equation}
    w_c=\frac{a_c}{\mathcal E_c^0},
    \qquad c\in\{\mathrm G,\mathrm F\},
    \qquad
    A=\frac{a_{\mathrm F}}{a_{\mathrm G}},
    \label{eq:channel_weights}
\end{equation}
where $A$ denotes the relative weight for combined observations.

For homogeneous identification, the reference misfits $\mathcal E_c^0$ are evaluated using a prescribed bilinear law with parameters different from those of the generating model.
For heterogeneous identification, they are evaluated using the uniform initial latent field.
In either case, the reference state is fixed across initializations, ensuring comparable normalized objectives within each observation configuration.

For single-channel identification, the active channel receives unit weight.
For combined identification, we use $A=1$, corresponding to equal weights on the normalized channel misfits, except for the three-fiber composite in \cref{sec:meta_dd_comp3fiber}, where $A=50$ emphasizes the strain observations.
The relative weighting does not directly determine the influence of individual measurements, which also depends on their normalization and aggregation.

\paragraph{Offline prior training.}
The shared network parameters $\boldsymbol\Theta$ and reference latent coordinates $\{\boldsymbol z_m\}$ are optimized jointly using \cref{eq:meta_loss_gauge}.
Each training history is evaluated sequentially to account for damage evolution.
The prior in \cref{sec:meta_data_driven} is trained on nine paired Mode~I and Mode~II cohesive behaviors (18 histories) for $3{,}000$ Adam epochs.
Constant learning rates of $0.003$ and $0.036$ are used for the network parameters and latent coordinates, respectively.
The latent regularization scale is $\sigma_z=25.5$, and the global gradient-clipping norm is $10$.
Following training, the shared network parameters are fixed during inverse identification.

\paragraph{Online optimization.}
Inverse identification uses Adam~\citep{diederik2014adam} followed by L-BFGS refinement~\citep{liu1989limited}.
The optimization settings are summarized in \cref{tab:identification_settings}.
Unless otherwise specified, the Adam learning rate follows cosine annealing with warm restarts~\citep{loshchilov2016sgdr}, with each cycle spanning half the Adam budget and $\eta_{\min}=0.2\eta_0$.
The homogeneous composite uses $\eta_{\min}=0.03\eta_0$.

For homogeneous identification, the learning rate is increased linearly during the first ten Adam iterations to reduce the risk of premature convergence to an undamaged response.
The heterogeneous bilinear baseline uses $\eta_0=0.03$, rather than $0.1$ for the learned prior, because the larger initial step leads to poor optimization progress.
Global gradient clipping~\citep{pascanu2013difficulty} is applied during both Adam and L-BFGS optimization using the thresholds in \cref{tab:identification_settings}.
L-BFGS employs a history size of 50 and a strong-Wolfe line search, with gradient and step tolerances of $10^{-12}$ and $10^{-14}$, respectively.

\begin{table}[htbp]
    \centering
    \small
    \caption{Online identification settings.
    $\eta_0$ denotes the initial Adam learning rate.
    Epoch counts are outer optimization iterations; L-BFGS may perform multiple objective evaluations per epoch.}
    \label{tab:identification_settings}
    \begin{tabular}{p{0.43\textwidth}rrrr}
        \toprule
        Example
        & \shortstack{Adam\\epochs}
        & $\eta_0$
        & \shortstack{L-BFGS\\epochs}
        & \shortstack{Clipping\\norm}\\
        \midrule
        Homogeneous DCB
            & 500 & 0.03 & 25 & 1\\
        Homogeneous composite
            & 500 & 0.01 & 30 & 1\\
        Heterogeneous DCB with prior
            & 300 & 0.10 & 20 & 10\\
        Three-fiber composite with prior
            & 150 & 0.10 & 20 & 10\\
        \bottomrule
    \end{tabular}
\end{table}

Newton iterations employ a backtracking line search with up to six step halvings and a convergence tolerance of $10^{-5}$~N/m for the reduced residual, including stabilization.
If convergence is not achieved, the load increment is reduced by half, subject to the cutback limits in \cref{tab:solver_settings}.
The maximum Newton iterations and cutbacks are adjusted according to the numerical difficulty of each example.
The corresponding L-BFGS settings are also summarized in the table.

\begin{table}[htbp]
    \centering
    \small
    \caption{L-BFGS and Newton solver settings.
    The L-BFGS step denotes the initial trial step in the line search.}
    \label{tab:solver_settings}
    \begin{tabular}{p{0.40\textwidth}rrrr}
        \toprule
        Example
        & \shortstack{L-BFGS\\step}
        & \shortstack{L-BFGS\\iter./epoch}
        & \shortstack{Newton\\iter.}
        & \shortstack{Max.\\cutbacks}\\
        \midrule
        Homogeneous DCB
            & 0.03 & 20 & 120 & 10\\
        Homogeneous composite
            & 0.01 & 20 & 120 & 10\\
        Heterogeneous DCB with prior
            & 1 & 5 & 120 & 15\\
        Three-fiber composite with prior and bilinear baseline
            & 1 & 5 & 40 & 5\\
        \bottomrule
    \end{tabular}
\end{table}

The homogeneous examples use five initializations for each observation configuration.
For the heterogeneous DCB in \cref{sec:meta_dd_dcb}, 23 latent coordinates corresponding to the cohesive elements traversed by the crack front and a few elements ahead of it are optimized, initialized uniformly at $z_e=0.1$, with Adam momentum parameter $\beta_1=0.5$.
The remaining interface coordinates are held fixed.
For the three-fiber composite, all 144 latent coordinates are initialized uniformly at zero.

\paragraph{Forward-solver stabilization.}
The numerical stabilization settings are summarized in \cref{tab:stabilization_settings}.
The pseudo-time $\tau$ in \cref{eq:viscous_damage_main,eq:reduced_residual} is parameterized by the prescribed loading displacement $U$.
Thus, $\Delta\tau_n=|U_n-U_{n-1}|$ denotes the applied displacement increment, including any reduced increment following load-step cutbacks.
Consequently, $\tau_{\mathrm v}$ is a numerical regularization length rather than a physical relaxation time.
With forces expressed per unit out-of-plane thickness, $c_u$ has units of N/m and $\gamma_n=c_u/\Delta\tau_n$ has units of N/m$^2$.

The discrete adjoint is derived from the same stabilized equilibrium and constitutive-update equations used in the forward solver, including the displacement-stabilization contribution and the damage-relaxation factor $\rho_n$.

\begin{table}[htbp]
    \centering
    \small
    \caption{Forward-solver stabilization parameters in
    \cref{eq:viscous_damage_main,eq:reduced_residual}.}
    \label{tab:stabilization_settings}
    \begin{tabular}{lrr}
        \toprule
        Example & $\tau_{\mathrm v}$ (m) & $c_u$ (N/m)\\
        \midrule
        Homogeneous DCB
            & $2\times10^{-8}$ & 0.2\\
        Homogeneous composite
            & $3\times10^{-8}$ & 5.0\\
        Heterogeneous DCB with prior
            & $5\times10^{-8}$ & 0.5\\
        Three-fiber composite with prior
            & $10^{-7}$ & 1.0\\
        \bottomrule
    \end{tabular}
\end{table}

\section{Sensitivity of the Structural Response to Cohesive Strength}
\label{app:precrack_sensitivity}

To investigate how the cohesive process-zone size influences the information contained in the global force response, we perform additional forward simulations for the DCB configurations considered in \cref{sec:precrack}.
Within each configuration, the cohesive peak traction $\sigma_n^0$ is varied while the initial stiffness $k_n$ and fracture energy $G_{Ic}$ are held fixed.
Consequently, the cohesive laws have the same initial slope and fracture energy but differ in their peak traction, failure separation, and softening behavior.
Four normalized peak tractions, $\sigma_n^0/\bar{\sigma}_n^0 \in \{0.40,0.50,0.71,1.00\}$, are considered, where $\bar{\sigma}_n^0$ denotes the baseline cohesive strength of each configuration.

\begin{figure}[htbp]
    \centering
    \begin{subfigure}[b]{0.34\textwidth}
        \centering
        \includegraphics[width=\textwidth]{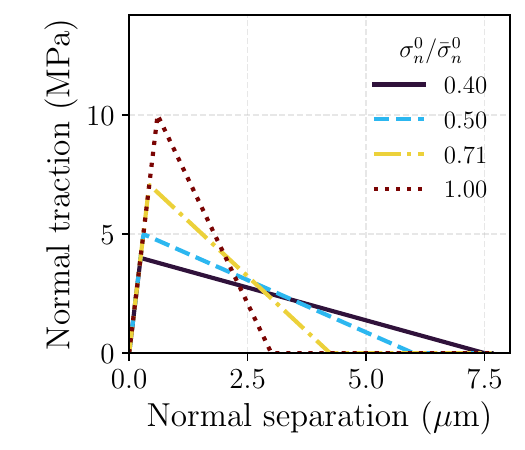}
        \caption{}
        \label{fig:precrack_sens_ssb_ts}
    \end{subfigure}
    \hspace{0.04\textwidth}
    \begin{subfigure}[b]{0.34\textwidth}
        \centering
        \includegraphics[width=\textwidth]{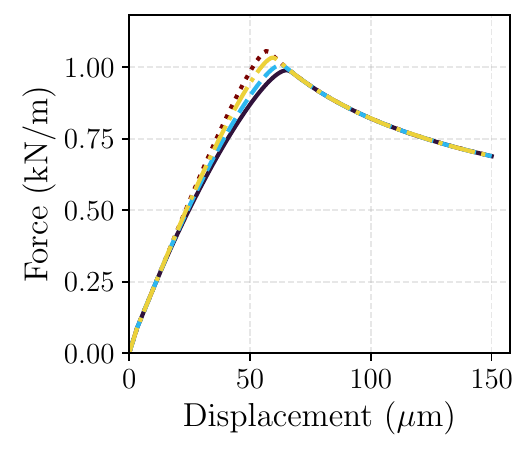}
        \caption{}
        \label{fig:precrack_sens_ssb_fd}
    \end{subfigure}

    \vspace{0.02\textwidth}

    \begin{subfigure}[b]{0.34\textwidth}
        \centering
        \includegraphics[width=\textwidth]{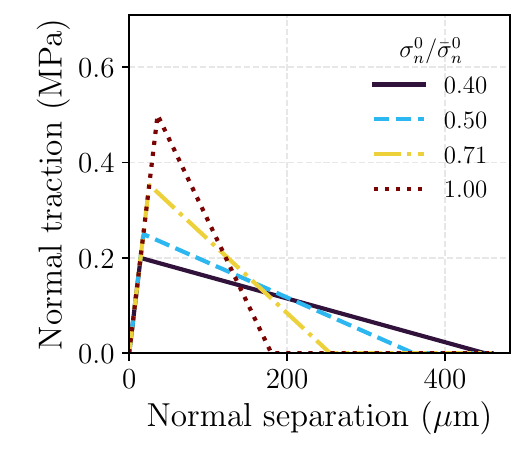}
        \caption{}
        \label{fig:precrack_sens_lsb_ts}
    \end{subfigure}
    \hspace{0.04\textwidth}
    \begin{subfigure}[b]{0.34\textwidth}
        \centering
        \includegraphics[width=\textwidth]{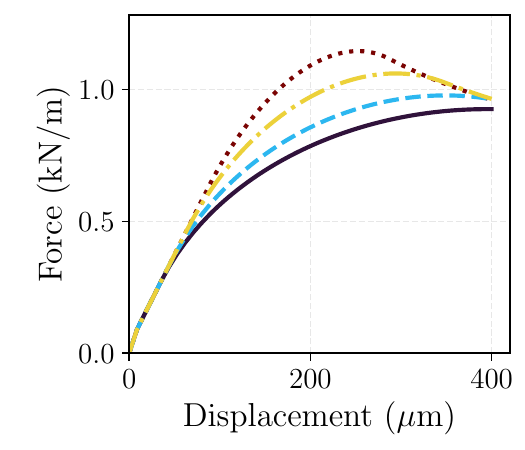}
        \caption{}
        \label{fig:precrack_sens_lsb_fd}
    \end{subfigure}

    \caption{Sensitivity of the DCB response to cohesive peak traction at fixed fracture energy and initial stiffness.
    Panels (a) and (b) show the traction--separation laws and corresponding force--displacement responses for the SSB configuration, while panels (c) and (d) present the corresponding results for the LSB configuration.
    The peak traction is normalized by its baseline value $\bar{\sigma}_n^0$ for each configuration.}
    \label{fig:precrack_sensitivity}
\end{figure}

The resulting traction--separation laws and force--displacement responses are compared in \cref{fig:precrack_sensitivity}.
For the SSB configuration, the force responses exhibit relatively small differences near the peak and nearly coincide during subsequent crack propagation, despite substantial differences in the underlying cohesive laws.
In contrast, the LSB configuration exhibits more pronounced differences in the force response over a broader portion of the loading history.
These results indicate that the global force response is less sensitive to variations in cohesive strength in the shorter-zone configuration, whereas the longer-zone configuration provides greater sensitivity to the constitutive response within the investigated family of cohesive laws.

\section{Reference Cohesive Models}
\label[appendix]{app:reference_czm}

This appendix summarizes the analytical cohesive laws used to generate the homogeneous reference responses 
of the DCB and composite examples in \cref{sec:examples:homo}.
The bilinear law also serves as the conventional cohesive-model baseline in \cref{sec:meta_dd_comparison}.

We adopt the local separations and tractions defined in \cref{eq:local_separation,eq:cohesive_components}, the positive-part operator of \cref{eq:active_separation}, and the damage variable $d\in[0,1]$.
Defining the remaining stiffness fraction as $g=1-d$, both cohesive laws employ the traction relations
\begin{equation}
    t_n=g\,k_n\langle\delta_n\rangle_+
        -k_{\mathrm{pen}}\langle-\delta_n\rangle_+,
    \qquad
    t_t=g\,k_t\delta_t,
    \label{eq:ref_common_tractions}
\end{equation}
where $k_n,k_t>0$ are the undamaged normal and tangential stiffnesses.
The compressive contact response remains unaffected by damage, consistent with \cref{eq:traction_state_law}.
The peak normal and tangential tractions are denoted by $\sigma_n^0$ and $\sigma_t^0$, while $\delta_i^0$ and $\delta_i^f$ denote the corresponding pure-mode damage-initiation and complete-failure separations for $i\in\{n,t\}$.

\subsection{Mixed-Mode Bilinear Law}
\label[appendix]{app:ref_bilinear}

For the bilinear model, the pure-mode fracture energies $G_{Ic}$ and $G_{IIc}$ correspond to the areas under the Mode~I and Mode~II traction--separation curves.
The corresponding failure separations are therefore
\begin{equation}
    \delta_n^f=\frac{2G_{Ic}}{\sigma_n^0},
    \qquad
    \delta_t^f=\frac{2G_{IIc}}{\sigma_t^0}.
    \label{eq:ref_bilinear_deltac}
\end{equation}
A common initiation-to-failure separation ratio $\lambda_c\in(0,1)$ is prescribed for both modes, giving
\begin{equation}
    \delta_n^0=\lambda_c\delta_n^f,
    \qquad
    \delta_t^0=\lambda_c\delta_t^f,
    \qquad
    k_n=\frac{\sigma_n^0}{\delta_n^0},
    \qquad
    k_t=\frac{\sigma_t^0}{\delta_t^0}.
    \label{eq:ref_bilinear_t}
\end{equation}
Thus, the initial stiffnesses are determined by the fracture energies, peak tractions, and $\lambda_c$, rather than treated as independent parameters.

Mixed-mode damage evolution is governed by the dimensionless effective separation
\begin{equation}
    \lambda(\boldsymbol\delta)
    =\sqrt{
        \left(\frac{\langle\delta_n\rangle_+}{\delta_n^f}\right)^2
        +\left(\frac{\delta_t}{\delta_t^f}\right)^2
    },
    \qquad
    \lambda_{\max}(\tau)
    =\max_{0\leq s\leq\tau}
        \lambda\bigl(\boldsymbol\delta(s)\bigr).
    \label{eq:ref_bilinear_lambda}
\end{equation}
Damage initiates when $\lambda_{\max}=\lambda_c$ and reaches completion at $\lambda_{\max}=1$.
The remaining stiffness fraction is defined as
\begin{equation}
    g(\lambda_{\max})=
    \begin{cases}
        1,
        & \lambda_{\max}\leq\lambda_c,\\[4pt]
        \dfrac{\lambda_c(1-\lambda_{\max})}
              {(1-\lambda_c)\lambda_{\max}},
        & \lambda_c<\lambda_{\max}<1,\\[8pt]
        0,
        & \lambda_{\max}\geq1.
    \end{cases}
    \label{eq:ref_bilinear_g}
\end{equation}
Because $\lambda_{\max}$ is non-decreasing throughout loading, the damage variable $d=1-g$ evolves irreversibly.
Unloading and reloading below the previously attained maximum occur at fixed damage according to \cref{eq:ref_common_tractions}.

\subsection{Mixed-Mode Law with Exponential Softening}
\label[appendix]{app:ref_exponential}

The second reference model combines an initial linear elastic response with a truncated exponential softening branch, following the displacement-based exponential degradation form commonly used in cohesive-zone modeling.
Unlike the bilinear model, the initial stiffnesses $k_n$ and $k_t$ are prescribed independently, giving
\begin{equation*}
    \delta_n^0=\frac{\sigma_n^0}{k_n},
    \qquad
    \delta_t^0=\frac{\sigma_t^0}{k_t}.
\end{equation*}

Mixed-mode evolution is described using the stiffness-weighted effective separation $\delta_m$ and the corresponding elastic energy-based mode fraction $\beta_e$, defined as
\begin{equation}
    \delta_m
    =\sqrt{\langle\delta_n\rangle_+^2
           +\frac{k_t}{k_n}\delta_t^2},
    \qquad
    \beta_e
    =\frac{k_t\delta_t^2}
           {k_n\langle\delta_n\rangle_+^2+k_t\delta_t^2}.
    \label{eq:ref_exp_deltam}
\end{equation}
We set $\beta_e=0$ when the active separation vanishes.
Unlike the separation-based mode mixity $\beta$ introduced in \cref{eq:mode_mixity}, $\beta_e$ accounts for the relative contributions of the normal and tangential elastic energies.
For nonzero active separation, the two measures are related by
\begin{equation}
    \beta_e
    =\frac{k_t\beta^2}
           {k_n(1-\beta)^2+k_t\beta^2}.
    \label{eq:ref_mixity_relation}
\end{equation}

Damage initiation follows the quadratic nominal-traction criterion
\begin{equation}
    \left(
        \frac{k_n\langle\delta_n\rangle_+}{\sigma_n^0}
    \right)^2
    +
    \left(
        \frac{k_t\delta_t}{\sigma_t^0}
    \right)^2
    =1,
    \label{eq:ref_exp_initiation}
\end{equation}
which excludes compressive normal contact.
The corresponding mixed-mode initiation and failure separations are
\begin{equation}
    \delta_m^0(\beta_e)
    =\left[
        \frac{1-\beta_e}{(\delta_n^0)^2}
        +\frac{k_n\beta_e}{k_t(\delta_t^0)^2}
    \right]^{-1/2},
    \qquad
    \delta_m^f(\beta_e)
    =\frac{2G_{\mathrm c}^{\mathrm{tri}}(\beta_e)}
           {k_n\delta_m^0(\beta_e)},
    \label{eq:ref_exp_envelopes}
\end{equation}
with $\delta_m^f>\delta_m^0$.
The failure envelope is constructed using a Benzeggagh--Kenane-type interpolation \citep{benzeggagh1996measurement},
\begin{equation}
    G_{\mathrm c}^{\mathrm{tri}}(\beta_e)
    =G_{Ic}^{\mathrm{tri}}
    +\bigl(G_{IIc}^{\mathrm{tri}}-G_{Ic}^{\mathrm{tri}}\bigr)
        \beta_e^\eta,
    \qquad
    G_{Ic}^{\mathrm{tri}}
    =\frac12\sigma_n^0\delta_n^f,
    \qquad
    G_{IIc}^{\mathrm{tri}}
    =\frac12\sigma_t^0\delta_t^f.
    \label{eq:ref_exp_nominal_energy}
\end{equation}
The superscript $\mathrm{tri}$ denotes the area of a reference triangular traction--separation envelope with the same peak traction and failure separation.
These quantities determine the mixed-mode failure envelope but should not be interpreted as the actual fracture energies of the exponential law.

The history-dependent effective separation and normalized softening coordinate are defined as
\begin{equation}
    \delta_m^{\max}(\tau)
    =\max_{0\leq s\leq\tau}\delta_m(s),
    \qquad
    x
    =\frac{\delta_m^{\max}-\delta_m^0}
           {\delta_m^f-\delta_m^0}.
    \label{eq:ref_exp_history}
\end{equation}
For a softening parameter $\alpha>0$, the remaining stiffness fraction follows
\begin{equation}
    g=
    \begin{cases}
        1,
        & \delta_m^{\max}\leq\delta_m^0,\\[4pt]
        \dfrac{\delta_m^0}{\delta_m^{\max}}
        \dfrac{e^{-\alpha x}-e^{-\alpha}}
              {1-e^{-\alpha}},
        & \delta_m^0<\delta_m^{\max}<\delta_m^f,\\[8pt]
        0,
        & \delta_m^{\max}\geq\delta_m^f.
    \end{cases}
    \label{eq:ref_exp_g}
\end{equation}
The corresponding tractions are obtained from \cref{eq:ref_common_tractions}.
For fixed mode mixity, this formulation ensures irreversible damage evolution through the history maximum $\delta_m^{\max}$.
In the limit $\alpha\to0$, the exponential softening factor approaches $1-x$, recovering a linear softening branch with the same initiation and failure separations.

An important distinction from the bilinear model concerns the actual work of separation.
For monotonic proportional loading at fixed $\beta_e$, integration of the traction--separation response gives
\begin{equation}
    G_{\mathrm c}^{\mathrm{exp}}(\beta_e)
    =k_n\delta_m^0
    \left[
        \frac{\delta_m^0}{2}
        +(\delta_m^f-\delta_m^0)C_\alpha
    \right],
    \qquad
    C_\alpha
    =\frac{1}{\alpha}-\frac{1}{e^\alpha-1}.
    \label{eq:ref_exp_work}
\end{equation}
Since $C_\alpha<1/2$ for $\alpha>0$, the actual work of separation is smaller than the corresponding triangular reference area $G_{\mathrm c}^{\mathrm{tri}}$.
As $\alpha\to0$, $C_\alpha\to1/2$, and the triangular area is recovered.
For the parameters used in \cref{sec:composite}, the nominal triangular areas are $50$ and $72$~J/m$^2$, whereas the exponential laws with $\alpha=4$ yield actual pure-mode fracture energies of approximately $25.8$ and $37.2$~J/m$^2$, respectively.

Finally, the mixed-mode history dependence requires a qualification.
Following the formulation of \citet{camanho2003numerical}, the initiation and failure envelopes depend on the current mode fraction $\beta_e$, while $\delta_m^{\max}$ is the sole history variable.
The corresponding initiation criterion and envelopes reduce to the classical formulation when $k_n=k_t$.
Although this construction ensures irreversible damage under fixed mode mixity, changes in $\beta_e$ during softening may cause the calculated damage variable to decrease.
Consequently, irreversible evolution and thermodynamic consistency are not guaranteed for arbitrary nonproportional mixed-mode loading \citep{turon2006damage}.

For the loading histories considered here, the variation in $\beta_e$ during softening remains below $0.05$, and the damage variable was verified to be non-decreasing at every integration point and load step.
Explicitly enforcing non-decreasing damage would therefore leave the reported reference responses unchanged.

\subsection{Finite Element Operators and Assembly}
\label[appendix]{app:fe_operators}

We employ a two-dimensional, small-strain, plane-strain finite element formulation, with forces expressed per unit out-of-plane thickness.
The bulk domain $\Omega\setminus\Gamma_{\mathrm{c}}$ is discretized using four-node bilinear quadrilateral elements with $2\times2$ Gauss quadrature, while the interfaces are represented by four-node zero-thickness cohesive elements with two Gauss points.
Each node has two displacement degrees of freedom.

For a bulk element, let $\boldsymbol{u}_e\in\mathbb{R}^{8}$ denote the element displacement vector.
Using engineering shear strain, the strain--displacement relation is
\begin{equation}
    \boldsymbol{\varepsilon}
    =
    \begin{pmatrix}
        \varepsilon_{xx} &
        \varepsilon_{yy} &
        \gamma_{xy}
    \end{pmatrix}^{\top}
    =
    \boldsymbol{B}_{b}\boldsymbol{u}_e,
    \qquad
    \gamma_{xy}=2\varepsilon_{xy}.
\end{equation}
For isotropic linear elasticity under plane strain, the constitutive matrix is
\begin{equation}
    \boldsymbol{C}_e
    =
    \frac{E_e}
         {(1+\nu_e)(1-2\nu_e)}
    \begin{bmatrix}
        1-\nu_e & \nu_e & 0\\
        \nu_e & 1-\nu_e & 0\\
        0 & 0 & \dfrac{1-2\nu_e}{2}
    \end{bmatrix},
\end{equation}
where $E_e$ and $\nu_e$ denote the elastic properties of the corresponding bulk material.
The element stiffness matrix is obtained through numerical integration as
\begin{equation}
    \boldsymbol{k}^{b}_{e}
    =
    \sum_{g}
    w_g\,
    \det\boldsymbol{J}_{e,g}\,
    \boldsymbol{B}_{b,e,g}^{\top}
    \boldsymbol{C}_e
    \boldsymbol{B}_{b,e,g},
    \label{eq:bulk_stiffness}
\end{equation}
where $\boldsymbol{J}_{e,g}$ is the isoparametric Jacobian and $w_g$ is the quadrature weight.
Since the bulk response is linear elastic under the small-strain assumption, its stiffness matrix is independent of the displacement field and cohesive constitutive parameters and can be assembled once.

For a cohesive element, the linear shape functions along the interface are
\begin{equation}
    N_1(\xi)=\frac{1-\xi}{2},
    \qquad
    N_2(\xi)=\frac{1+\xi}{2}.
\end{equation}
We number the nodes on the two opposing faces in opposite directions, with nodes $1$ and $4$ forming one coincident pair and nodes $2$ and $3$ forming the other.
The displacement jump between the opposing faces, expressed in the global coordinate system, is
\begin{equation}
    \llbracket\boldsymbol{u}\rrbracket
    =
    \boldsymbol{B}_{\mathrm{jmp}}(\xi)\boldsymbol{u}_e,
\end{equation}
where
\begin{equation}
    \boldsymbol{B}_{\mathrm{jmp}}(\xi)
    =
    \begin{bmatrix}
        -N_1 & 0 & -N_2 & 0 & N_2 & 0 & N_1 & 0\\
        0 & -N_1 & 0 & -N_2 & 0 & N_2 & 0 & N_1
    \end{bmatrix}.
    \label{eq:cohesive_jump_operator}
\end{equation}
The cohesive constitutive response is evaluated in the local normal--tangential coordinate system.
Introducing the rotation matrix
\begin{equation}
    \boldsymbol{Q}_e
    =
    \begin{bmatrix}
        \boldsymbol{e}_n^{\top}\\
        \boldsymbol{e}_t^{\top}
    \end{bmatrix},
\end{equation}
the local separation vector follows as
\begin{equation}
    \boldsymbol{\delta}^{\mathrm{loc}}
    =
    \begin{pmatrix}
        \delta_n\\
        \delta_t
    \end{pmatrix}
    =
    \boldsymbol{Q}_e
    \boldsymbol{B}_{\mathrm{jmp}}
    \boldsymbol{u}_e
    =
    \boldsymbol{B}_{\mathrm{c},e}
    \boldsymbol{u}_e.
    \label{eq:cohesive_rotation}
\end{equation}
Under the small-deformation assumption, the local frame is defined in the reference configuration and remains fixed throughout loading.

For a cohesive element of reference length $\ell_e$, two-point Gauss quadrature gives the internal force vector
\begin{equation}
    \boldsymbol{f}_{e}^{c}
    =
    \frac{\ell_e}{2}
    \sum_g
    w_g\,
    \boldsymbol{B}_{\mathrm{c},e,g}^{\top}
    \boldsymbol{t}_{\mathrm{c},g}^{\mathrm{loc}},
    \label{eq:cohesive_fint}
\end{equation}
where the local traction at each integration point is evaluated as
\begin{equation}
    \boldsymbol{t}_{\mathrm{c},g}^{\mathrm{loc}}
    =
    \boldsymbol{t}_{\mathrm{c}}^{\mathrm{loc}}
    \left(
        \boldsymbol{\delta}_{g}^{\mathrm{loc}},
        d_g;
        \boldsymbol{z}_{e}
    \right).
\end{equation}
Here, $d_g$ denotes the local damage state and $\boldsymbol{z}_e$ represents the constitutive parameters or latent coordinates associated with the cohesive element.
The corresponding consistent element tangent is
\begin{equation}
    \boldsymbol{k}_{e}^{c}
    =
    \frac{\ell_e}{2}
    \sum_g
    w_g\,
    \boldsymbol{B}_{\mathrm{c},e,g}^{\top}
    \boldsymbol{D}_{\mathrm{alg},g}
    \boldsymbol{B}_{\mathrm{c},e,g},
    \label{eq:cohesive_ktan}
\end{equation}
where $\boldsymbol{D}_{\mathrm{alg},g}$ is the consistent algorithmic tangent defined in \cref{eq:cohesive_tangent_main}, accounting for the constitutive state update.

Finally, let $\boldsymbol{L}_e\in\{0,1\}^{8\times N_{\mathrm{dof}}}$ denote the Boolean connectivity matrix that extracts the element displacement vector from the global displacement vector $\boldsymbol{U}$ through $\boldsymbol{u}_e=\boldsymbol{L}_e\boldsymbol{U}$,
and let $\mathcal E_{\mathrm{bulk}}$ and $\mathcal E_{\mathrm{coh}}$ denote the sets of bulk and cohesive elements.
The assembled internal force vector is
\begin{equation}
    \boldsymbol{f}_{\mathrm{int}}
    =
    \sum_{e\in\mathcal{E}_{\mathrm{bulk}}}
    \boldsymbol{L}_e^{\top}
    \boldsymbol{k}_{e}^{b}
    \boldsymbol{L}_e
    \boldsymbol{U}
    \nonumber
    +
    \sum_{e\in\mathcal{E}_{\mathrm{coh}}}
    \boldsymbol{L}_e^{\top}
    \boldsymbol{f}_{e}^{c},
    \label{eq:global_fint}
\end{equation}
with the corresponding global tangent matrix
\begin{equation}
    \boldsymbol{K}
    =
    \sum_{e\in\mathcal{E}_{\mathrm{bulk}}}
    \boldsymbol{L}_e^{\top}
    \boldsymbol{k}_{e}^{b}
    \boldsymbol{L}_e
    +
    \sum_{e\in\mathcal{E}_{\mathrm{coh}}}
    \boldsymbol{L}_e^{\top}
    \boldsymbol{k}_{e}^{c}
    \boldsymbol{L}_e.
    \label{eq:global_tangent}
\end{equation}

For computational efficiency, cohesive constitutive responses and their consistent tangents are evaluated in batches across all interface quadrature points.
The corresponding element contributions are assembled using vectorized scatter operations with precomputed connectivity arrays, avoiding explicit Python-level loops over cohesive elements.

\end{document}